\documentclass[twoside,12pt]{article}%
\usepackage{epsfig}
\usepackage{mathrsfs}
\usepackage{epsf}
\usepackage{array}
\usepackage{subfigure}
\usepackage{amsmath}
\usepackage{amssymb}
\usepackage[american]{babel}
\usepackage{journals}
\usepackage{eso-pic}
\usepackage{dcolumn}
\usepackage[pagewise, mathlines, displaymath]{lineno}
\usepackage{amsfonts}
\usepackage{graphicx}%
\providecommand{\U}[1]{\protect\rule{.1in}{.1in}}
\def\Journal#1#2#3#4{{#1} {#2} (#4) #3 }

\def\NPA{{\em Nucl. Phys.} A}

\def\PLB{{\em Phys. Lett.} B}

\def\PRL{\em Phys. Rev. Lett.}

\def\PREP{\em Phys. Rep.}

\def\PRD{{\em Phys. Rev.} D}
\def\PRC{{\em Phys. Rev.} C}

\newcommand{\be}{\begin{equation}}
\newcommand{\ee}{\end{equation}}
\newcommand{\bea}{\begin{eqnarray}}
\newcommand{\eea}{\end{eqnarray}}

\begin{document}

\title{ \vspace{1cm} Jets in Hadron-Hadron Collisions}
\author{S.\ D.\ Ellis,$^{1}$ 
  J.\ Huston,$^{2}$ 
  K.\ Hatakeyama,$^{3}$ 
  P.\ Loch,$^{4}$
  M.\ T\"onnesmann,$^{5}$\\ \\
  $^{1}$University of Washington, Seattle, Washington 98195\\
  $^{2}$Michigan State University, East Lansing, Michigan 48824\\
  $^{3}$Rockefeller University, New York, New York 10021\\
  $^{4}$University of Arizona, Tucson, Arizona 85721\\
  $^{5}$Max Planck Institute fur Physics, Munich, Germany\\
}
\maketitle

\begin{abstract}
In this article, we review some of the complexities of jet algorithms and of
the resultant comparisons of data to theory. 
We review the extensive experience with jet measurements at
the Tevatron, the extrapolation of this acquired wisdom to the LHC and the
differences between the Tevatron and LHC environments.
We also describe a framework (SpartyJet) for the convenient comparison
of results using different jet algorithms. 
\end{abstract}
\tableofcontents


\section{Introduction}

\label{sec:intro}

Most of the interesting physics signatures at the Tevatron and LHC involve
final states with jets of hadrons. A jet is reconstructed from energy
depositions in calorimeter cells and/or from charged particle track momenta,
and ideally is corrected for detector response and resolution effects so that
the resultant 4-vector corresponds to that of the sum of the original hadrons
comprising the jet. The jets can also be further corrected, for hadronization
effects, back to the parton(s) from which the jet originated. The resultant
measurements can be compared to predictions from parton shower Monte Carlos.
\ If further corrections are made to account for the showers, or if these
corrections are assumed to be small, comparisons can be made directly to the
short distance partons described by fixed order perturbative calculations.

In order to actually reconstruct a jet, and make comparisons between data
and theoretical predictions, a precise definition of the jet is
required.  The definition is presented in the form of a jet algorithm.
Jet algorithms
cluster partons, or particles or calorimeter towers based on proximity in
coordinate space (as for example in cone algorithms) or proximity in momentum
space (as for example in $k_{T}$ algorithms). For a precise comparison of
experiment to theory, it is advantageous for a jet algorithm to provide a
similar description of a hard scatter event regardless if it is applied at the
detector, hadron or parton level.

In this article, we will review some of the complexities of jet algorithms
and of the resultant comparisons of data to theory. We will review the
extensive experience with jet measurements at the Tevatron,
the extrapolation of this acquired wisdom to the LHC and the
differences between the Tevatron and LHC environments.
We will also discuss ways in which the jet algorithm systematics can
be reduced to the percent level, an important goal for LHC
analyses. Finally we will describe a framework (SpartyJet) for the
convenient comparison of results using different jetalgorithms.
Several of the authors are members of CDF and ATLAS and we apologize
in advance for our concentration on those experiments. Given the
restrictions of space, we will not try a comprehensive review of
physics with jets at hadron-hadron colliders, but instead refer the
reader to a recent review~\cite{Campbell:2006wx}.

\begin{figure}[th]
\centering
\includegraphics[width=.60\hsize]
{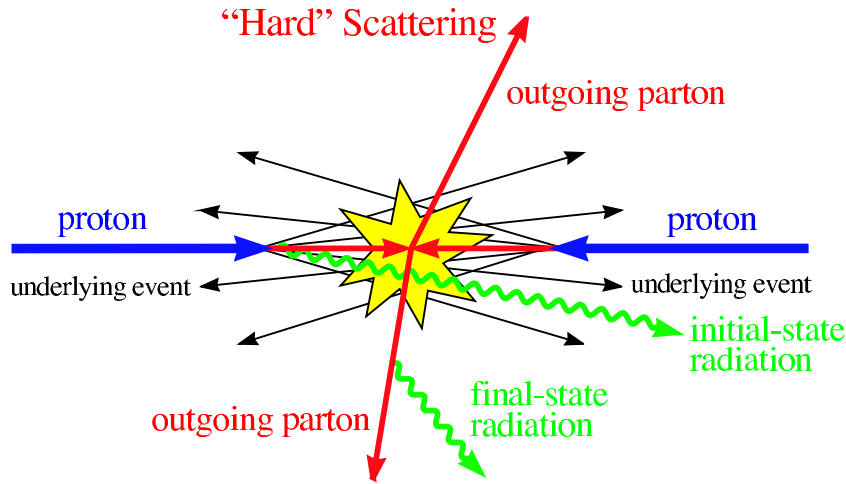} \caption{Pictorial representation of a hard
scattering event~\cite{Affolder:2001xt}.}
\label{scatter}%
\end{figure}


\section{Factorization}

\label{sec:factorization}

The fundamental challenge when trying to make theoretical predictions or
interpret experimentally observed final states is that the theory of the
strong interactions (QCD) is most easily applied to the short distance ($\ll$
1 fermi) degrees of freedom, \textit{i.e}., to the color-charged quarks and
gluons, while the long distance degrees of freedom seen in the detectors are
color singlet bound states of these degrees of freedom. \ We picture the
overall scattering process, as pictorially displayed in Fig.~\ref{scatter}
~\cite{Affolder:2001xt}, evolving from the incoming long distance hadrons in
the beams to the short-distance scattering process to the long distance
outgoing states, as occurring in several (approximately) distinct steps. \ The
separation, or factorization, of these steps is essential both conceptually
and calculationally. \ It is based on the distinct distance (or momentum)
scales inherent at each step. \ 

We imagine as a first step picking out from the incident beam particles the
short distance partons (defined by an appropriate factorization scale) that
participate in the short distance scattering.
\begin{figure}[th]
\centering
\includegraphics[width=.90\hsize]
{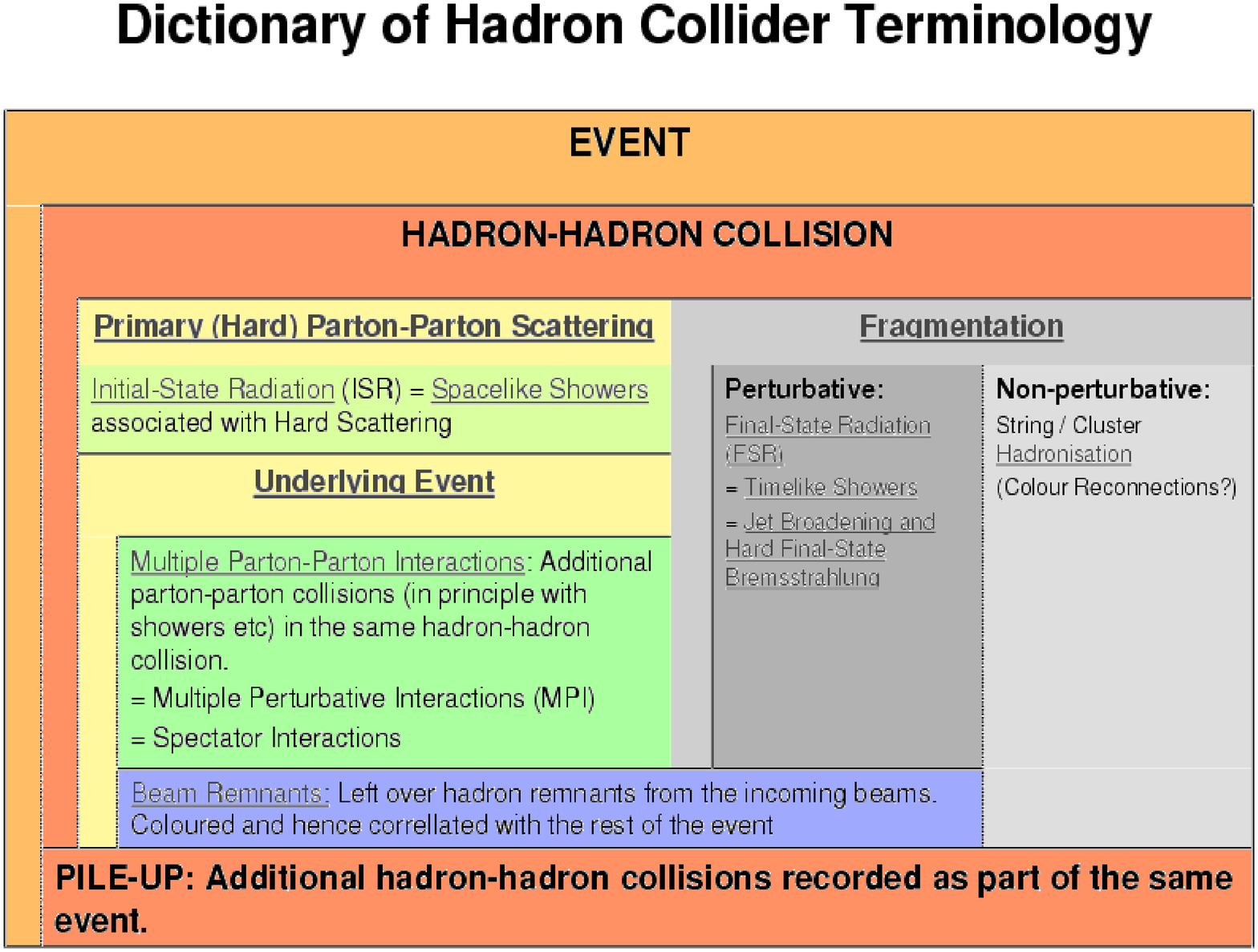} \caption{A dictionary of hadron collider terms relating
to jet measurements~\cite{Albrow:2006rt}.}%
\label{dictionary}%
\end{figure}
The relative probability to find the scattering partons at this step is
provided by the parton distribution functions (pdfs), which are functions of
the partons' color and flavor, the longitudinal momentum fractions $x_{k}$
carried by the partons and the factorization scale $\mu$, all of which serve
to uniquely define the desired partons. \ The pdfs are themselves determined
from global fits~\cite{Pumplin:2002vw} to a wide variety of data, all of which
can be analyzed in the context of perturbative QCD (pQCD) essentially as
outlined here. The partons selected in this way can emit radiation prior to
the short distance scattering yielding the possibility of initial state
radiation (ISR). \ The remnants of the original hadrons, with one parton
removed, are no longer color singlet states and will interact, presumably
softly, with each other generating (approximately incoherently from the hard
scattering) an underlying distribution of soft partons, the beginning of the
underlying event (UE).

Next comes the short distance, large momentum transfer scattering process that
may change the character of the scattering partons and/or produce more partons
(or other interesting particles). \ The cross section for this step is
evaluated at fixed order in pQCD, presumably to next-to-leading-order (NLO),
or higher. \ Then comes another color radiation step, when many new gluons and
quark pairs are added to the state (final state radiation or FSR), dominated
by partons that have low energy and/or are nearly collinear with the scattered
short distance partons. The FSR, like the ISR, is described by the showers in
the Monte Carlo programs~\cite{Sjostrand:2006za,Corcella:2002jc} and
calculated probabilistically in terms of summed leading logarithm perturbation theory.

The final step in the evolution to long distance states involves a
non-perturbative hadronization process that organizes the colored degrees of
freedom from the showering and from the softer interactions of other initial
state partons (the UE is simulated in terms of 
beam-beam remnants and multiple parton interactions)
into color-singlet hadrons with physical
masses. 
The non-perturbative hadronization process yields a collection of ground
state hadrons (primarily pions) and resonances ($A_1,A_2$,etc), and the
resonances then decay into lighter hadrons such as pions. The masses of the 
resonances result in the decay pions being produced with non-zero transverse
momenta with respect to the momentum direction of the original resonance
(but still
small compared to typical jet momenta). The details are determined from data
but the decay modes for some of the higher mass resonances are not
well-understood 
leading to uncertainties (and differences between Monte Carlo programs) in
the details of the production of the final state hadrons.
A convenient dictionary of the terms described above is summarized in
Fig.~\ref{dictionary}~\cite{Albrow:2006rt}.

\begin{figure}[th]
\centering
\includegraphics[width=.40\hsize]
{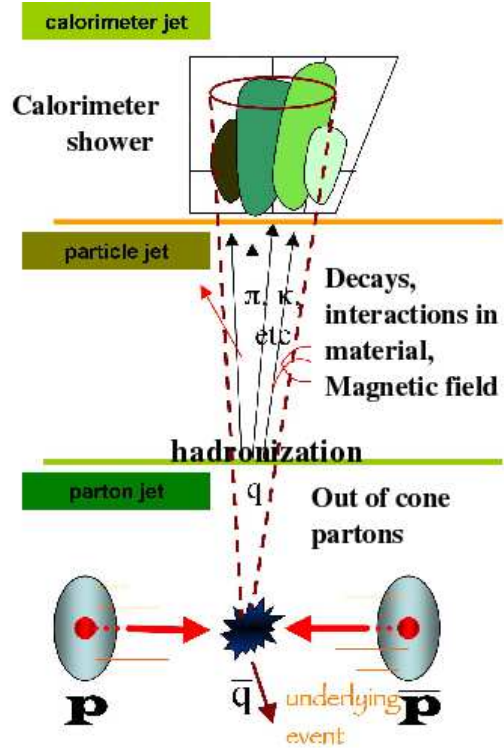} \caption{A representation of the stages of jet
production and measurement.}
\label{jet_stages}%
\end{figure}
The hadronization step is accomplished in a model dependent fashion
(\textit{i.e}., a different fashion) in different Monte Carlos.

The union of the showering and the hadronization steps is what has
historically been labelled as fragmentation, as in fragmentation functions
describing the longitudinal distribution of hadrons within final state jets.
\ In practice, both the radiation and hadronization steps tend to smear out
the energy that was originally localized in the individual short distance
partons (a \textquotedblleft splash-out\textquotedblright\ effect), while the
contributions from the underlying event (and any \textquotedblleft
pile-up\textquotedblright\ from multiple hadron collisions in the same
data-taking time interval) add to the energy originally present in the short
distance scattering (a \textquotedblleft splash-in\textquotedblright\ effect).
\ Finally the hadrons, and their decay products, are detected with finite
precision in a detector.
This final step of the jet components interacting in the detector is
represented in Fig.~\ref{jet_stages}.

The separation of this complicated scattering process into distinct steps is
not strictly valid in quantum mechanics where interference plays a role; we
must sum the amplitudes before squaring them and not just sum the squares.
However, some features of this factorization can be rigorously established
~\cite{Collins:2004nx}, and the numerical dominance of collinear QCD radiation
ensures that the simple picture presented here and quantified by Monte Carlo
generated events, with interference only approximately represented in the
structure of the showers, \textit{e.g}., the angles of emission of partons in
the showers are monotonically ordered, provides a reliable first
approximation. 

In order to interpret the experimentally detected long distance objects, the
charged particles and energy depositions in calorimeter cells, in terms of the
underlying short distance physics, the partons, jet algorithms are employed to
associate \textquotedblleft nearby\textquotedblright\ experimental objects
into jets. \ The underlying assumption is that the kinematics (energy and
momentum) of the resulting cluster or jet provides a useful measure of the
kinematics (energy and momentum) of the underlying, short-distance partons.
\ The goal is to characterize the short-distance physics, event-by-event, in
terms of the discrete jets found by the algorithm. \ In particular, we assume
that the basic mismatch between colored short-distance theory objects and the
colorless long-distance experimental objects does not present an important
numerical limitation, compared to our goal of percent level accuracy for jet
algorithm effects. \ We assume that we are either insensitive to or can
reliably correct for the effects of the UE, ISR, FSR and hadronization. We
assume that most relevant features of a jet can be described by the (up to) 2
partons per jet present in a NLO calculation. We will investigate the features
that are not so well-described.

As noted, jet algorithms rely on the merging of objects that are, by some
measure, nearby each other. \ This feature is essential in perturbation
theory, where the divergent contributions from virtual\ diagrams must
contribute in exactly the same way, \textit{i.e}., contribute to the same
kinematic bins, as the divergent contributions from soft and collinear real
emissions, in order that these contributions can cancel. It is only through
this cancellation that jet algorithms serve to define an IR-safe (finite)
quantity, \textit{i.e.}, a quantity that is insensitive to the emission of
extra soft and/or collinear partons. The standard measures of
\textquotedblleft nearness\textquotedblright\ (see~\cite{Blazey:2000qt})
include pair-wise relative transverse momenta, as in the $k_{T}$ algorithm, or
angles relative to a jet axis, as in the cone algorithm. \ By definition a
\textquotedblleft good\textquotedblright\ algorithm yields stable
(\textit{i.e}., very similar) results whether it is applied to a state with
just a few partons (as in NLO\ perturbation theory), a state with many partons
(after the parton shower as simulated in a Monte Carlo), a state with hadrons
(as simulated in a Monte Carlo including a model for the hadronization step
and the underlying event), or applied to the observed tracks and energy
deposition in a real detector. As we will see, this requirement constitutes a
substantial challenge. \ Further, it is highly desirable that the
identification of jets be insensitive to the contributions from the
simultaneous uncorrelated soft collisions that occur during pile-up at high
luminosity. Finally, we want to be able to apply the \textit{same} algorithm
(in detail) at each level in the evolution of the hadronic final state. This
implies that we must avoid components in the algorithm that make sense when
applied to data but not to fixed order perturbation theory, or vice versa.
This constraint will play an important role in our subsequent discussion.

In practice, we can think of the jet algorithm as a set of mathematical rules
that detail how to carry out two distinct steps. \ The first step operates,
event-by-event, on the list of 4-vectors, which describes either the
perturbative final state, the final-state hadrons in the MC simulated event or
the output from the detector, to turn the original list into a set of
sublists, one sublist for each jet (plus the beam jets). \ The second step
specified by the algorithm tells us how to construct appropriate kinematic
quantities from each sublist in order to describe the kinematic properties of
the individual jets. \ Both steps depend on the specific jet algorithm, as
will be illustrated in detail in the next sections for specific jet
algorithms, and have varied over time. \ In particular, early applications of
jets did not employ true 4-vector arithmetic and largely ignored the
information carried by the invariant mass of the jet. \ In Run II at the
Tevatron and at the LHC true 4-vector arithmetic is and will be employed (the
so-called \textquotedblleft E-scheme\textquotedblright\ as recommended for Run
II in~\cite{Blazey:2000qt}). \ The corresponding kinematic variables
describing the jets include the (true) transverse momentum, $p_{T}$, the
rapidity, $y=\frac{1}{2}\log{\frac{E+p_Z}{E-p_Z}}$, the azimuthal angle,
$\phi$, and the invariant mass of the jet, $M_{J}$. \ We will have
more to say about the history of different choices of kinematic
variables in the next section.

\begin{figure}[h]
\centerline{
\includegraphics[
width=5.00in
]{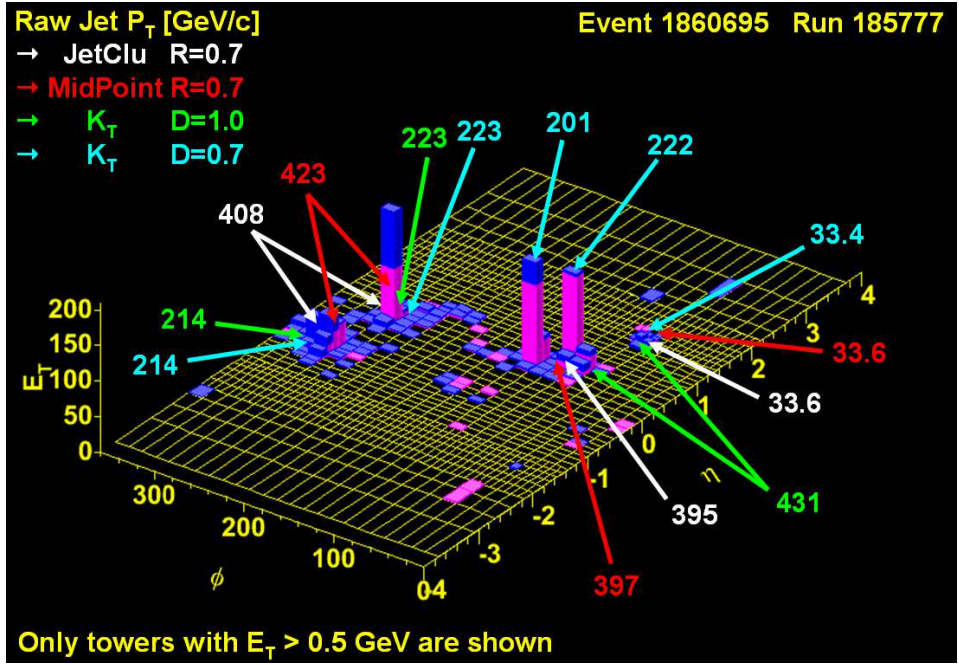}} \caption{Impact of different jet clustering
algorithms on an interesting CDF event taken in Run II. \ The segmentation in
$\eta$ and $\phi$ shown the lego plot corresponds to the calorimeter
segmentation. \ Energy depositions in the electromagnetic portion of the
calorimeter are colored red and those in the hadronic section are colored
blue. The numbers in the figure are transverse momenta of the jets
pointed to by the arrows, and different colors represent jets clustered
by different algorithms.}%
\label{lego3}%
\end{figure}

For many events, the jet structure is clear and the jets, into which the
individual calorimeter towers should be assigned, are fairly unambiguous.
However, in other events, such as the lego plot of a CDF event shown in
Fig.~\ref{lego3}, the complexity of the energy depositions means that
different algorithms will result in different assignments of towers to the
various jets. This is not a problem if a similar complexity is exhibited by
the theoretical calculation which is to be compared to the data. \ However,
the most precise and thoroughly understood theoretical calculations arise in
fixed order perturbation theory, which can exhibit only limited complexity,
\textit{e.g}., at most 2 partons per jet at NLO. \ On the other hand, for
events simulated with parton shower Monte Carlos the complexity of the final
state is more realistic, but the intrinsic theoretical uncertainty is larger.
\ Correspondingly the jets identified by the algorithms vary if we compare at
the perturbative, shower, hadron and detector levels. Thus it is essential to
understand these limitations of jet algorithms and, as much as possible,
eliminate or correct for them to approach our percent level goal. \ It is the
aim of the following review to highlight the issues that arose during Runs I
and II at the Tevatron, discuss their current understanding, and outline
possible preventative measures for the LHC~\cite{Albrow:2006rt}.

\section{Jets: Parton Level vs Experiment}

\label{sec:jets}

\subsection{Iterative Cone Algorithm}

\subsubsection{Definitions}

\begin{figure}[htp]
\centerline{
\includegraphics[
width=6.0in
]{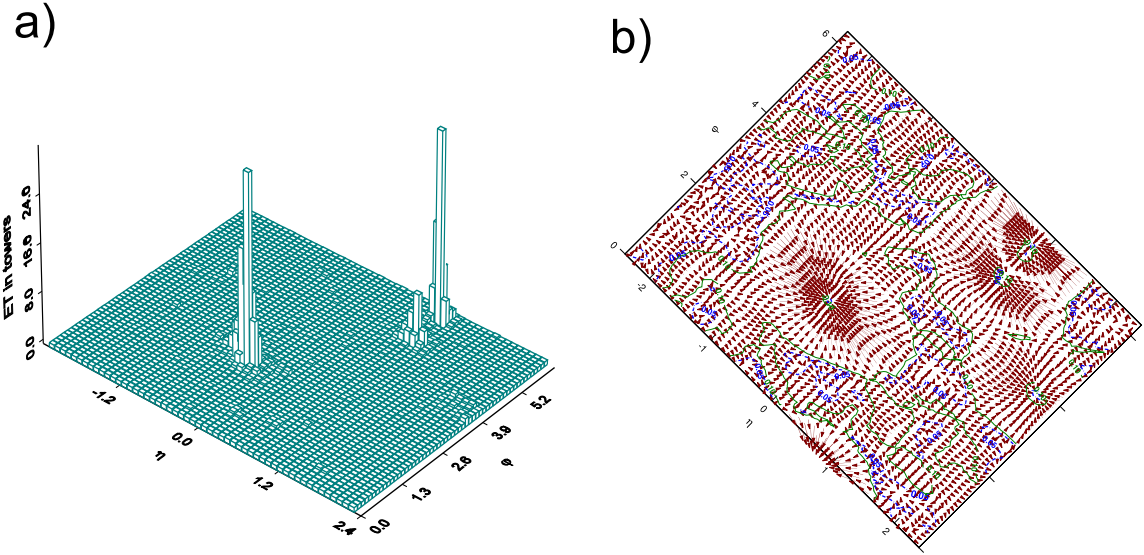}}
\caption{(An ideal) Monte Carlo generated event with 2
large energy jets and 1 small energy jet in the LEGO\ plot a), and the
corresponding flow structure of the trial cones in b). }%
\label{towerflow}%
\end{figure}

To illustrate the behavior of jet algorithms consider first the original
Snowmass implementation of the iterative cone algorithm~\cite{Huth:1990mi}.
The first step in the algorithm, \textit{i.e}., the identification of the
sublists of objects corresponding to the jets, is defined in terms of a simple
sum over all (short distance or long distance) objects within a cone centered
at rapidity (the original version used the pseudorapidity $\eta=\frac{1}%
{2}\log(\cot\frac{\theta}{2})$) and azimuthal angle $\left(  y_{C},\phi
_{C}\right)  $. Using the objects in the cone we can define a $p_{T}$-weighted
centroid via%
\begin{align*}
k  &  \subset C\text{ iff }\sqrt{\left(  y_{k}-y_{C}\right)  ^{2}+\left(
\phi_{k}-\phi_{C}\right)  ^{2}}\leq R_{cone},\\
\overline{y}_{C}  &  \equiv\frac{\sum_{k\subset C}y_{k}\cdot p_{T,k}}%
{\sum_{l\subset C}p_{T,l}},\text{ }\overline{\phi}_{C}\equiv\frac
{\sum_{k\subset C}\phi_{k}\cdot p_{T,k}}{\sum_{l\subset C}p_{T,l}}.
\end{align*}
If the $p_{T}$-weighted centroid does not coincide with the geometric center
of the cone, $\left(  \overline{y}_{C},\overline{\phi}_{C}\right)  \neq\left(
y_{C},\phi_{C}\right)  $, a cone is placed at the $p_{T}$-weighted centroid
and the calculation is repeated. \ This simple calculation is \textit{iterated
}until a \textquotedblleft stable\textquotedblright\ cone is found, $\left(
\overline{y}_{C},\overline{\phi}_{C}\right)  =\left(  y_{C},\phi_{C}\right)
$, which serves to define the jet (and the name of this algorithm as the
iterative cone algorithm). \ Thus, at least in principle, one can think in
terms of placing trial cones everywhere in $\left(  y,\phi\right)  $ and
allowing them to \textquotedblleft flow\textquotedblright\ until a stable cone
or jet is found. \ This flow idea is illustrated in Fig.~\ref{towerflow},
where a) illustrates the LEGO plot for a simple (quiet) Monte Carlo generated
event with 3 apparent jets and b) shows the corresponding flows of the trial
cones towards the (obvious) final jets. Compared to the event in
Fig.~\ref{lego3} there is little ambiguity in this event concerning the jet structure.

To facilitate the subsequent discussion and provide some mathematical
structure for this image of \textquotedblleft flow\textquotedblright\ we can
define the \textquotedblleft Snowmass potential\textquotedblright\ in terms of
the 2-dimensional vector $\overrightarrow{r}=\left(  y,\phi\right)  $ via%
\[
V\left(  \overrightarrow{r}\right)  =-\frac{1}{2}\sum_{k}p_{T,k}\left(
R_{cone}^{2}-\left(  \overrightarrow{r}_{k}-\overrightarrow{r}\right)
^{2}\right)  \Theta\left(  R_{cone}^{2}-\left(  \overrightarrow{r}%
_{k}-\overrightarrow{r}\right)  ^{2}\right)  ,
\]
where the $\Theta$ function defines the objects inside the cone centered at
$r$.
The flow described by the iteration process is driven by the \textquotedblleft
force\textquotedblright%
\begin{align*}
\overrightarrow{F}\left(  \overrightarrow{r}\right)   &  =-\overrightarrow
{\nabla}V\left(  \overrightarrow{r}\right)  =\sum_{k}p_{T,k}\left(
\overrightarrow{r}_{k}-\overrightarrow{r}\right)  \Theta\left(  R_{cone}%
^{2}-\left(  \overrightarrow{r}_{k}-\overrightarrow{r}\right)  ^{2}\right) \\
&  =\left(  \overrightarrow{\overline{r}}_{C\left(  \overrightarrow{r}\right)
}-\overrightarrow{r}\right)  \sum_{k\subset C\left(  r\right)  }p_{T,k},
\end{align*}
where $\overrightarrow{\overline{r}}_{C\left(  \overrightarrow{r}\right)
}=\left(  \overline{y}_{C\left(  \overrightarrow{r}\right)  },\overline{\phi
}_{C\left(  \overrightarrow{r}\right)  }\right)  $ and $k\subset C\left(
\overrightarrow{r}\right)  $ is defined by $\sqrt{\left(  y_{k}-y\right)
^{2}+\left(  \phi_{k}-\phi\right)  ^{2}}$ $\leq R_{cone}$. \ As desired, this
\textquotedblleft force\textquotedblright\ pushes the cone to the stable cone
position, \textit{i.e}., the minimum of the Snowmass potential. \ As noted
above the current Run II analyses and those expected at the LHC, described in
more detail below, 4-vector techniques are used and the corresponding E-scheme
centroid is given instead by
\begin{align*}
k  &  \subset C\text{ iff }\sqrt{\left(  y_{k}-y_{C}\right)  ^{2}+\left(
\phi_{k}-\phi_{C}\right)  ^{2}}\leq R_{cone},\\
p_{C}  &  =\left(  E_{C},\overrightarrow{p}_{C}\right)  ={\sum_{k\subset C}%
}\left(  E_{k},\overrightarrow{p_{k}}\right)  ,\text{ }\overline{y}_{C}%
\equiv\frac{1}{2}\ln\frac{E_{C}+p_{z,C}}{E_{C}-p_{z,C}},\text{ }\overline
{\phi}_{C}\equiv\tan^{-1}\frac{p_{y,C}}{p_{x,C}}.
\end{align*}
In the NLO perturbative calculation these changes in definitions result in
only tiny numerical changes, compared to the Snowmass definition
numbers.

\begin{figure}[ht]
\centerline{
\includegraphics[
width=6.0in
]{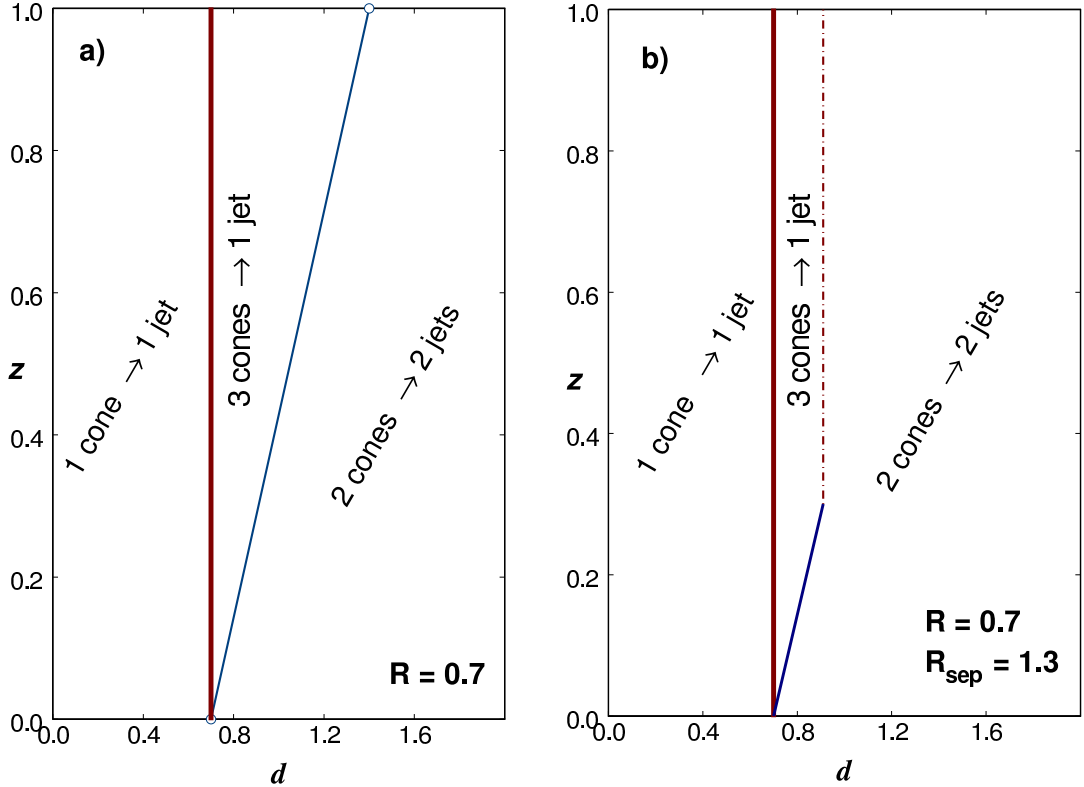}} \caption{Perturbative 2-parton phase space:
$z=p_{T,2}/p_{T,1}\left(  p_{T,1}\geq p_{T,2}\right)  $,
$d=\sqrt{\left(
y_{1}-y_{2}\right)  ^{2}+\left(  \phi_{1}-\phi_{2}\right)  ^{2}}$ for a) the
naive $R_{sep}=2$ case and b) for $R_{sep}=1.3$ case suggested by data. }%
\label{pertthy}%
\end{figure}

\begin{figure}[ht]
\centerline{
\includegraphics[
width=6.0in
]{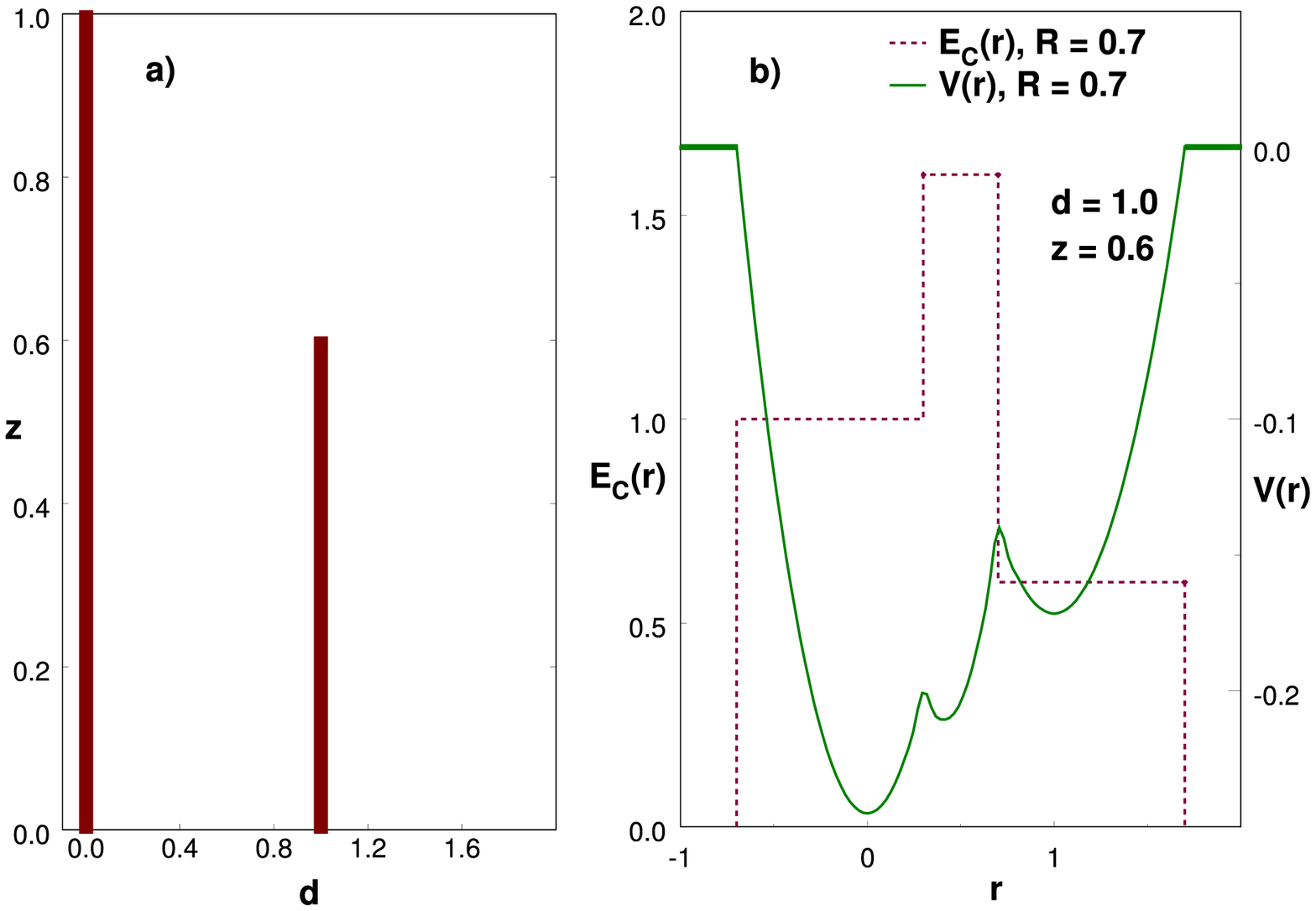}}
\caption{2-parton distibution in $\left(  d,z\right)  $ in
a) with $d=1.0$, $z=0.6$ and the corresponding energy-in-cone, $E_{C}\left(
r\right)  $, and potential, $V\left(  r\right)  $. }%
\label{edist}%
\end{figure}

As an introduction to how the iterative cone algorithm works, consider first
its application to NLO level in perturbation theory 
(see, \textit{e.g.},~\cite{Ellis:1992en}), where there are at most 2 partons in a cone. \ As
defined above, the cone algorithm specifies that two partons are included in
the same jet (\textit{i.e}., form a stable cone) if they are both within
$R_{cone}$ (\textit{e.g}., 0.7 in $\left(  y,\phi\right)  $ space) of the
centroid, which they themselves define. \ This means that 2 partons of equal
$p_{T}$ can form a single jet as long as their pair-wise angular separation
does not exceed the boundaries of the cone, $\Delta R=2R_{cone}$. \ On the
other hand, as long as $\Delta R>R_{cone}$, there will also be stable cones
centered around each of the partons. \ The corresponding 2-parton phase space
for $R_{cone}=0.7$ is illustrated in Fig.~\ref{pertthy} a) in terms of the
ratio $z=p_{T,2}/p_{T,1}\left(  p_{T,1}\geq p_{T,2}\right)  $ and the angular
separation variable $d=\sqrt{\left(  y_{1}-y_{2}\right)  ^{2}+\left(  \phi
_{1}-\phi_{2}\right)  ^{2}}$. \ To the left of the line $d=R_{cone}$ the two
partons always form a single stable cone and jet, while to the far right,
$d>2R_{cone}$, there are always two distinct stable cones and jets, with a
single parton in each jet. \ More interesting is the central region,
$R_{cone}<d<2R_{cone}$, which exhibits both the case of two stable cones (one
of the partons in each cone) and the case of three stable cones (the previous
two cones plus a third cone that includes both partons). \ The precise outcome
depends on the specific values of $z$ and $d$. \ (Note that the exactly
straight diagonal boundary in the figure corresponds to the $p_{T}$-weighted
definition of the Snowmass algorithm, but is only slightly distorted, $<2\%$,
when full 4-vector kinematics is used in the Run II algorithms.) \ \ 
To see the three stable cone structure in terms of the 2-parton ``Snowmass
potential'' consider the point $z = 0.6$ 
and $d = 1.0$, which is in the 3 cones $\rightarrow$ 1 jet region.  This configuration,
in terms of the 2-parton coordinates 
$z$ and $d$, is illustrated in Fig.~\ref{edist}a.  The 
corresponding energy in a cone, $E_c$, normalized to the energy of the more
energetic parton, is illustrated by 
the (red) dashed line in Fig.~\ref{edist}b,  where the location of the cone center,
$r$, is constrained to lie in the plane 
of the 2 partons with the origin at the location of the more energetic
parton.  The solid (green) curve in Fig.~\ref{edist}b
shows the corresponding ``Snowmass potential'', again normalized to the
energy of the more energetic parton, versus 
the same 1-D location radius $r$.  This potential exhibits the expected 3
minima corresponding to a stable cone at 
each parton ($r = 0$ and $r = d = 1.0$) and a third stable cone, central
between the other two, that includes the 
contributions from both partons.
A relevant point
is that the central minimum is not nearly as deep (\textit{i.e.}, as robust)
as the other two. As we shall see, this minimum often does not survive the
smearing inherent in the transition from the short distances of fixed order
perturbation theory to the long distances of the physical final state. \ As
indicated by the labeling in Fig.~\ref{pertthy}, in the 3 stable cone region
the original perturbative calculation~\cite{Ellis:1992en} kept as the jet the
2-in-1 stable cone, maximum $p_{T}$ configuration, \textit{i.e}., the cone
that included all of the energy in the other two cones consistent with the
merging discussion below. \ 

As we will see, much of the concern and confusion about the cone algorithm
arises from the treatment of this 3 stable cone region. \ \ It is intuitively
obvious that, as the energy in the short-distance partons is smeared out by
subsequent showering and hadronization, the detailed structure in this region
is likely to change. \ In particular, while two nearly equal $p_{T}$ partons
separated by nearly $2R_{cone}$ may define a stable cone in fixed order pQCD,
this configuration is unlikely to yield a stable cone after showering and hadronization.

Having performed the first step in the algorithm to identify the particles in
the cone $C$, the second step is to define the kinematic variables describing
the jet. \ As suggested above the the Snowmass definition of the iterative
cone algorithm used angular variables defined by the $p_{T}$ weighted
expressions
\[
\overline{y}_{J}\equiv\frac{\sum_{k\subset C}y_{k}\cdot p_{T,k}}{\sum_{l\subset
C}p_{T,l}},\text{ }\overline{\phi}_{J}\equiv\frac{\sum_{k\subset C}\phi
_{k}\cdot p_{T,k}}{\sum_{l\subset C}p_{T,l}},
\]
and, instead of the true transverse momentum, the scalar transverse momentum%
\[
P_{J}=p_{T,\text{Snowmass}}=\sum_{l\subset C}p_{T,l}.
\]
In the (recommended) case of 4-vector arithmetic we have instead (as suggested
as above)%
\begin{align*}
p_{C}  &  =\left(  E_{C},\overrightarrow{p}_{C}\right)  ={\sum_{k\subset C}%
}\left(  E_{k},\overrightarrow{p_{k}}\right)  ,\text{ }y_{J}\equiv\frac{1}%
{2}\ln\frac{E_{C}+p_{z,C}}{E_{C}-p_{z,C}},\text{ }\phi_{J}\equiv\tan^{-1}%
\frac{p_{y,C}}{p_{x,C}},\\
P_{J}  &  =\left\vert \overrightarrow{p}_{C,T}\right\vert =\left\vert
{\sum_{k\subset C}}\overrightarrow{p}_{k,T}\right\vert ,M_{J}^{2}=p_{C}%
^{2}=E_{C}^{2}-\overrightarrow{p}_{C}^{2}.
\end{align*}
During Run I CDF used an intermediate set of definitions with E-scheme
(4-vector) angles and the jet \textquotedblleft transverse
momentum\textquotedblright\ given by the transverse energy%
\[
P_{J}=E_{T}=E_{C}\sin\theta_{C}.
\]
In the next section we will briefly explore the quantitative differences
between\ these definitions in the context of NLO perturbation theory.

\subsubsection{$R_{sep}$, Seeds and IR-Sensitivity}

Iterative cone algorithms similar to the ones described in the previous
section were employed by both the CDF and D\O \ collaborations during Run I
with considerable success. \ There was fairly good agreement with NLO pQCD for
the inclusive jet cross section over a dynamic range of order $10^{8}$.
\ During Run I, the data were corrected primarily for detector effects and for
the contributions of the underlying event. \ In fact, a positive feature of
the cone algorithm is that, since the cone's geometry in $\left(
y,\phi\right)  $ space is (meant to be) simple, the correction for the
\textquotedblleft splash-in\textquotedblright\ contribution of the (largely
uncorrelated) underlying event (and pile-up) is straightforward. \ (As we will
see below, the corrections being used in Run II are more sophisticated.) \ The
uncertainties in both the data and the theory were $10\%$ or greater,
depending on the kinematic regime, and helped to ensure agreement. \ However,
as cone jets were studied in more detail, various troubling issues arose.
\ For example, it was noted long ago~\cite{Abe:1991ui} that, when using the
experimental cone algorithms implemented at the Tevatron, two jets of
comparable energy (taken from 2 different events in the data) are not merged
into a single jet if they are separated by an angular distance greater than
approximately 1.3 times the cone radius, while the simple picture of Fig.
\ref{pertthy} a) suggests that merging should occur out to an angular
separation of $2R_{cone}$. \ Independently it was also noted that the
dependence of the experimental inclusive jet cross section on the cone radius
$R_{cone}$~\cite{Abe:1991ea} and the shape of the energy distribution within a
jet~\cite{Abe:1992wv} both differed discernibly from the NLO predictions (the
data were less $R_{cone}$ dependent and exhibited less energy near the edge of
the cone). \ All three of these issues seemed to be associated with the
contribution from the perturbative configuration of two partons with
comparable $p_{T}$ at opposite sides of the cone ($z\simeq1$, $d\simeq
2R_{cone}=1.4$ in Fig.~\ref{pertthy} a)) and the data suggested a lower
contribution from this configuration than present in the perturbative result.
\ To simulate this feature in the perturbative analysis, a phenomenological
parameter $R_{sep}$ was added to the NLO implementation of the cone
algorithm~\cite{Ellis:1992qq}. \ In this "experiment-aware" version of the
perturbative cone algorithm, two partons are not merged into a single jet if
they are separated by more than $R_{sep}\cdot R_{cone}$ from each other,
independent of their individual distance from the $p_{T}$-weighted jet
centroid. \ Thus, the two partons are merged into the same jet if they are
within $R_{cone}$ of the $p_{T}$-weighted jet centroid and within $R_{sep}\cdot
R_{cone}$ of each other; otherwise the two partons are identified as separate
jets. In order to describe the observed $R_{cone}$ dependence of the cross
section and the observed energy profile of jets, the specific value
$R_{sep}=1.3$ was chosen (along with a \textquotedblleft
smallish\textquotedblright\ renormalization/factorization scale $\mu=p_{T}%
/4$), which was subsequently noted to be in good agreement with the
aforementioned (independent) jet separation study. \ The resulting 2 parton
phase space is indicated in Fig.~\ref{pertthy} b). \ In the perturbative
calculation, this redefinition, leading to a slightly lower average $p_{T}$
for the leading jet, lowers the NLO jet cross section by about 5\% (for
$R=0.7$ and $p_{T}=100$ GeV$/c$). \ \ It is important to recognize that the
fractional contribution to the inclusive jet cross section of the merged 2
parton configurations in the entire wedge to the right of $d=R_{cone}$ is only
of order $10\%$ for jet $p_{T}$ of order 100 GeV$/c$, and, being proportional to
$\alpha_{s}\left(  p_{T}\right)  $, decreases with increasing $p_{T}$. \ Thus
it is no surprise that, although this region was apparently treated
differently (in the NLO theory comparisons) by the cone algorithm
implementations of CDF and D\O \ during Run I as discussed below, there were
no relevant cross section disagreements above the $>10\%$ uncertainties.
\ Further, as we will discuss below, it is the variation in the treatment of
this (effectively 10\%) region of the 2-parton phase space that drives many of
the differences between jet algorithms.

\begin{figure}[h]
\centerline{
\includegraphics[
width=7.00in
]{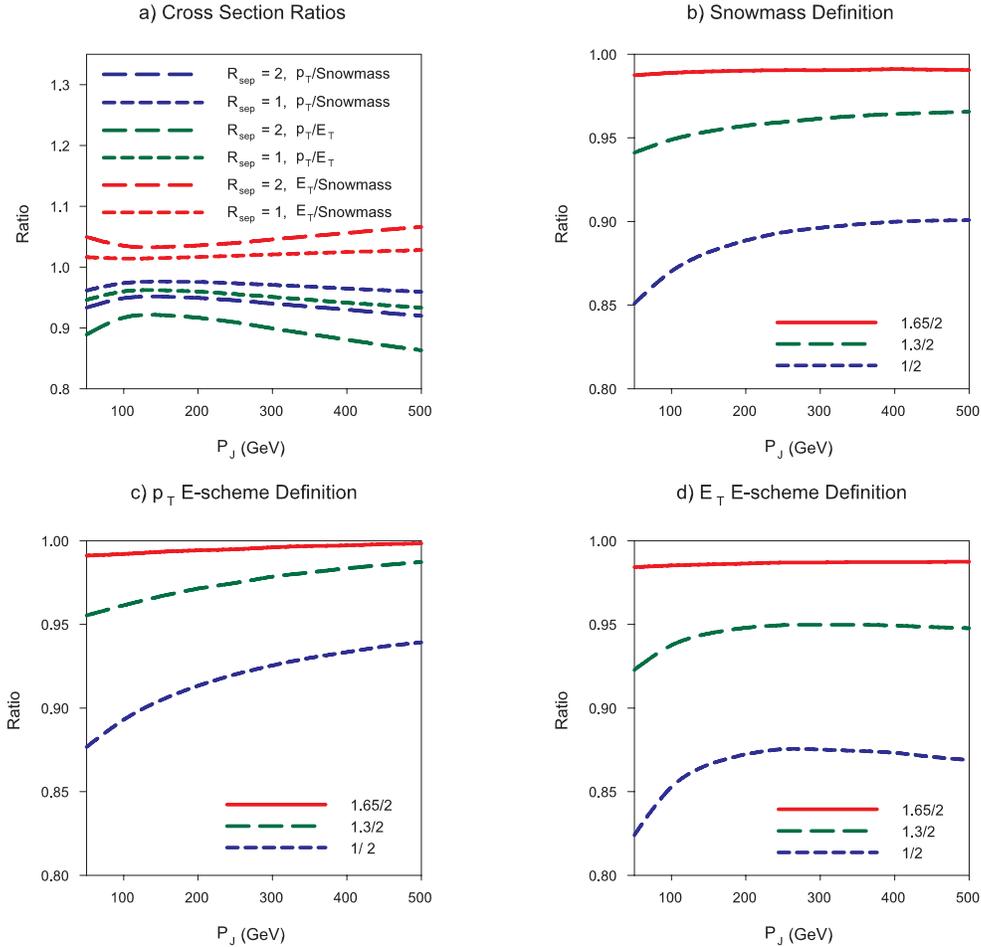}}\caption{Ratios of the NLO inclusive cone jet cross section
versus the jet momentum for 3 definitions of the kinematics for various values
of $R_{sep}$.}%
\label{fig:allET}%
\end{figure}

While the parameter $R_{sep}$ is \textit{ad hoc} and thus an undesirable
complication in the perturbative jet analysis, it will serve as a useful
pedagogical tool in the following discussions. To illustrate this point
quantitatively, Fig.~\ref{fig:allET} shows the dependence on $R_{sep}$ for
various choices of the jet momentum $P_{J}$ at NLO in perturbation theory. The
curves labeled Snowmass use the $p_{T}$ weighted kinematics described above
with $P_{J}$ given by the scalar sum of the transverse momenta of the partons
in the cone. The two E-scheme algorithms use full 4-vector kinematics and
$P_{J}$ equal to either the magnitude of the true (vector sum) transverse
momentum (the recommended choice), or the \textquotedblleft transverse
energy\textquotedblright\ defined by $P_{J}=E_{T}=E\sin\theta$ (as defined by
CDF in Run I). Thus this last variable knows about both the momentum and the
invariant mass of the set of partons in the cone, which can be sizable for
well separated parton pairs. The differences in the various ratios for
different values of $R_{sep}$ tell us about how the 2-parton configurations
contribute. For example, Fig.~\ref{fig:allET} a) tells us that, since, for a
given configuration of 2 partons in a cone, $E_{T}>p_{T,Snowmass}>p_{T}$, the
cross sections at a given value of $P_{J}$ will show the same ordering.
Further, as expected, the differences are reduced if we keep only
configurations with small angular separation, $R_{sep}=1$. From
Fig.~\ref{fig:allET} b) we confirm the earlier statement that lowering
$R_{sep}$ from 2 to 1.3 yields a $5\%$ change for the Snowmass algorithm cross
section with $P_{J}=100$ GeV, while lowering it all the way to $R_{sep}=1$,
\textit{i.e.}, removing all of the triangular region, lowers the 100 GeV
Snowmass jet cross section by approximately $12\%$. Figs.~\ref{fig:allET} c)
and d) confirm that 4-vector kinematics with $P_{J}=p_{T}$ exhibits the
smallest sensitivity to $R_{sep}$, \textit{i.e.}, to the 2-parton
configurations in the triangle. The choice $P_{J}=E_{T}$, with its dependence
on the mass of the pair, exhibits the largest sensitivity to $R_{sep}$. These
are all good reasons to use the recommended E-scheme kinematics with
$P_{J}=p_{T}$. \ The move to employ 4-vector kinematics for jet analyses in
Run II and at the LHC is a positive step. \ It will allow the meaningful
investigation of jet masses, which will likely be very useful at the LHC.
\ The decay of large mass (few TeV) new particles will lead to highly boosted
$W^{\prime}$s, $Z^{\prime}$s and top quarks that will be observed as single
jets. \ Thus the mass of such jets may constitute a useful selection tool.

\begin{figure}[h]
\centerline{
\includegraphics[
width=4.3in
]{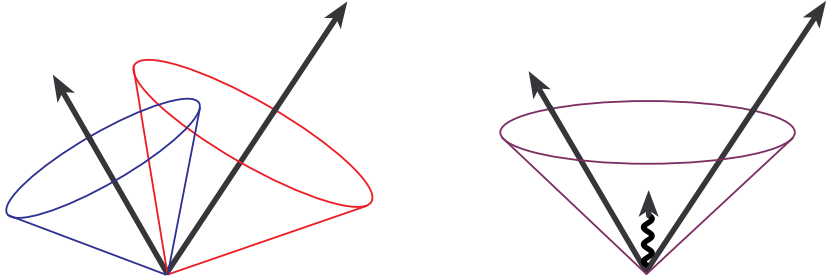}} \caption{Two partons in two cones or in one cone with a
(soft) seed present~\cite{Blazey:2000qt}.}%
\label{2inone}%
\end{figure}

The difference between the perturbative implementation of the iterative cone
algorithm and the experimental implementation at the Tevatron, which is
simulated by $R_{sep}$, is thought to arise from several sources. \ While the
perturbative version (with $R_{sep}=2$) analytically includes \emph{all}
2-parton configurations that satisfy the algorithm (recall Fig.~\ref{pertthy}
a)), the experiments employ the concept of \textit{seeds} to reduce the
analysis time and place trial\ cones only in regions of the detector where
there are seeds, \textit{i.e}., pre-clusters with substantial energy. \ This
procedure introduces another parameter, the lower $p_{T}$ threshold defining
the seeds, and also decreases the likelihood of finding the
2-showers-in-one-jet configurations corresponding to the upper right-hand
corner of the 3 cones $\rightarrow$ 1 jet region of Fig.~\ref{pertthy} a) and
the middle minimum in Fig.~\ref{edist} b). \ Thus, the use of seeds
contributes to the need for $R_{sep}<2$. \ Perhaps more importantly, the
desire to match the theoretical algorithm with the experimental one means that
we should include seeds in the perturbative algorithm. \ This is accomplished
by placing trial cones only at the locations of each parton and testing to see
if any other partons are inside of these cones. \ Thus at NLO, 2 partons will
be merged into a single jet only if they are closer than $R_{cone}$ in
$\left(  y,\phi\right)  $ space. \ This corresponds to $R_{sep}=1.0$ in the
language of Fig.~\ref{pertthy} and produces a larger numerical change in the
analysis than observed, \textit{i.e.}, we wanted $R_{sep}\simeq1.3$. \ More
importantly at the next order in perturbation theory, NNLO, there are extra
partons that can play the role of low energy seeds. \ The corresponding parton
configurations are illustrated in Fig.~\ref{2inone}. \ At NLO, or in the
virtual correction in NNLO, the absence of any extra partons to serve as a
seed leads to two distinct cones as on the left, while a (soft) real emission
at NNLO can lead to the configuration on the right where the soft gluon
\textquotedblleft seeds\textquotedblright\ the middle cone that includes all
of the partons. \ The resulting separation between the NNLO virtual
contribution and the NNLO soft real emission contribution (\textit{i.e}., they
contribute to different jet configurations) leads to an undesirable
logarithmic dependence on the seed $p_{T}$ threshold~\cite{Seymour:1997kj}.
\ In the limit of an arbitrarily soft seed $p_{T}$ cutoff, the cone algorithm
with seeds is no longer IR-safe. \ By introducing seeds in the algorithm, we
have introduced exactly what we want to avoid, sensitivity to soft emissions.
\ From the theory perspective, seeds are an undesirable component in the
algorithm and should be eliminated.

\subsubsection{Seedless and Midpoint Algorithms}
\label{sec:seedless_midpoint}

\ The labeling of the Run I cone algorithm with seeds as Infrared Unsafe has
led some theorists to suggest that the corresponding analyses should be
disregarded. \ This is too extreme a reaction, especially since the numerical
difference between the jet cross section found in data using seeds is expected
to be less than a few percent different from an analysis using a seedless algorithm.
\ A more useful approach will be to either avoid the use of seeds\footnote{The
Run II recommendations~\cite{Blazey:2000qt} did include the suggestion of a
seedless algorithm.}, or to correct for them in the analysis of the data,
which can then be compared to a perturbative analysis without seeds. \ Note
that it may seem surprising that an algorithm, which is Infrared Unsafe due to
the use of seeds, leads to experimental results that differ from a Infrared
Safe seedless algorithm by only a few percent. The essential point is that the
lack of IR-safety is a property of the fixed-order perturbative application of
the algorithm with seeds, not of the experimental application. In real data
the additional soft components of the event (initial state radiation, final
state radiation and the underlying event) ensure that there are seeds
\textquotedblleft nearly\textquotedblright\ everywhere. Thus there is only a
small change from the situation where seeds are assumed to be everywhere (the
seedless algorithm). In stark contrast the NLO perturbative application of an
algorithm with seeds has only the energetic partons themselves to act as
seeds. Thus there is a dramatic change at NNLO where the extra parton can
serve as a seed, as in Fig.~\ref{2inone}, changing the found jet structure of
the event even when the extra parton is quite low energy. \ This is the source
of the perturbative Infrared sensitivity.

One of the main problems with the use of a seedless cone algorithm has been
its slow speed with respect to the seeded cone algorithms. This has made its
use in reconstruction of a large number of events difficult. Combined with the
fact that, for inclusive distributions, the differences between the results
from a seeded cone algorithm like Midpoint (defined below) and a seedless
algorithm tend to be on the order of a percent or less\footnote{In
Ref.~\cite{Salam:2007xv}, a statement is made that the impact may be larger
for some exclusive final state observables.}, there was no strong motivation
for its use. Recently, a new seedless algorithm (SISCone)~\cite{Salam:2007xv}
that has speeds comparable to the seeded cone algorithms has been developed,
removing this difficulty. For this reason, the SISCone algorithm is being
adopted by the experiments at both the Tevatron and LHC\footnote{A
$\mathit{streamlined}$ (faster) version of the seedless algorithm was used
during the early stages of CDF in Run II, but was dropped because of the near
equivalence of the results obtained with the Midpoint cone algorithm.}. Note
that the problems with dark towers and the smearing of stable solution
points (discussed later in Section~\ref{sec:smearing_dark_towers})
still remain with a seedless algorithm.

To address the issue of seeds on the experimental side and the $R_{sep}$
parameter on the phenomenological side, the Run II study~\cite{Blazey:2000qt}
recommended using the Midpoint cone algorithm, in which, having identified 2
nearby jets, one always checks for a stable cone with its center at the
midpoint between the 2 found cones. \ Thus, in the imagery of Fig.
\ref{2inone}, the central stable cone is now always looked for, whether there
is an actual seed there or not. \ It was hoped that this would remove the
sensitivity to the use of seeds and remove the need for the $R_{sep}$
parameter. \ While this expectation is fully justified with the localized,
short distance configuration indicated in Fig.~\ref{2inone}, more recent
studies suggest that at least part of the difficulty with the missing stable
cones at the midpoint position is due to the (real) smearing effects on the
energy distribution in $\left(  y,\phi\right)  $ of showering and
hadronization. \ Also it is important to note that, in principle, IR-safety
issues due to seeds will reappear in perturbation theory at order NNNLO, where
the midpoint is not the only problem configuration (for example, a seed at the
center of a triangular array of 3 hard and merge-able partons can lead to
IR-sensitivity). \ Eliminating the use of seeds remains the most attractive option.

\subsubsection{Merging}
\label{sec:merging}

Before proceeding we must consider another important issue that arises when
comparing the cone algorithm applied in perturbation theory with its
application to data at the Tevatron. \ The practical definition of jets
equaling stable cones does not eliminate the possibility that the stable cones
can overlap, \textit{i.e}., share a subset (or even all) of their calorimeter
towers. \ To address this ambiguity, experimental decisions had to be made as
to when to completely merge the two cones (based on the level of overlap), or,
if not merging, how to split the shared energy. \ \ Note that there is only a
weak analogy to this overlap issue in the NLO calculation. As described in
Fig.~\ref{pertthy} a), there is no overlap in either the left-hand (1 cone
$\rightarrow$ 1 jet) or right-hand (2 cones $\rightarrow$ 2 jets) regions,
while in the middle (3 cones $\rightarrow$ 1 jet) region the overlap between
the 3 cones is 100\% and the cones are always merged. \ Arguably the
phenomenological parameter $R_{sep}$ also serves to approximately simulate not
only seeds but also the role of non-complete merging in the experimental
analysis. \ In practice in Run I, CDF and D\O \ chose to use slightly
different merging parameters. \ Thus, largely unknown to most of the theory
community, the two experiments used somewhat different cone jet algorithms in
Run I. \ The CDF collaboration cone algorithm, JETCLU~\cite{Abe:1991ui}, also
employed another \textquotedblleft feature\textquotedblright\ called
ratcheting, that was likewise under-documented. \ Ratcheting ensured that any
calorimeter tower in an initial seed was always retained in the corresponding
final jet. \ Stability of the final cones was partially driven by this
\textquotedblleft no-tower-left-behind\textquotedblright\ feature.
\ Presumably the two experiments reported compatible jet physics results in
Run I due to the\ substantial $\left(  \geq10\%\right)  $ uncertainties.
\ Note also that, after the splitting/merging step, the resulting cone jets
will not always have a simple, symmetric shape in $\left(  y,\phi\right)  $,
which complicates the task of correcting for the underlying event and leads to
larger uncertainties. \ In any case the plan for Run II as outlined in the Run
II\ studies~\cite{Blazey:2000qt}, called for cone jet algorithms in the 2
collaborations as similar as possible. Unfortunately, during Run II the
collaborations have evolved to employing somewhat different cone algorithms.
\ On the merging question, CDF in Run II merges two overlapping cones when
more than 75\% of the smaller cone's energy overlaps with the larger jet.
\ When the overlap is less, the overlap towers are assigned to the nearest
jet. \ D\O , on the other hand, uses a criterion of a 50\% overlap in order to
merge. There is anecdotal evidence from several studies that a merging
criterion of 75\% may be \textquotedblleft safer\textquotedblright\ in
high-density final states~\cite{gavin}, where there is a tendency for
over-merging to occur. \ While it is not necessary that all
analyses use the same jet algorithm, for purposes of comparison it would be
very useful for the experiments to have one truly common algorithm. \ There is
certainly a lesson to be learned here for the collaborations at the LHC.

\subsubsection{Summary}

In summary, the iterative cone algorithm benefits, in principle, from a simple
geometric definition that allows simple corrections for the UE contributions.
At the same time it suffers from the experimental use of seeds and the need to
implement a split/merge stage. The inclusion of the $R_{sep}$ parameter in the
perturbative calculation, while allowing more detailed comparisons to data,
has also served to confuse the situation. Before discussing the role of
fragmentation and smearing in the cone algorithm, we turn now to the other
primary jet algorithm, the $k_{T}$ algorithm.

\subsection{k$_{T}$ Algorithm}

With the mixed history of success for the cone algorithm, the (as yet) less
well studied $k_{T}$ algorithm~\cite{Ellis:1993tq,Catani:1992zp} offers the
possibility of more nearly identical analyses in both experiments and in
perturbation theory. \ This algorithm, which was first used at
electron-positron colliders, is based on a pair-wise recombination scheme
intended, in some sense, to \textquotedblleft undo\textquotedblright\ the
splitting that occurs during the fragmentation stage. Two
partons/particles/calorimeter towers are combined if their relative transverse
momentum is less than a given measure. \ To illustrate the clustering process,
consider a multi-parton final state. \ Initially each parton is considered as
a proto-jet. The quantities%
\begin{align*}
k_{T,i}^{2}  &  =p_{T,i}^{2},\\
k_{T,(i,j)}^{2}  &  =\min(p_{T,i}^{2},p_{T,j}^{2})\frac{\Delta R_{i,j}^{2}%
}{D^{2}}%
\end{align*}
are computed for each proto-jet $i$ and each pair of proto-jets $ij$,
respectively. As earlier, $p_{T,i}$ is the transverse momentum of the $i^{th}$
proto-jet and $\Delta R_{i.j}$ is the distance (in $y,\phi$ space, $\Delta
R_{i.j}=\sqrt{\left(  y_{i}-y_{j}\right)  ^{2}+\left(  \phi_{i}-\phi
_{j}\right)  ^{2}}$) between each pair of proto-jets. \ $D$ is the parameter
that controls the size of the jet (analogous to $R_{cone}$). If the smallest
of the above quantities is a $k_{T,i}^{2}$, then that proto-jet becomes a jet
and is removed from the proto-jet list. \ If the smallest quantity is a
$k_{T,(i,j)}^{2}$, then the two proto-jets $\left(  i,j\right)  $ are merged
into a single proto-jet by summing their four-vector components, and the two
original entries in the proto-jet list are replaced by this single merged
entry. \ This process is iterated with the corrected proto-jet list until all
the proto-jets have become jets, \textit{i.e}., at the last step the
$k_{T,(i,j)}^{2}$ for all pairs of proto-jets are larger than all $k_{T,i}%
^{2}$ for the proto-jets individually (\textit{i.e}., the remaining proto-jets
are well separated) and the latter all become jets. \ 

Note that in the pQCD NLO inclusive $k_{T}$ jet calculation, the parton pair
with the smallest $k_{T}^{2}$ may or may not be combined into a single jet,
depending on the $k_{T,i}^{2}$ of the individual partons. Thus the final state
can consist of either 2 or 3 jets, as was also the case for the cone
algorithm. \ In fact, the pQCD NLO result for the inclusive $k_{T}$ jet cross
section~\cite{Ellis:1993tq} suggests near equality with the cone jet cross
section in the case that $D\simeq1.35R_{cone}$ (with no seeds, $R_{sep}=2$).
\ Thus the inclusive cone jet cross section with $R_{cone}=0.7$ ($R_{sep}=2$)
is comparable in magnitude to the inclusive $k_{T}$ jet cross section with
$D=0.9$, at least at NLO. \ In the NLO language illustrated in Fig.
\ref{pertthy} the condition that the partons be merged in the $k_{T}$
algorithm is that $z^{2}\left(  d^{2}/D^{2}\right)  <z^{2}$ or $d<D$. \ Thus,
at NLO, the $k_{T}$ algorithm corresponds to the cone algorithm with
$R_{cone}=D$, $R_{sep}=1$. \ The earlier result, $D\simeq1.35R_{cone}$ (with
$R_{sep}=2$), is just the NLO statement that the contribution of the
rectangular region $0\leq d\leq1.35R_{cone}$, $0\leq z\leq1$ is numerically
approximately equal to the contribution of the rectangular region $0\leq d\leq
R_{cone}$, $0\leq z\leq1$ plus the (3 stable cone) triangular region
$R_{cone}\leq d\leq\left(  1+z\right)  R_{cone}$, $0\leq z\leq1$.

In contrast to the cone case, the $k_{T}$ algorithm has no problems with
overlapping jets and, less positively, every calorimeter tower is assigned to
some jet. \ While this last result made some sense in the $e^{+}e^{-}$
collider case, where every final state particle arose from the short-distance
process, it is less obviously desirable in the hadron collider case. \ While
the $k_{T}$ algorithm tends to automatically correct for the splash-out effect
by re-merging the energy distribution smeared by showering and hadronization
into a single jet, this same feature leads to a tendency to enhance the
splash-in effect by ``vacuuming up'' the contributions from the underlying event
and including them in the large $k_{T,i}^{2}$ jets.\ \ This issue is
exacerbated when the luminosity reaches the point that there is more than one
collision per beam bunch crossing and pile-up is significant. \ This is now
true at the Tevatron and will certainly be true eventually at the LHC. \ Thus
while the (splash-out) fragmentation corrections for the $k_{T}$ algorithm are
expected to be smaller than for cone algorithms, the (splash-in) underlying
event corrections will be larger.
\ A test of our understanding of these corrections will be provided by the
comparison of the $D$ and $R_{cone}$ parameter values that yield comparable
experimental jet cross sections. \ If we can reliably correct back to the
fixed order perturbative level for both the cone and $k_{T}$ algorithms, we
should see $D\simeq1.35R_{cone}$. Note that this result assumes that the cone
jet cross section has been corrected to the value corresponding to $R_{sep}%
=2$.\ On the other hand, under-corrected splash-in contributions in the
$k_{T}$ algorithm will require $D<1.35R_{cone}$ for comparable jet cross
section values (still assuming that $R_{sep}=2$ describes the cone results).
If the cone algorithm jet cross section has under-corrected splash-out effects
($R_{sep}<2$), we expect that an even smaller ratio of $D$ to $R_{cone}$ will
be required to obtain comparable jet cross sections (crudely we expect
$D<(1+0.35(R_{sep}-1))R_{cone}$ for $1\leq R_{sep}\leq2$). \ As we will
discuss below, the current studies at the Tevatron suggest that $D<R_{cone}$
for comparable cross sections implying that indeed the $k_{T}$ algorithm is
efficiently vacuuming up extra particles. However, once corrected back
to the parton level, the $k_T$ algorithm cross section is smaller than
the cone result for $D=R_{cone}$ as expected.

Another concern with the $k_{T}$ algorithm is the computer time needed to
perform multiple evaluations of the list of pairs of proto-jets as 1 pair is
merged with each pass, leading to a time that grows as $N^{3}$, where $N$ is
the number of initial proto-jets in the event.
\ Recently~\cite{Cacciari:2005hq} a faster version of the $k_{T}$ algorithm,
\textquotedblleft Fastjet\textquotedblright, has been defined that
recalculates only an intelligently chosen sub-list with each pass and the time
grows only as $N\ln N$, for large $N$. \ The software in
\cite{Cacciari:2005hq} also provides alternative versions of the $k_{T}$
algorithm. \ As defined above, the algorithm is the \textit{inclusive}
version, keeping all possible jets defined by the parameter $D$. \ There is
also an \textit{exclusive }version \cite{Catani:1992zp} where another
parameter $d_{cut}$ is introduced. \ Merging stops when all remaining
$k_{T,\left(  i,j\right)  }^{2}$ and $k_{T,i}^{2}$ exceed $d_{cut}$ and the
remaining $k_{T,i}^{2}$ define the exclusively defined jets (\textit{i.e}.,
the previously removed $k_{T,i}^{2}<d_{cut}$ jets are discarded). \ The Fastjet
code also includes the so-called Cambridge/Aachen $k_{T}$ algorithm
\cite{Dokshitzer:1997in}\cite{Wobisch:1998wt}, the inclusive version of which
is defined similarly to the algorithm above except that the prefactor
$\min(p_{T,i}^{2},p_{T,j}^{2})$ is absent from $k_{T,(i,j)}^{2}$ and
$k_{T,i}^{2}=1$. \ Finally the Fastjet code for the $k_{T}$ algorithm also
includes an innovative technique for defining the \textquotedblleft
area\textquotedblright\ of the jet and allowing a correcting for the UE
contribution. \ Fake or \textquotedblleft ghost\textquotedblright\ particles
with exponentially tiny energies are added to each event on an essentially
uniform grid in $\left(  y,\phi\right)  $ (\textit{i.e.}, each ghost particle
is representative of a fixed area in $\left(  y,\phi\right)  $). \ The final
jets found by the algorithm will then contain some of the ghost particles.
\ While the kinematics of the jet is unchanged by the presence of the ghost
particles (since they have such tiny energies), their number gives a measure
of the area in $\left(  y,\phi\right)  $ of the jet.

It should also be noted that, although it would appear that the inclusive
$k_{T}$ algorithm is defined by a single parameter $D$, the suggested code for
the $k_{T}$ algorithm \cite{webpage} includes several switches and parameters
to fully define the specific implementation of the algorithm. \ Thus, as is
the case for the cone algorithm, the $k_{T}$ algorithm also exhibits
opportunities for divergence between the implementation of the algorithm in
the various experiments, and care should be taken to avoid this outcome.

\subsection{Jet Masses for Jets at NLO}

\label{sec:jetmass}

\begin{figure}[tp]
\begin{center}
\includegraphics[width=14cm]{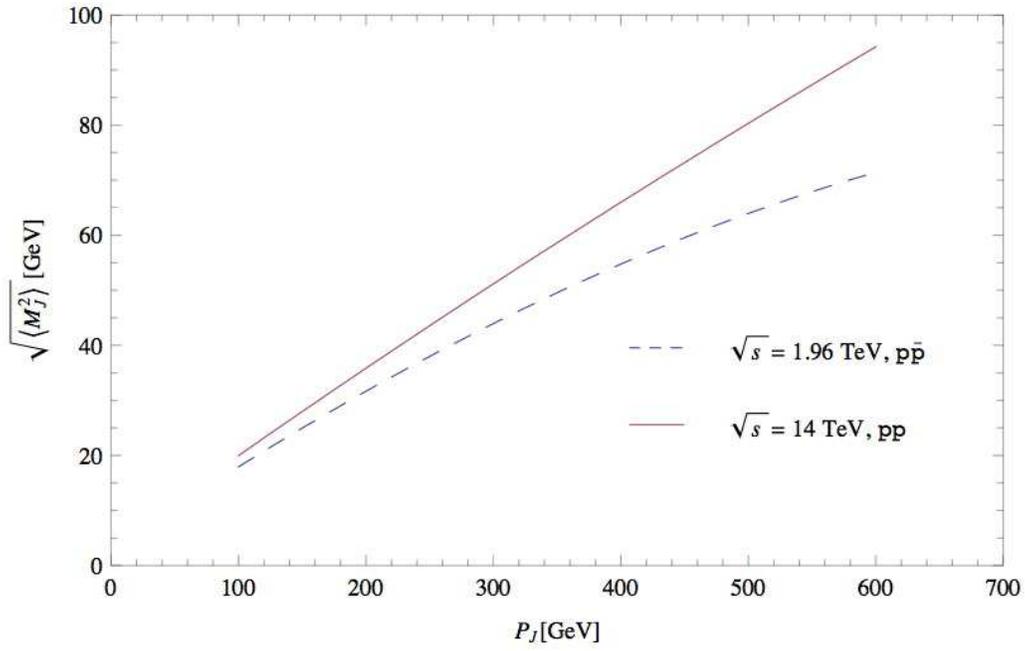}%
\end{center}
\par
\vspace*{-0.5cm}\caption{NLO jet masses for a cone jet with $R=0.7$
  and $R_{sep}=1.3$ with Tevatron and LHC energies and beams.}%
\label{mass1a}%
\end{figure}

As has already been suggested, the invariant masses of jets are expected to play an increasingly important role at the LHC as a useful jet property.  For example, jet masses can help to isolate events where individual jets correspond to essentially all of the decay products of boosted heavy objects (say top quarks or $W$ bosons).  It is helpful to set the stage for these analyses by discussing first the expected magnitude of masses of jets arising
from perturbative QCD interactions only.  We will return to the question of jet masses in Section~\ref{sec:Sparty} where showering, hadronization and UE effects will be
included.  The NLO configurations with 2 partons in a jet, corresponding to $z>0$ and $d>0$, 1 cone $\rightarrow$1 jet and 3 cones $\rightarrow$ 1 jet in Fig.~\ref{pertthy},  will result in perturbative jets with nonzero masses.  While the
distribution of jet masses will be singular at the origin in perturbation theory, due to the soft and collinear singularities, the average jet mass (squared) is an infrared safe quantity that can be evaluated order-by-order in
perturbation theory.  In particular, to find the average jet mass squared for a fixed jet $p_{J}$ at NLO we simply evaluate the NLO inclusive jet cross
section at that fixed $p_{J}$ weighted by the corresponding jet mass squared, where the evaluation involves a sum over all 2 partons-in-a-jet configurations
with the required $p_{J}$ .  We then divide this sum by the
corresponding Born level inclusive jet cross section. Note that the notation for the perturbative order used here corresponds to
the 
order of the jet cross section.  Thus at LO jets correspond to single
partons with 
vanishing jet mass.  As noted above, NLO jets receive contributions from 2
parton 
configurations with nonzero jet masses, with the masses proportional to a
single 
power of $\alpha_s$. 
 In detail
this calculation is complicated due to the large number of perturbative processes that contribute and the fact that the available phase is restricted by the pdfs in a way that varies with $p_{J}/\sqrt{s}$.  On the other hand, the expected general form
of the perturbative jet mass is straightforward to motivate.  By dimensional analysis the dominant contribution to the NLO jet mass squared will scale with
$p_{J}^{2}$, will scale with the algorithm defined \textquotedblleft
size\textquotedblright\ of the jet, $R^{2}$ or $D^{2}$ and exhibit a factor of $\alpha_{s}$.  Choosing the factorization/renormalization scale in the
running coupling to be $\mu=p_{J}/2$ we are led to expect for a cone algorithm%
\[
\left\langle M_{J}^{2}\right\rangle _{NLO} \simeq\overline{C}\left(
\frac{p_{J}}{\sqrt{s}}\right)  \alpha_{s}\left(  \frac{p_{J}}{2}\right)
p_{J}^{2}R^{2},
\]
where the prefactor (prefunction) $\overline{C}$ has a magnitude of
order unity and decreases slowly with increasing $p_{J}/\sqrt{s}$.
Using the CTEQ 6.2 pdfs, averaging over jet rapidities in the range
$\left\vert y_{J}\right\vert \leq2.5$ and employing the EKS NLO
inclusive jet 
code~\cite{Ellis:1992en} for the cone jet
algorithm, we find the average NLO jet masses illustrated in the
following figures.  Due to the expected simple linear momentum
dependence we will focus on the linear jet mass,
$\sqrt{\left\langle
    M_{J}^{2}\right\rangle
  _{NLO}}\simeq\sqrt{\overline{C}\alpha_{s}}p_{J}R$.
The simplest
dimensionful NLO predictions for the jet mass at both Tevatron and LHC
energies are illustrated
in Fig.~\ref{mass1a}.%

\begin{figure}[thp]
\begin{center}
\includegraphics[width=14cm]{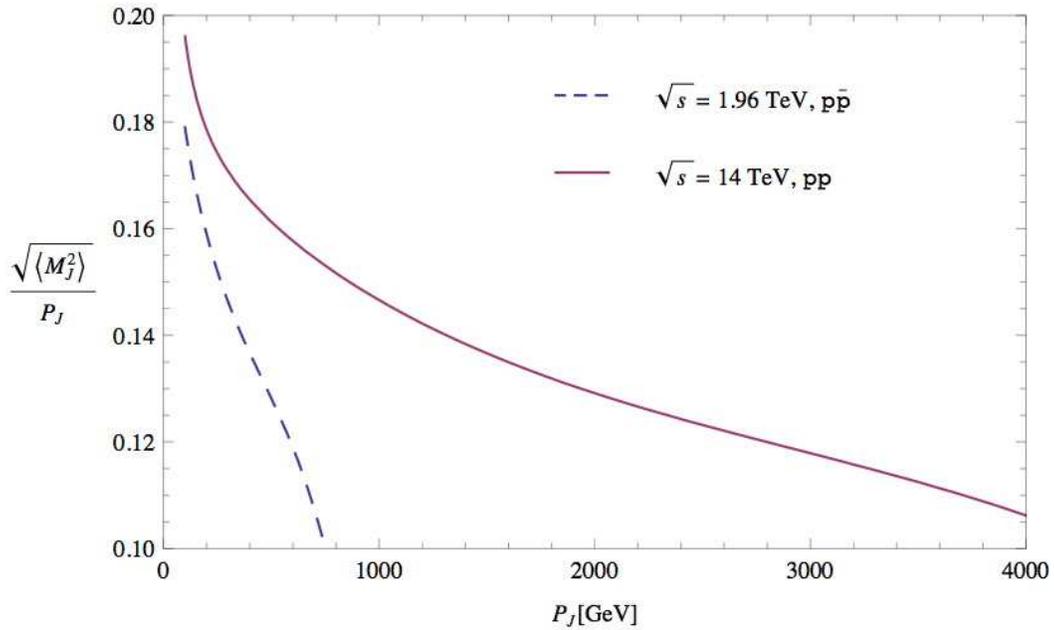}%
\end{center}
\par
\vspace*{-0.5cm}\caption{NLO jet mass for a cone jet with $R=0.7$ and
  $R_{sep}=1.3$ scaled by
  the jet momentum $p_{J}$ with Tevatron and LHC energies and beams.}%
\label{mass1b}%
\end{figure}

\begin{figure}[pth]
\begin{center}
\includegraphics[width=14cm]{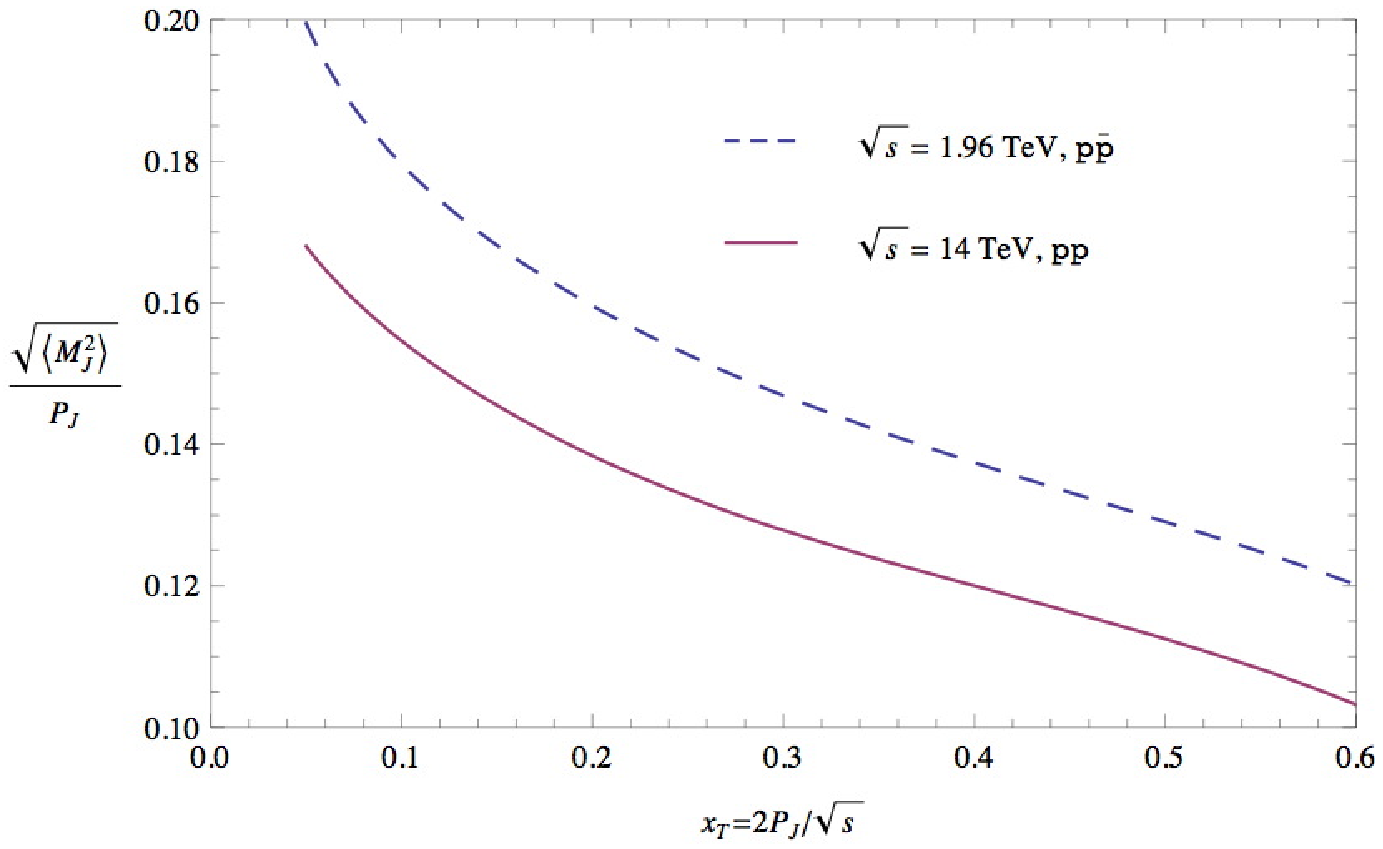}%
\end{center}
\par
\vspace*{-0.5cm}\caption{NLO jet mass for a cone jet with $R=0.7$ and
  $R_{sep}=1.3$ scaled to
  the jet momentum $p_{J}$ plotted versus $x_{T}=2p_{J}/\sqrt{s}$ with
  Tevatron and LHC energies and beams.}%
\label{mass2}%
\end{figure}

\begin{figure}[t]
\begin{center}
\includegraphics[width=14cm]{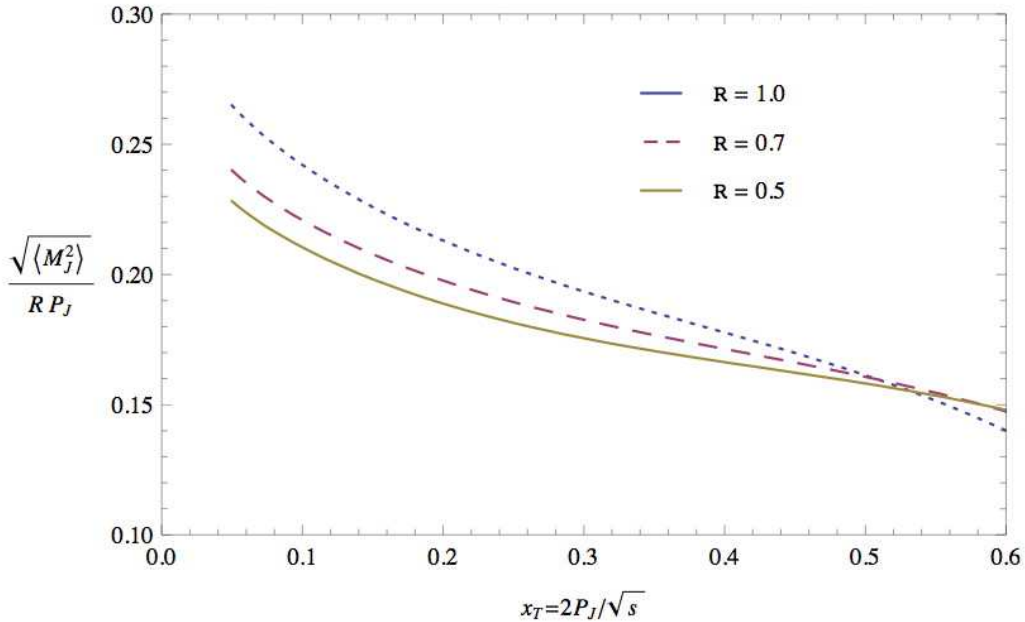}%
\end{center}
\par
\vspace*{-0.5cm}\caption{NLO jet mass for a cone jet with $R_{sep}=1.3$, $\sqrt{s}=14$ TeV
scaled by the jet momentum $p_{J}$ and the cone radius $R$ plotted versus
$x_{T}=2p_{J}/\sqrt{s}$ for various values of the cone radius $R$.}%
\label{mass3}%
\end{figure}

We see that for low momentum jets, $p_{J}\sim100$ GeV, the jet mass is
relatively independent of the overall energy $s$. \ On the other hand, for
substantially larger momenta the reduced phase space at the Tevatron leads to
smaller predicted jet masses at the same momentum. \ We can simplify the
discussion by first scaling out the (necessary dimensionful) linear dependence
on $p_{J}$ as displayed in Fig.~\ref{mass1b}.%
As suggested earlier we can largely correct for the overall energy difference
by plotting instead versus the momentum \textit{fraction}, $x_{T}=2p_{J}%
/\sqrt{s}$, as illustrated in Fig.~\ref{mass2}.%
Now the impact of the differing beam energies (and beam flavor) is much
reduced with scaled jet mass distributions of very similar shape, and the scaled LHC jet mass smaller than the scaled Tevatron jet mass at the same $x_{T}$ by approximately 10 to 15\%. \ The common falling shape of the distributions as functions of $x_{T}$ can be understood as arising from the falling coupling, the falling pdfs and the transition from dominantly gluons at small $x_{T}$ with $\overline{C}\propto C_{A}=3$ to dominantly quarks (and anti-quarks) at large $x_{T}$ with $\overline{C}\propto C_{F}=4/3.$  The difference in the magnitude of the two distributions arises from the scale breaking in the theory ($p_{J}$ is larger at the LHC at the same $x_{T}$) with
both the running coupling and the running pdfs being smaller at the LHC.  We can verify the approximately linear dependence on the cone radius $R$ by both varying the radius and scaling it out as in 
Fig.\ref{mass3} for LHC beam energies.

While the dependences on $p_{J}$ and $R$ are not exactly linear, the NLO jet
mass is remarkably well described by the simple rule-of-thumb%
\[
\sqrt{\left\langle M_{J}^{2}\right\rangle_{NLO} }\approx0.2p_{J}R,
\]
where the numerical prefactor of $0.2$ (approximately) includes the dependence
on $\sqrt{\alpha_{s}}$, the color charges (a mix of $C_{F}$ and $C_{A}$) and
the pdfs. \ For better than 25\% accuracy
the still simple prefactor $\left(
0.2+\left(  0.3-x_{T}\right)  /6\right)  $ suffices. \ 

\begin{figure}[t]
\begin{center}
\includegraphics[width=14cm]{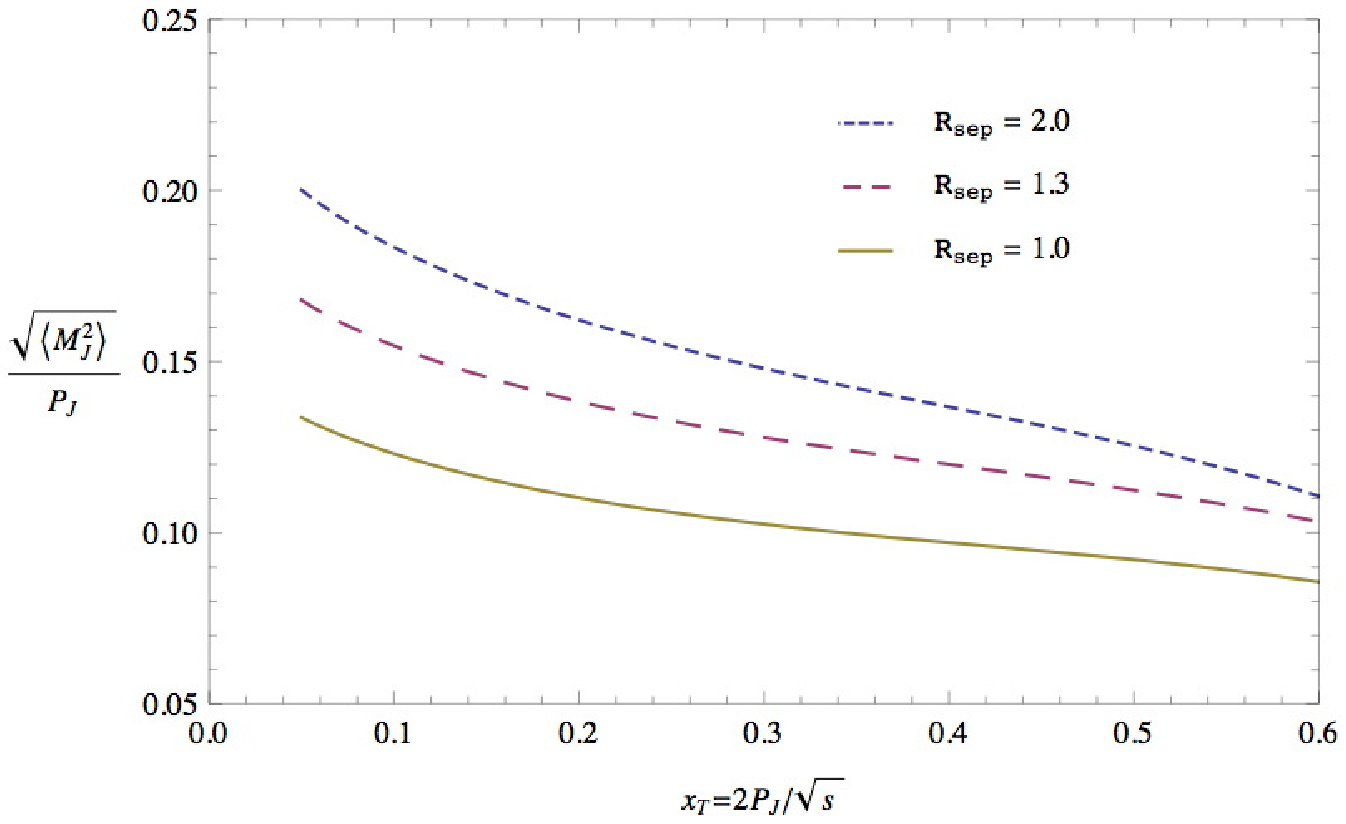}%
\end{center}
\par
\vspace*{-0.5cm}\caption{NLO jet mass for a cone jet with $R=0.7$, $\sqrt{s}=14$ TeV scaled by
the jet momentum $p_{J}$ plotted versus $x_{T}=2p_{J}/\sqrt{s}$ for various
values of the cone radius $R_{sep}$.}%
\label{mass4}%
\end{figure}
Finally consider the dependence on the \textit{ad hoc} parameter $R_{sep}$.
This is illustrated in Fig.~\ref{mass4}.
We see the expected monotonic dependence on $R_{sep}$, where the variation
is somewhat less rapid than linear.  In the language introduced earlier we see 
that the mass for a naive Snowmass cone jet ($R_{sep}=2.0$) is expected 
(in NLO perturbation theory) to be approximately 30 to 50\% larger than the 
mass of a $k_{T}$ jet ($R_{sep}=1.0$) with $D=R$.  Qualitatively we
anticipate 
that, compared to NLO perturbation theory, the inclusion
of showering, hadronization and the underlying event will lead to
``splash-out'' 
effects and smaller jet masses for the cone algorithm, but ``splash-in'' and
larger jet 
masses for the $k_T$ algorithm.  Hence the masses of jets found in realistic
environments 
by different algorithms, as discussed in Section~\ref{sec:Sparty}), are
expected to be more
similar than suggested by NLO perturbation theory and Fig.~\ref{mass4}.

In summary, perturbative QCD (alone) leads us to expect jet masses at
the LHC that grow slightly more slowly than linearly with the jet
momentum, scale linearly with the jet algorithm defined jet size, $R$
or $D$, and have a magnitude approximately $\left(  \pm25\%\right)  $
described by $\sqrt{\left\langle M_{J}^{2}\right\rangle_{NLO}}\approx0.2p_{J}R$.
The explicit prefactor ($0.2$) corresponds to $R_{sep} = 1.3$ cone jets with
Snowmass jets ($R_{sep} = 2.0$) having masses larger by $10\%$ and NLO $k_T$
jets ($R_{sep}=1.0$) having masses smaller by $20\%$.

\subsection{Recent Cone Algorithm Issues}

\subsubsection{Jets at the \textquotedblleft Smeared\textquotedblright\ Parton
Level and Dark Towers}
\label{sec:smearing_dark_towers}

\begin{figure}[htp]
\centerline{
\includegraphics[
width=4.3007in
]{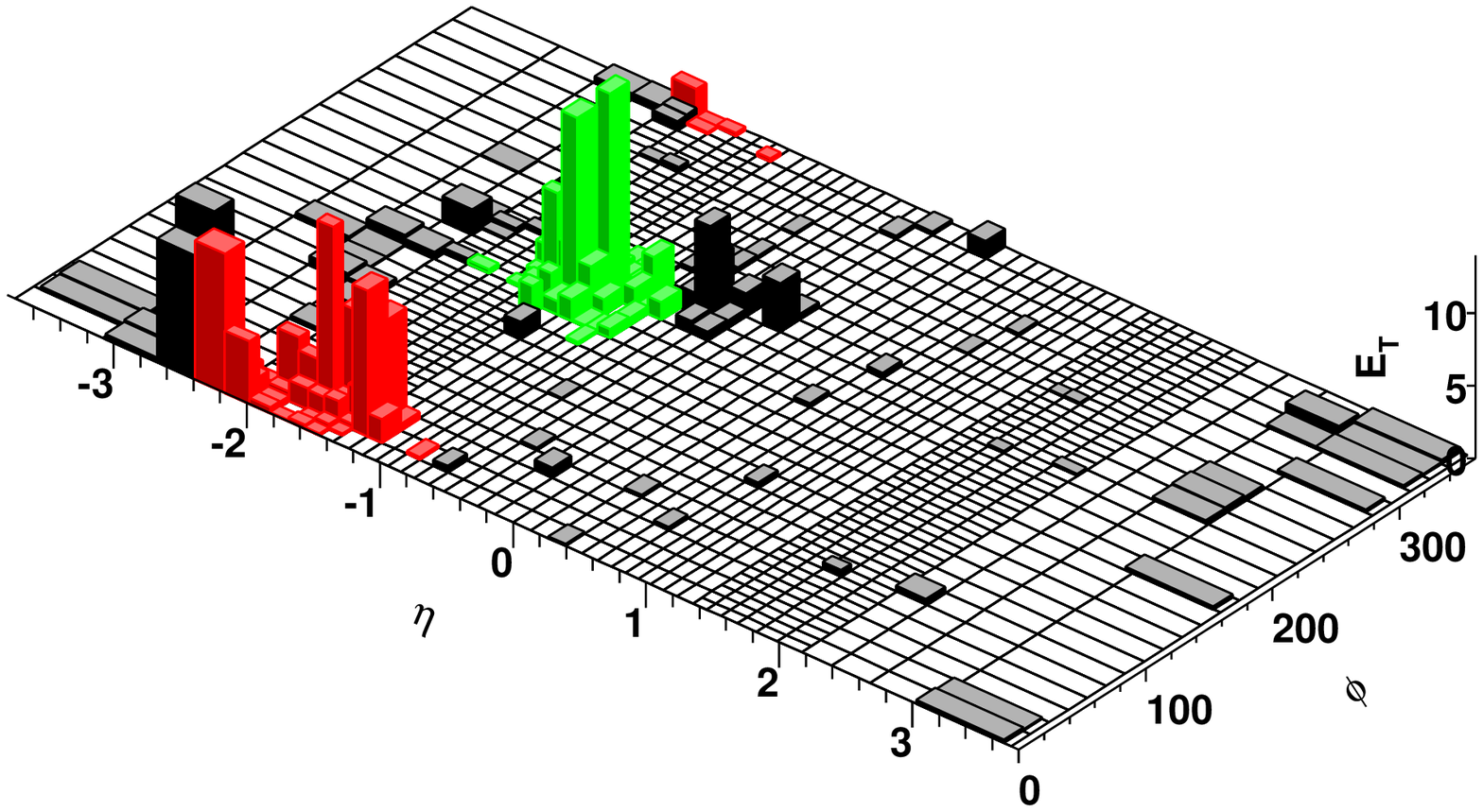}} \centerline{
}\caption{An example of a Monte Carlo inclusive jet event where the Midpoint
algorithm has left substantial energy unclustered. }%
\label{dark_towers}%
\end{figure}

In studies of the Run II\ iterative cone algorithms a previously unnoticed
problem has been identified~\cite{Ellis:2001aa} at the particle and
calorimeter level, which is explicitly not present at the NLO parton level.
\ It is observed in a (relatively small) fraction of the events that some
energetic particles/calorimeter towers remain unclustered in any jet. \ This
effect is understood to arise in configurations of two nearby (\textit{i.e}.,
nearby on the scale of the cone size) showers, where one shower is of
substantially larger energy. \ Any \ trial cone at the location of the lower
energy shower will include contributions from the larger energy shower, and
the resulting centroid will migrate towards the larger energy peak. This
feature is labeled \textquotedblleft dark towers\textquotedblright\ in Ref.
~\cite{Ellis:2001aa}, \textit{i.e.,} clusters that have a transverse momentum
large enough to be designated either a separate jet or to be included in an
existing nearby jet, but which are not clustered into either. \ A Monte Carlo
event with this structure is shown in Fig.~\ref{dark_towers}, where the towers
unclustered into any jet are shaded black.

\begin{figure}[htp]
\centerline{
\includegraphics[
width=4.3007in
]{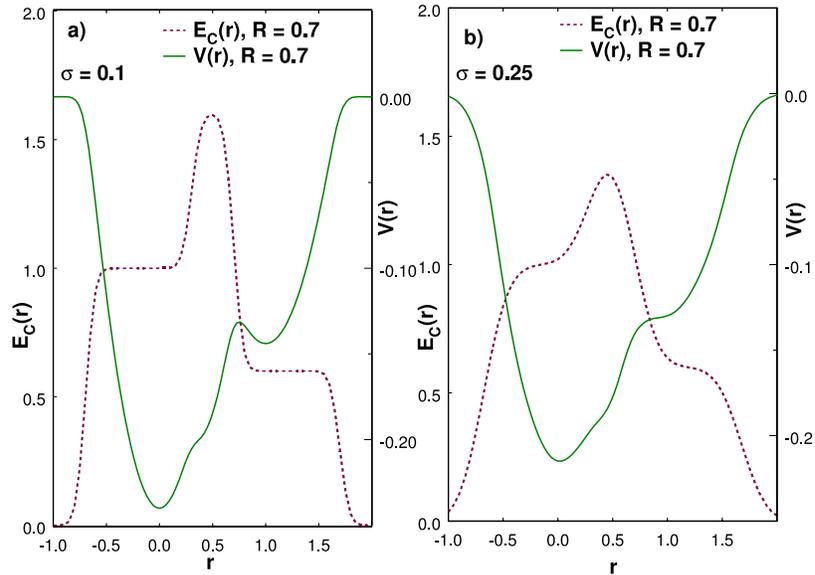}} \centerline{
}\caption{Energy-in-cone and potential distributions corresponding to Gaussian
smearing with a) $\sigma=0.1$ and b) $\sigma=0.25$ for $d=1.0$ and $z=0.6$. }%
\label{smearpt}%
\end{figure}

A simple way of understanding the dark towers can be motivated by returning to
Fig.~\ref{edist}, where the only smearing in $\left(  y,\phi\right)  $ between
the localized energy distribution of panel a) (the individual partons) and the
\textquotedblleft potential\textquotedblright\ of panel b) arises from the size
of the cone itself. \ On the other hand, we know that showering and
hadronization will lead to further smearing of the energy distribution and
thus of the potential. \ Sketched in Fig.~\ref{smearpt} is the potential (and
the energy-in-cone) distributions that results from Gaussian smearing with a
width of a) $\sigma=0.1$ and b) $\sigma=0.25$ (in the same angular units as
$R=0.7$). \ Note that the smearing of the potential occurs in 2 dimensions and
that here we are considering only a 1 dimensional slice. \ In both panels, as
in Fig.~\ref{edist}, the partons have $p_{T}$ ratio $z=0.6$ and angular
separation $d=1.0$. \ 
Note that as the smearing increases from zero as in panel b) of
Fig.~\ref{edist} to the smeared results in Fig.~\ref{smearpt},
we first lose the (not so deep) minimum corresponding to the
midpoint stable cone (and jet), providing another piece of the explanation for
why showers more than $1.3\cdot R_{cone}$ apart are not observed to merge by
the experiments. \ \ In panel b), with even more smearing, the minimum in the
potential near the shower from the lower energy parton also vanishes, meaning
this (lower energy) shower is part of no stable cone or jet, \textit{i.e}.,
leading to dark towers. \ Any attempt to place the center of a trial cone at
the position of the right parton will result in the centroid \textquotedblleft
flowing\textquotedblright\ to the position of the left parton and the energy
corresponding to the right parton remaining unclustered in any jet. Note that
the Run I CDF algorithm, JETCLU with ratcheting, limited the role of dark
towers by never allowing a trial cone to leave the seed towers, the potential
dark towers, behind. \ The effective smearing in the data is expected to lie
between $\sigma$ values of 0.1 and 0.25 (with shower-to-shower fluctuations
and some energy dependence, being larger for smaller $p_{T}$ jets) making this
discussion relevant, but this question awaits further studies. \ Note that Fig.
\ref{smearpt} also suggests that the Midpoint algorithm will not entirely fix
the original issue of missing merged jets. \ Due to the presence of (real)
smearing this middle cone is often unstable and the merged jet configuration
will not be found even though we explicitly look for it with the Midpoint
cone. \ Thus, even using the recommended Midpoint algorithm (with seeds),
there may remain a phenomenological need for the parameter value $R_{sep}<2$,
\textit{i.e}., a continuing mismatch between data and pQCD that requires correction.

\subsubsection{The Search Cone Algorithm}
\label{sec:searchcone}

A potential solution for the dark towers problem is described in 
Ref.~\cite{Ellis:2001aa}. \ The idea is to decouple the jet finding step from the
jet construction step. \ In particular, the stable cone finding procedure is
performed with a cone of radius half that of the final jet radius,
\textit{i.e}., the radius of the search cone, $R_{search}=R_{cone}/2$. \ This
procedure reduces the smearing in Figs. \ref{edist} and \ref{smearpt}, and
reduces the phase space for configurations that lead to dark towers (and
missing merged jets). \begin{figure}[h]
\centerline{
\includegraphics[
width=4.3007in
]{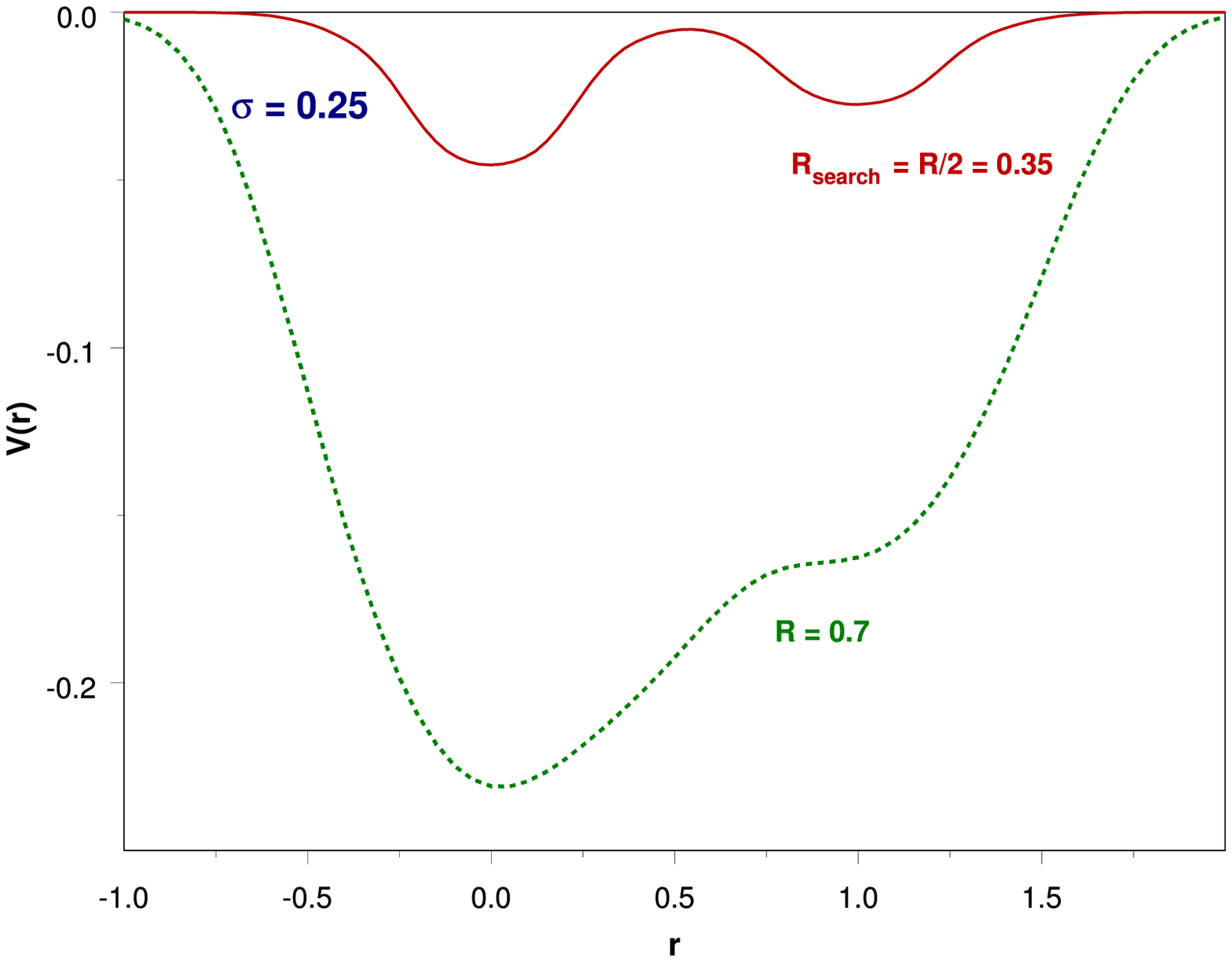}}
\centerline{
}\caption{The stable cone finding potential with the reduced search cone,
$R_{search}=R_{cone}/2$. \ The original potential from Fig.~\ref{smearpt},
panel b) with $R_{search}=R_{cone}$ is indicated as the dashed curve. }%
\label{smearpotfix}%
\end{figure}This point is illustrated in Fig.~\ref{smearpotfix}, which shows
the potential of Fig.~\ref{smearpt}, panel b) corresponding to the reduced
radius search cone. \ Note, in particular, that there is again a minimum at
the location of the second parton. \ Seeds placed at each parton will yield a
stable cone at each location even after the smearing. \ Using the smaller
search cone size means there is less influence from the (smeared) energy of
nearby partons. After identifying the locations of stable cones, the larger
cone size, \textit{e.g.}, $R_{jet}=R_{cone}=0.7$, is used to sum all objects
inside and construct the energy and momentum of the jet (with no iteration).
All pairs of stable cones separated by less than $2R_{cone}$ are then used to
define midpoint seeds as in the usual Midpoint cone algorithm. A trial cone of
size $R_{cone}$ is placed at each such seed and iterated to test for
stability. (Note that this Midpoint cone is iterated with cone size $R_{cone}%
$, not the smaller $R_{search}$, contrary to what is described in the
literature~\cite{Acosta:2005ix}.) Thus, just as in the Midpoint cone algorithm, stable midpoint
cones will be found by the CDF Search cone algorithm. However, as already
discussed, we expect that there will be no stable midpoint cone due to the
smearing. Note that, even with the reduced smearing when using the smaller
search cone radius, there is still no central stable cone in the potential of
Fig.~\ref{smearpotfix}. On the other hand, as applied to NLO perturbation
theory without smearing, the Search cone algorithm should act like the usual
Midpoint cone algorithm and yield the na\"{\i}ve result of Fig.~\ref{pertthy}
a). The net impact of adding the step with the smaller initial search cone as
applied to data is an approximately 5\% increase in the inclusive jet cross
section. \ In fact, as applied to data the Search cone algorithm identifies so
many more stable cones, that the CDF collaboration has decided to use the
Search cone algorithm with the merging parameter $f_{merge}=0.75$ (instead of
$0.5$) to limit the level of merging. \ 

Unfortunately, a disturbing feature of the Search Cone Algorithm arises when
it is applied at higher orders in perturbation theory as was pointed out
during the TeV4LHC Workshop. At NNLO in perturbation theory, the Search cone
algorithm can act much like the seeds discussed earlier. \ In particular, the
Search cone algorithm can identify a (small radius $R_{search}$) stable
(soft) cone between two energetic cones arising from the soft gluon between 2
energetic partons configuration discussed earlier. \ The soft search cone is
stable exactly because it \textquotedblleft fits\textquotedblright\ between
the two energetic partons without including either; the spacing between the
two energetic partons can be in the range $2R_{search}=R_{cone}<\Delta
R<2R_{cone}$. \ Then, when the radius of the (stable, soft) search cone is
increased to $R_{cone}$, the resulting full size cone will envelop, and serve
to merge, the two energetic partons. \ This can occur even when the two
energetic partons do not constitute a stable combined cone in the standard
cone algorithm. Thus at NNLO, the Search cone algorithm can exhibit an
IR-sensitivity very similar to, and just as undesirable as, the seed-induced
problem discussed earlier. \ The conclusion is that the Search cone algorithm,
while it does address the dark tower issue, creates its own set of issues and
is not considered to be a real solution of the dark tower problem.

\subsubsection{The Midpoint Cone Algorithm with a \textquotedblleft Second
Pass\textquotedblright}

The recommendation adopted in the TeV4LHC workshop, and which we also endorse,
is to remove the Search Cone from the Midpoint cone algorithm and to
substitute a ``second-pass'' step. After all stable cones are found in an
event, the towers corresponding to those jets are removed from the list, and
the Midpoint cone algorithm is run again. The ``dark towers'' are then
reconstructed as jets in their own right, as the attractive influence of
nearby larger jets is no longer present. The question remains as to what to do
with the second-pass jets. According to the criteria discussed above, some of
them should be merged with the larger nearby jet, while others should remain
as separate jets. As separate jets, the second-pass jets have an insignificant
impact on inclusive jet cross sections. If added to the nearby larger jets,
they can have an effect of the order of up to 5\%. The correct treatment is
still under investigation. Note that the second-pass jets may also be
important in the accurate reconstruction of a complicated multi-jet final
state, such as in $t\overline{t}$ events.

It is also important to keep in mind that the experimental reality in which
jet algorithms are applied is more complex than the 1-D slice shown in
Fig.~\ref{smearpt}. The final-state hadrons are distributed in two dimensions
($y$ and $\phi$) and the resultant Snowmass potential has fluctuations in
these two dimensions. This has implications regarding the presence or not of
dark towers and of the effect of the distribution of the jet energies on the
stability of cone centroids.

\subsubsection{Summary}

In summary,
to compare fixed order perturbation theory with data there must be corrections
for detector effects, for the splash-in contributions of the underlying event
(and pile-up) and for the splash-out effects of hadronization (and for
showering when comparing to leading order calculations). \ It is the response
to both the splash-in and splash-out effects that 
distinguishes the various cone algorithms and
drives the issues we have just been discussing. \ 
It is also important to recognize that the splash-in and splash-out effects
(and corrections) come with opposite signs and there can be substantial
cancellation.
We will return to the question of Run
II\ corrections below. \ The conclusion from the previous discussion is that
it would be very helpful to include also a correction in the experimental
analysis that accounts for the use of seeds, or to use a seedless cone
algorithm such as SISCone~\cite{Salam:2007xv}. \ Then
these experimental results could be compared to perturbative results without
seeds, avoiding the inherent infrared problems caused by seeds in perturbative
analyses. At the same time, the analysis described above suggests that using
the Midpoint cone algorithm, to remove the impact of seeds at NLO , does not
fully eliminate the impact of the smearing due to showering and hadronization,
which serves to render the Midpoint cone of fixed order perturbation theory
unstable. The same may be true, but possibly to a lesser extent, with the use
of a seedless algorithm such as SISCone, which increases the number of
identified stable cones. \ Thus we should still not expect to be able to
compare data to NLO theory with $R_{sep}=2$, although the impact may be
somewhat reduced.
The possible downside of finding more stable cones is that the split/merge
step will play a larger role, with no analogue effect in the perturbative
analysis. \ Splitting the calorimeter towers with a host of low $E_{T}$ stable
cones can serve to lower the $E_{T}$ of a leading jet and introduce an
undesirable dependence on the minimum $p_{T}$ cut defining which stable cones
are included in the split/merge process. \ This is surely only a percent level
correction, but control at that level is the ultimate goal.

\subsection{Jets at the Hadron Level}

\label{sec:hadron}

\begin{figure}[th]
\centering
\includegraphics[width=.49\textwidth, trim= 0 50 0 100]{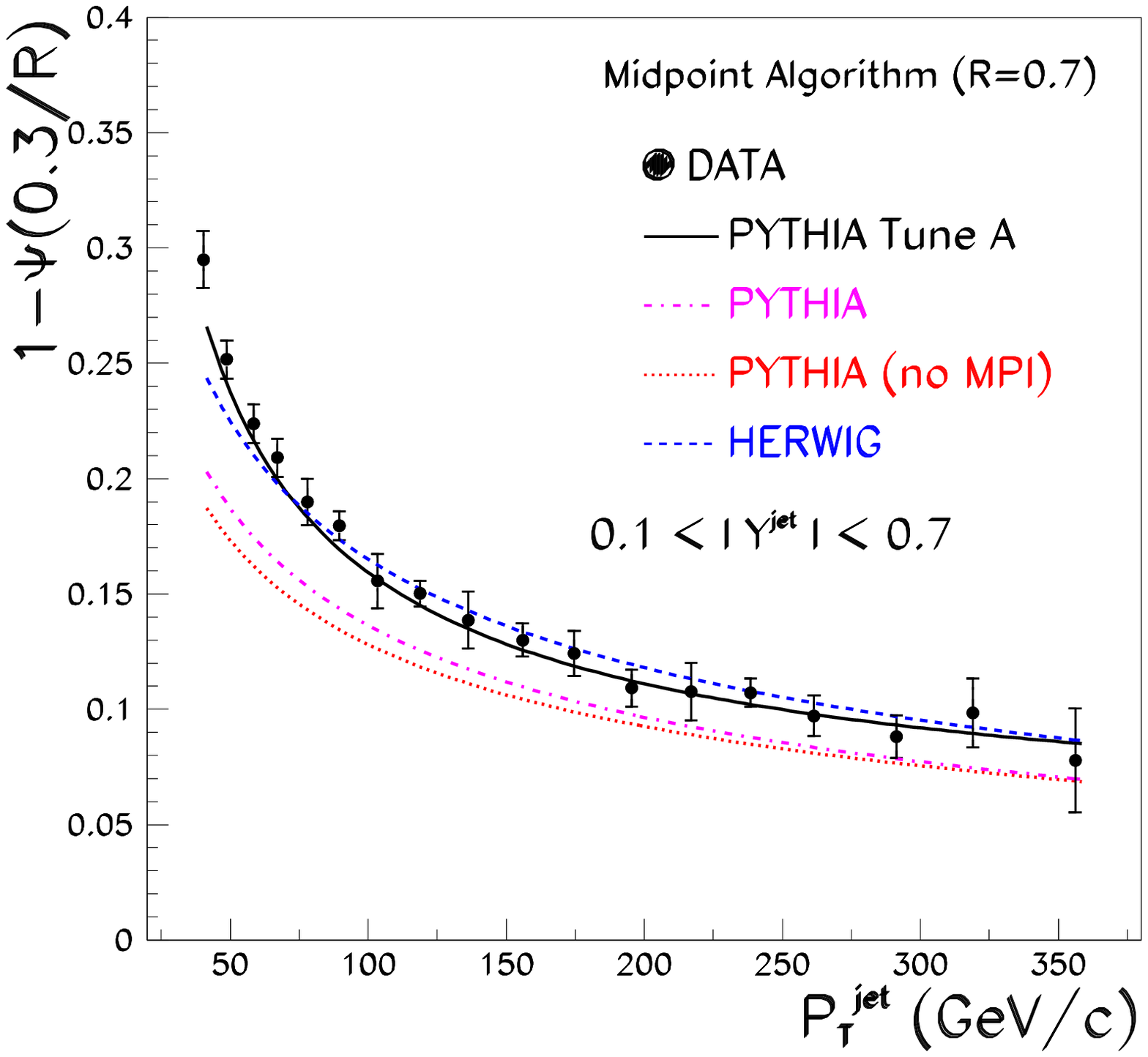}
\includegraphics[width=.49\textwidth, trim= 0 50 0 100]{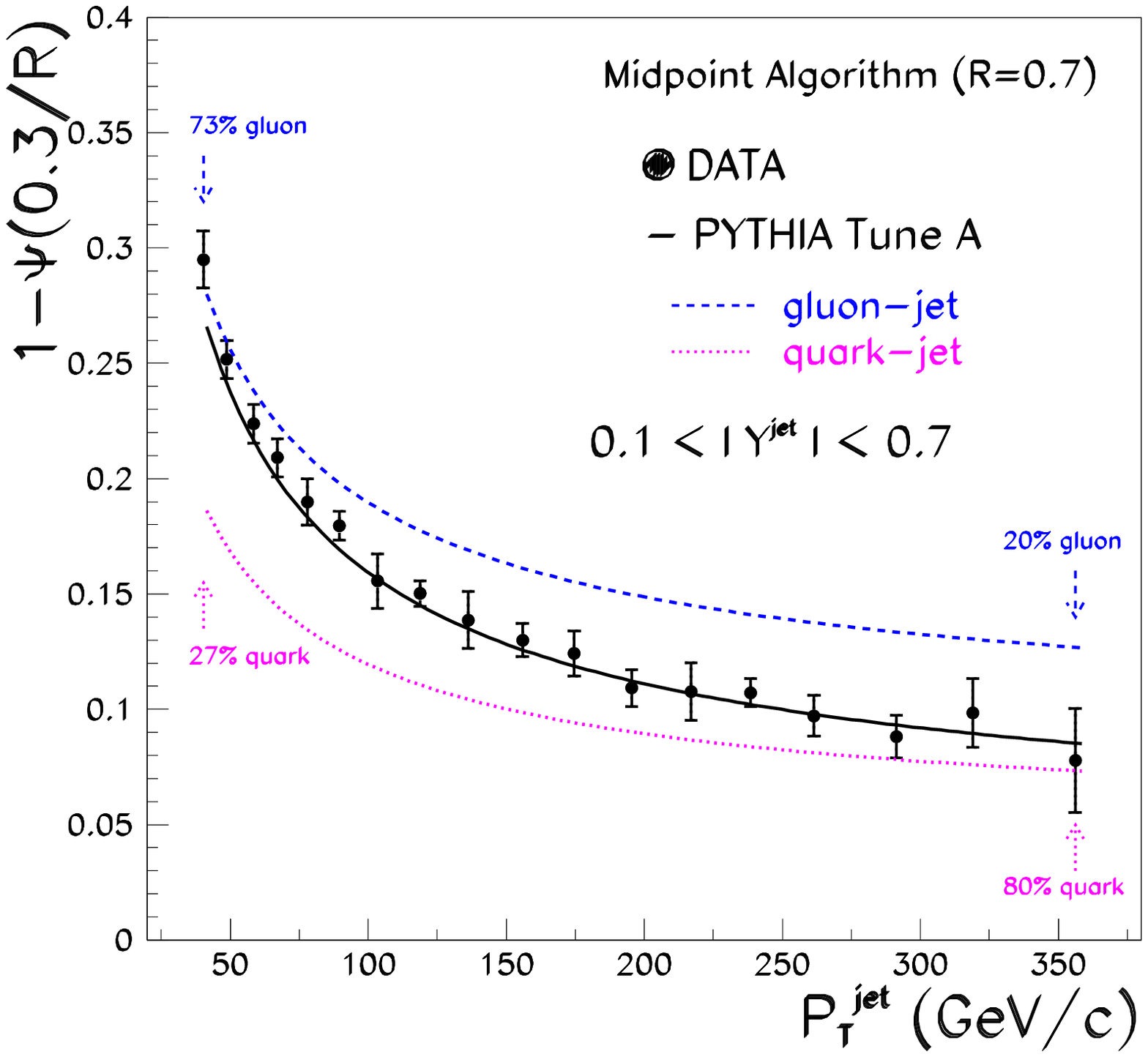}\caption{The
fraction of $p_{T}$ in a cone jet of radius 0.7 that lies in the annulus from
0.3 to 0.7 as a function of jet $p_{T}$. The CDF Run II
measurement 
is compared to predictions from Herwig and Pythia with
different set of parameters (left), and to the separate predictions for quark
and gluon jets (right). Figures from Ref.~\cite{Acosta:2005ix}.}
\label{fig:shapes}%
\end{figure}

The last step before the detector in our factorized, approximate picture of a
hard scattering event is the hadronization step. \ Here the colored partons
arising from the hard scattering itself, from ISR and FSR and from the UE are
organized into color singlet hadrons. \ Thus this step requires information
about the color flow within the event as the event evolves from short
to long distance, and that information must be maintained
during the showering process. \ The initial
color single hadrons will include resonances, which are then decayed into the
long lifetime, long distance final hadrons (primarily pions and kaons)
according to the known properties of the resonances. \ Essentially by
definition, this step represents non-perturbative physics, is model dependent
and therefore not well understood. \ On the other hand, by construction, it
looks for minimal mass color singlet final states while conserving energy and
momentum and so results in only a minimal kinematic rearrangement. \ Thus,
although there is considerable uncertainty about the details of hadronization,
the impact on jet reconstruction is small and is fairly easily corrected for
in both types of algorithms except perhaps at the lowest jet energies.
\ In experimental analyses, these corrections are estimated using leading-order
parton-shower Monte Carlo generators, by observing the variation of the
predicted jet cross sections after turning off the interaction between beam
remnants and the hadronization. This procedure relies on the Monte Carlo
providing a good description of those observables in the data that are most
sensitive to non-perturbative contributions such as, for example, the
underlying event energy away from jets and the internal structure of the jets.
Recent precise measurements on jet shapes~\cite{Acosta:2005ix}, as indicated
in Fig.~\ref{fig:shapes}, have allowed the detailed study of the models
employed to describe the underlying event in inclusive jet production at the
Tevatron (see also Ref.~\cite{Field:2005yw}). \ \ Besides the contribution
from the UE the jet shape is also sensitive to the changing character of the
scattered parton, \textit{i.e}., gluons at low $p_{T}$ and quarks at high
$p_{T}$, and to the perturbative scaling of the jet size with $\alpha
_{s}\left(  p_{T}\right)  $. \ The Tevatron studies, which will be discussed
in more detail below, suggest that these last effects are reasonably well
described both by the showers in Monte Carlo simulations and by pQCD analyses.
\ Future measurements of the underlying event in Run II, for different
hadronic final states, promise to play a major role in the understanding of
the measured jet cross sections at the LHC.

Overall the challenge for the future, as noted above, is to continue to reduce
the systematic uncertainties below the current 10\% to 50\% level, which
effectively guarantees agreement with theory. \ Indeed, Run II studies of the
corrections due to splash-in, \textit{i.e}., the underlying event (and
pile-up), and the splash-out corrections due to hadronization are much more
sophisticated than in Run I and presented in such a way that they can be
applied either to the data (corrected for the detector) or to a theoretical
(perturbative) calculation. \ The evaluation of these corrections is based on
data and the standard Monte Carlos, Pythia and Herwig, especially Tune A of
Pythia, which seems to accurately simulate the underlying event in terms of
multiple parton interactions, including the observed correlations with the
hard scattering process~\cite{Field:2005yw}. \ We turn now to a detailed
discussion of the experiences and lessoned learned at the Tevatron.

\section{Jets at the Tevatron}

\label{sec:tevatron_jets}

Most of the interesting physics processes in $p\bar p$ collisions at the
Tevatron include jets in the final state.
Therefore, the reconstruction and precise measurement of jets serve as key
ingredients for the better understanding of hard scattering processes and for
new physics searches at the Tevatron.

As an example, consider top quark production. \ The top quark was discovered
by the CDF and D\O \ experiments in 1995~\cite{Abe:1994xt, Abe:1995hr,
Abachi:1995iq}, and since then studies of the top quark production and
properties have been considered to be one of the primary physics goals for the
Tevatron experiments. Top quark decays always have at least one jet in the
final state
and the systematic uncertainty of the top quark mass measurement
is dominated by the jet energy scale uncertainty. The search for the Standard
Model Higgs boson is also an important aspect of the Tevatron physics program.
The Standard Model Higgs boson dominantly decays to $b\bar{b}$ for masses
$m_{H}\lesssim135$ GeV/\textit{c}. The $b\bar{b}$ dijet mass resolution has
been stressed as one of the critical factors in the search for a light Higgs
boson decay at the Tevatron~\cite{Carena:2000yx,higgs_sensitivity_tevatron}; a
20\% decrease of $\sigma_{m}/m_{b\bar{b}}$ is expected to be have a similar
effect as a 20\% increase in the accumulated luminosity. Inclusive jet cross
section measurements have also been undertaken at the Tevatron. Such
measurements are sensitive to new physics such as quark compositeness and the
presence of new heavy particles from Beyond-the-Standard-Model scenarios, and
also they provide crucial information on parton distribution functions for the
proton.

\subsection{Detectors}

\begin{figure}[th]
\centering\leavevmode
\begin{tabular}
[t]{c}%
\subfigure[]{ \includegraphics[width=0.48\hsize,bb=0 -150 567 547,clip=]
{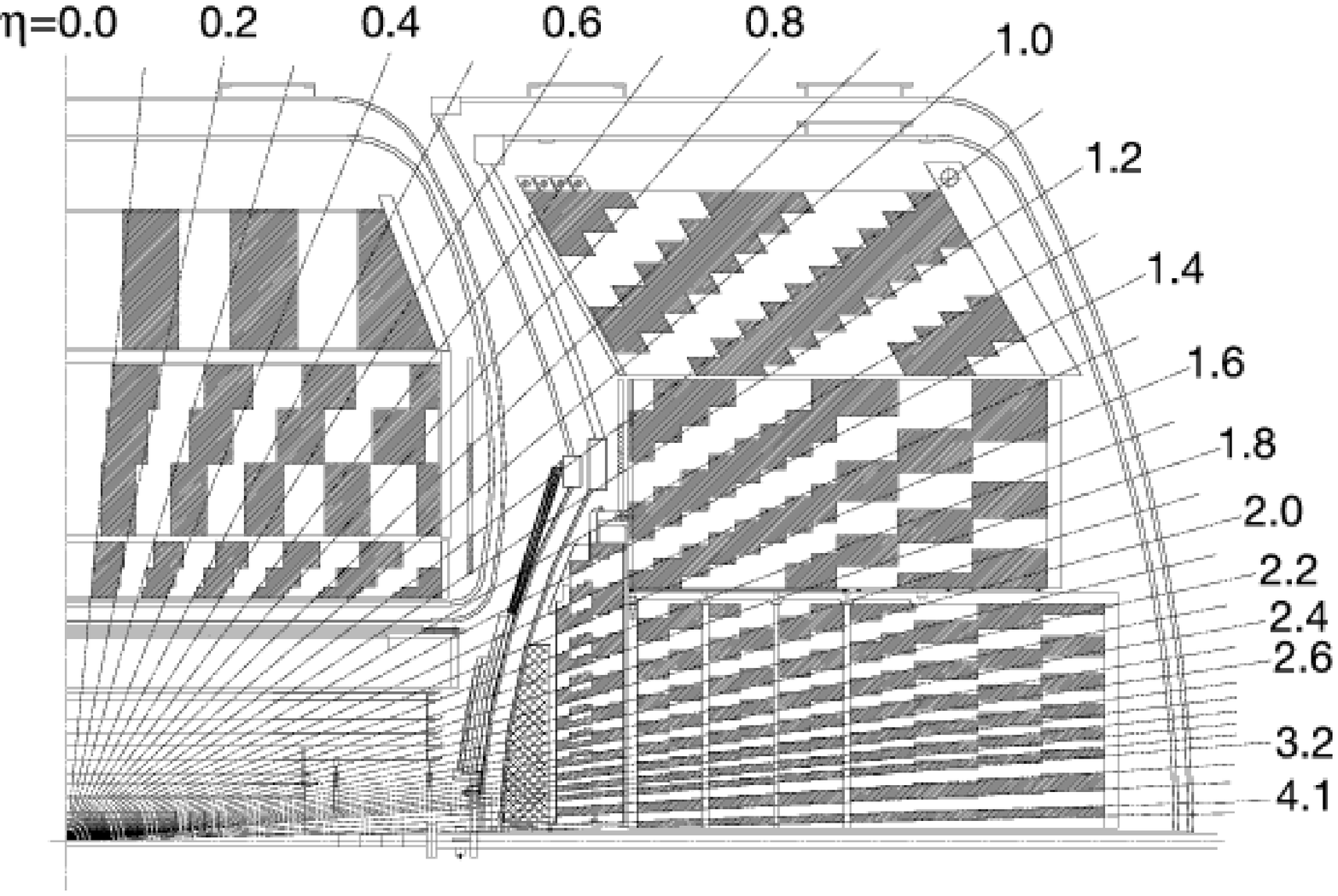} }
\subfigure[]{ \includegraphics[width=0.40\hsize,clip=]
{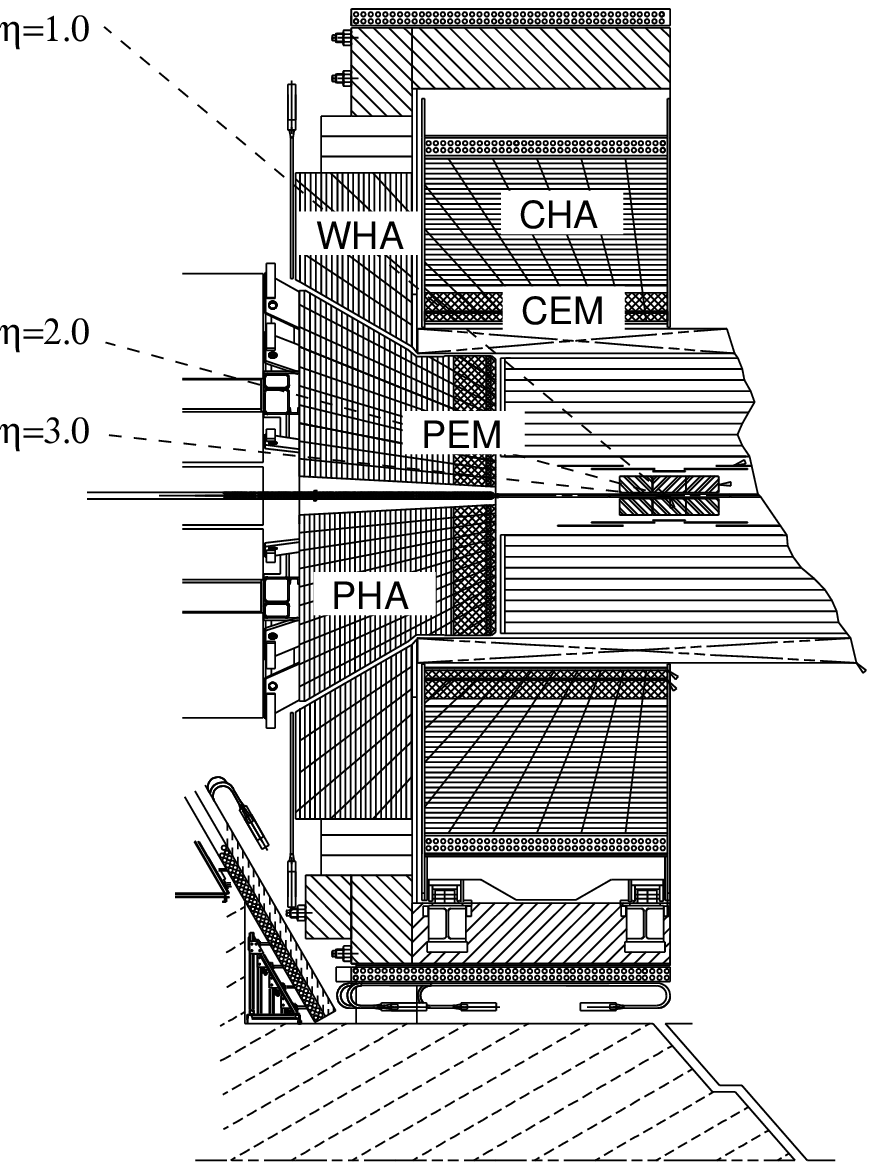} }
\end{tabular}
\caption{Schematic views of a portion of the D\O \ (a) and CDF (b)
calorimeters. Figures from Refs.~\cite{D0JES-Run2} and \cite{Bhatti:2005ai},
respectively.}%
\label{fig:cdf_dzero_calorimeters}%
\end{figure}

CDF and D\O \ are the two general-purpose detectors designed for measurement
of $p\bar{p}$ collisions at the Fermilab Tevatron collider. Both detectors are
composed of a solenoidal-magnet charged particle spectrometer surrounded by
(sampling) calorimetry and a system of muon chambers. The components most
relevant for jets are the calorimeters, which are used to measure the energy
and angle of particles produced in $p\bar{p}$ collisions. The CDF and
D\O \ calorimeters both have a projective tower geometry as shown in
Fig.~\ref{fig:cdf_dzero_calorimeters}.

The D\O \ calorimeters are uranium and liquid argon sampling calorimeters.
The D\O \ calorimetry consists of a central section covering $|\eta|\lesssim1$
and an endcap section extending the coverage to $|\eta|\sim4$. The calorimetry
has three sections: EM, fine hadronic (FH) and coarse hadronic (CH). In the CH
section, copper or steel is used instead of uranium as an absorber. The
D\O \ calorimeters are nearly compensating, with an $e/\pi$ ratio less than
1.05 above 30 GeV. The tower segmentation in $\eta$-$\phi$ space is
$0.1\times5.625^{\circ}$. The average noise per channel is on the order of 50 MeV.
The relatively long charge collection time for the liquid argon signal and the resultant electronic shaping of the signal results in the possibility of towers with (modest) negative energy. 

The CDF electromagnetic (EM) section consists of alternating layers of lead
and scintillator, while the hadronic (HAD) section consists of alternating
layers of steel and scintillator. The CDF calorimetry is divided into two main
pseudorapidity ($|\eta|$) regions; the central calorimeter covers $|\eta|<1.1$
and the plug region covers $1.1<|\eta|<3.6$. The region between the central
and plug regions is covered by the end-wall hadron calorimeter which has a
similar construction as the central hadron calorimeter. The tower size in the
central region is $\Delta\eta\times\Delta\phi\approx0.1\times15^{\circ}$, and
the segmentation in the plug region varies as a function of $\eta$
($\Delta\eta\times\Delta\phi\approx0.1\times7.5^{\circ}$ for $|\eta|<1.8$ and
with $\Delta\eta$ increasing for larger $|\eta|$ values). The noise level is
very low, with only $\sim1$ noise tower with $E_{T}>50$ MeV being present per
event. The $E_{T}$ threshold on each calorimeter tower used for jet
reconstruction is 100 MeV, and no special care is necessary for noise suppression.

\subsection{Jet Reconstruction and Energy Measurement}
\label{sec:jetcor}

In contrast to leptons or photons, jets are clusters of \textquotedblleft
objects\textquotedblright\ that have to be defined by a clustering algorithm,
as discussed previously in this paper.
In parton shower Monte Carlo events, a jet clustering algorithm can be
applied to particles produced after hadronization of the partons, producing
hadron-level jets; for pQCD parton-level predictions, jets are formed by
running the algorithm on partons from the fixed-order (typically
next-to-leading order) pQCD event generator. Experimentally, it is extremely
challenging to both identify and measure the kinematic properties of the large
variety of individual particles produced in high-energy hadronic collisions,
especially with the relatively coarse calorimeter segmentation present in the
Tevatron experiments. \ As a result jets are reconstructed by running the jet
algorithm on the \textquotedblleft raw\textquotedblright\ energy depositions
in the calorimeter towers, and not on the individual particles.

The reconstruction and energy measurement of these ``calorimeter'' jets are
affected by a variety of instrumental and physics effects. The instrumental
effects include (1) calorimeter non-uniformity, (2) resolution effects due to
large fluctuations in particle showering in the calorimeter, (3) non-linear
response of the calorimeter, especially to hadrons, and (4) low momentum
particles not reaching the calorimeter due to materials in front of the
calorimeter and the solenoidal magnetic field. Energy from additional $p\bar
p$ interactions occurring in the same bunch crossing also affects the jet
reconstruction and energy measurement. This has a non-negligible effect
especially for low transverse momentum jets and high instantaneous
luminosities. The situation is further complicated due to the fact that (a)
the underlying event contributes energy to the jet cone, and that (b) the jet
cone does not contain all the energy of the parent parton because of the
effects of parton showering and hadronization. All of these effects have to be
taken into account when comparing experimental data with theoretical
predictions or extracting physics quantities of interest from experimental data.

Analyses take these physics effects into account in different manners. For
example, in new physics searches or in measurements of top quark properties,
attempts are often made to correct the jets measured in the calorimeters back
to the parent parton from which the jet originated. Extraction of the physics
quantities of interest is done by comparing the data, after background
subtraction, with signal events implemented in the parton shower Monte Carlo.

Typically in jet cross section measurements, the data are corrected
to the hadron-level (\textit{i.e.}, corrected only for instrumental effects);
at this level, the results can be compared with hadron-level theoretical
predictions without any knowledge of the detectors. We strongly encourage
that, where possible, measurements at the Tevatron or LHC produce results at
this level; often, instead the predictions from Monte Carlos are themselves
passed through a detector simulation program and then directly compared to the
``raw'' data distributions.

Sometimes, the theoretical predictions from
perturbative QCD calculations are available only at the parton level and then
it is necessary to either correct the theory to the hadron level or the data
to the parton level. For inclusive jet cross sections, the best theoretical
predictions available, as of now, are from next-to-leading order perturbative
QCD calculations~\cite{Ellis:1992en, Giele:1994gf, Nagy:2001fj}. When making
comparisons between experimental measurements and next-to-leading (NLO) order
predictions, non-perturbative QCD effects from the underlying event and
hadronization (see Section~\ref{sec:MCtune} for details) are evaluated based on
tuned parton shower Monte Carlos. Note that a correction is not made for
``out-of-cone'' energy from hard gluon emission, as such effects are already
(at least partially) taken into account in the NLO calculations. In these
comparisons, next-to-leading order hard gluon emission is assumed to model the
whole parton shower perturbative QCD radiation process. NLO pQCD calculations
have been shown to provide a reasonable description of energy flow inside
jets~\cite{Ellis:1993ik}, so this is considered to be a reasonable
assumption\footnote{As $r\rightarrow0$ inside a jet, however, the energy
profile function develops large logarithms of the form $\alpha_{s}^{n}
\log^{2n-1}r$; for an accurate description of this region, these terms would
have to be re-summed. The pure NLO description should be accurate for $r$
values greater than 0.3.}; however, data-theory comparisons will benefit a
great deal from higher order QCD calculations or next-to-leading matrix
element calculations interfaced with parton shower models, as in
\textit{e.g.}, MC@NLO~\cite{Frixione:2002ik}.

The jet algorithms and procedures for jet
energy scale correction to the hadron-level (\textit{i.e.}, correction for
instrumental effects) employed by CDF and D\O \ are discussed below.

\subsubsection{Jet Algorithms}
In Run II, D\O\ uses the Midpoint iterative cone algorithm with
the cone radius $R_{cone}=0.5$ and $0.7$ and merging
criterion (see Section~\ref{sec:merging}) $f_{merge}=0.50$.
At CDF, several jet algorithms are in use.
In a large fraction of analyses, the JETCLU cone clustering has been
used, which was used also in Run~I at CDF, with the cone radius
$R_{cone}=0.4$ and $0.7$ and the merging fraction $f_{merge}=0.75$.
The Midpoint algorithm with the search cone step, \textit{i.e.}, the
Search cone algorithm (see Section~\ref{sec:searchcone}), with the
cone radius of $R_{cone}=0.7$ and the
merging fraction of $f_{merge}=0.75$ was also used in several jet
cross section measurements; however, because of the IR-sensitivity
introduced by the search cone step, the search cone step is being
removed from the clustering. The $k_T$ algorithm was also used in the
inclusive jet cross section measurements with $D=0.5$, $0.7$ and $1.0$.

\subsubsection{Jet Energy Scale at D\O }
\label{sec:jes_dzero}

The jet energy calibration procedure employed by D\O \ is based primarily on
data, exploiting the conservation of transverse momentum~\cite{Abbott:1998xw}.
The measured jet energy is corrected back to the true hadron-level jet energy
by:
\begin{equation}
E_{jet}^{hadron} = \frac{E_{jet}^{measured} - E_{0}} {R_{jet}\cdot S},
\end{equation}
where $E_{0}$ is an offset energy which includes the
uranium noise, energy from the previous bunch crossing, and additional $p\bar
p$ interactions. The offset $E_{0}$ is determined by measuring the transverse
energy density in the minimum-bias and zero-bias data. $R_{jet}$ represents
the calorimeter response to jets. This corrects both for the calorimeter
non-uniformity in $\eta$ and for the absolute energy scale. $S$ is the
showering correction, which corrects for the energy emitted outside the jet
cone due to detector effects.

\begin{figure}[th]
\centering\leavevmode
\begin{tabular}
[t]{c}%
\subfigure[]{ \includegraphics[width=0.49\hsize,clip=]
{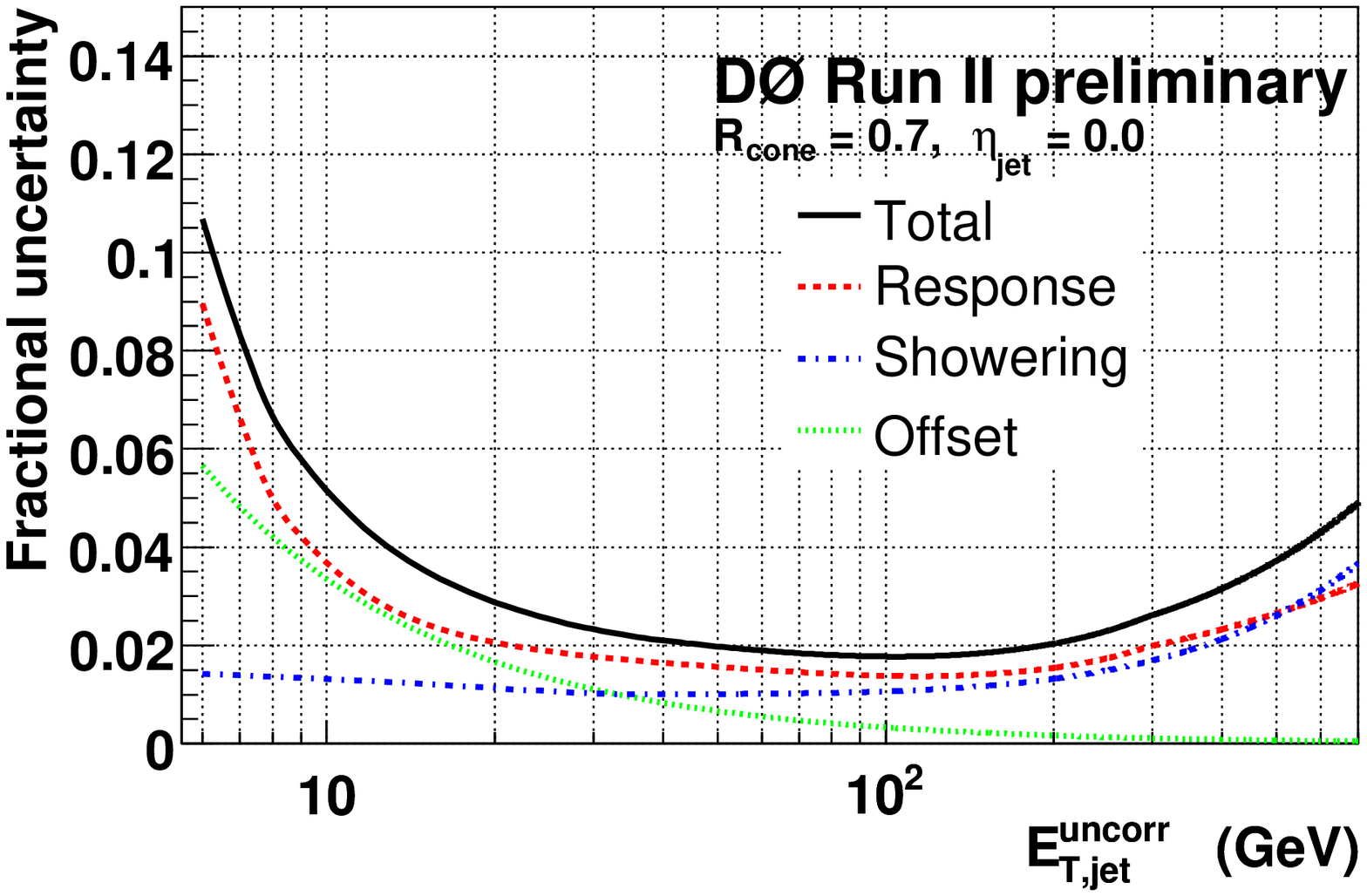} }
\subfigure[]{ \includegraphics[width=0.49\hsize,clip=]
{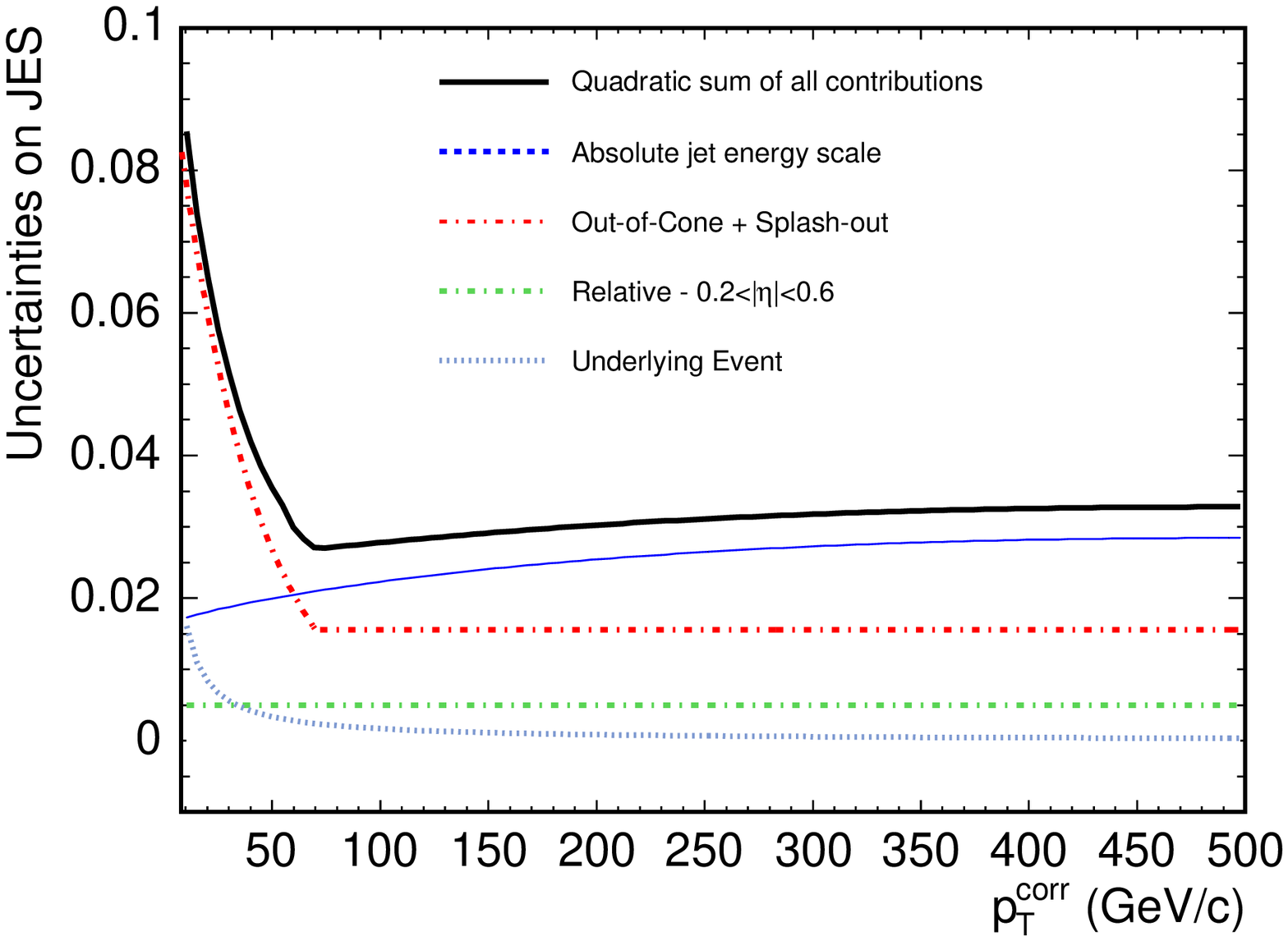} }
\end{tabular}
\caption{The jet energy scale systematic uncertainties for (a) central jets at
$\eta=0$ as a function of uncorrected jet energy at D\O \ and (b) central jets
in $0.2<|\eta|<0.6$ as a function of corrected jet $p_{T}$ at CDF.
Figures from Refs.~\cite{D0JES-Run2} and \cite{Bhatti:2005ai}.
}%
\label{fig:d0_cdf_sys}%
\end{figure}

D\O \ uses the missing $E_{T}$ projection fraction (MPF)
method~\cite{Abbott:1998xw}, 
which exploits the transverse momentum conservation in an event, to measure
the calorimeter response to jets ($R_{jet}$).
In photon+jet events, for example, the transverse energies of the real photon
and the other recoil particles at the hadron level should satisfy:
\begin{equation}
\vec E_{T}^{\gamma} + \vec E_{T}^{recoil}=0. \label{eq:mpf_method_1}%
\end{equation}
In general, the calorimeter response to both photons and recoils is less than
unity and the energy mis-measurement results in missing $E_{T}$ ($\not \!
\!E_{T}$) in events:
\begin{equation}
R_{\gamma}\vec E_{T}^{\gamma} + R_{recoil}\vec E_{T}^{recoil}=-{\not \!
\!\vec E_{T}}. \label{eq:mpf_method_2}%
\end{equation}
After the EM energy calibration, $R_{\gamma}=1$, and
Eqs.~(\ref{eq:mpf_method_1}), (\ref{eq:mpf_method_2}) transform to:
\begin{equation}
R_{recoil} = 1 + \frac{\not \!  \!\vec E_{T} \cdot\vec n_{T}^{\gamma}}
{E_{T}^{\gamma}},
\end{equation}
where $\vec n_{T}^{\gamma} = \vec E_{T}^{\gamma} / |\vec E_{T}^{\gamma}|$. In
back-to-back photon+jet events, $R_{recoil}$ can be considered as the response
of a jet ($R_{jet}$). The absolute jet response correction is determined and
applied after the response is equalized as a function of $\eta$.
Fig.~\ref{fig:d0_cdf_sys}(a) shows the jet energy scale systematic uncertainty
achieved for central jets as a function of jet energy at D\O .

\subsubsection{Jet Energy Scale at CDF}

At CDF, the determination of the jet energy absolute (response) correction at
the hadron-level relies primarily on a detector simulation and jet
fragmentation model~\cite{Bhatti:2005ai}. Therefore, the accuracy of the
calorimeter simulation is crucial for the precise jet energy scale
determination. The CDF calorimeter simulation response to single particles is
tuned to reproduce the response measured both in the test beam and in the
collision data taken with the real detector. The CDF detector simulation uses
a Geant-based detector simulation in which a parametrized shower simulation
(Gflash~\cite{Grindhammer:1989zg}) is used for the calorimeter response. The
use of Gflash was primarily motivated by its excellent performance in terms of
speed, but also the relative ease of tuning. After tuning, the absolute
correction for calorimeter-level jets to the corresponding hadron-level jets
is obtained on average from dijet Monte Carlo events
by matching the two leading hadron-level jets to the corresponding calorimeter
jets, and then taking the most probable value of calorimeter-jet $p_{T}$ for
hadron-level jets with a given $p_{T,jet}^{hadron}$.

This method can be used in the entire calorimeter coverage region; however, in
practice the tuning is limited in precision outside the central region
($0.2<|\eta|<0.6$) because of the limited tracking coverage.
Thus, the jet energy scale outside
the central region is rescaled to that of the central region
based on the $p_{T}$ balance of the leading two jets in exclusive
(\textit{i.e.}, stringent cuts are placed on the presence of any additional
jets in the event) dijet events; in such dijet events, two jets should have
the same $p_{T}$ to leading order and any imbalance is due to calorimeter non-uniformity.
This correction is called the relative correction.

Additional $p\bar p$ interactions, in the same bunch crossing as the
interaction which produced the jets, also contribute energy in the jet cone.
The number of reconstructed primary vertices ($N_{vtx}$) is a good estimator
of the number of interactions in the same bunch crossing. For jets
reconstructed with the cone algorithm, this multiple interaction correction,
$p_{T}^{MI}$, is derived by measuring the transverse momentum in a cone of the
same size as the jet cone in the central region as a function of
$N_{vtx}$ in minimum bias events. For a cone size $R_{cone}=0.7$, the
correction subtracts $p_{T}^{MI}\sim1$ GeV/\textit{c} for each additional
primary vertex.

The CDF jet energy corrections described above can be summarized as,
\begin{equation}
p_{T,jet}^{hadron} = 
\left[p_{T,jet}^{measured}\times f_{rel}- p_T^{
    MI}\times (N_{vtx}-1)\right]\times f_{abs}.
\end{equation}
where $f_{rel}$ is the relative jet energy correction factor as a function of
jet $\eta$ and $p_T$ and $f_{abs}$ is the absolute correction factor as a
function of jet $p_T$.

At CDF, the photon-jet and $Z$-jet $p_{T}$ balances are used to cross-check
the jet energy scale in data and Monte Carlo events. When photon-jet and
$Z$-jet balances are formed, tight cuts are made on the second jet $p_{T}$ and
$\Delta\phi$ between the photon/$Z$ and jet, $p_{T,2}<3$ GeV/\textit{c} and
$\Delta\phi>3$ (rad), to suppress gluon radiation in the events that may
affect the $p_{T}$ balance~\cite{Bhatti:2005ai}. With the cuts discussed
above, the Pythia and Herwig event simulation show differences in $p_{T}$
balance at the level of $\sim2-3$\%, a disagreement which is not well
understood yet. A better understanding of this difference will be extremely
useful for future improvements of the jet energy scale uncertainty. The
overall jet energy scale systematic uncertainty, evaluated as described above
for central jets at CDF, is shown in Fig.~\ref{fig:d0_cdf_sys}(b); it is
$\sim3$\% at high $p_{T}$ and increases at $p_{T} \lesssim50$ GeV.

The hadronic decays of resonances with well known masses, such as the $W$ and
$Z$ bosons, can also be useful to test and to calibrate the jet energy scale.
In most cases, the hadronic $W$ and $Z$ decays are swamped by QCD background
at hadron colliders; however in, \textit{e.g.},
$t\bar t \to W(\to l\nu) + \ge4$ jets events (referred to as lepton+jets
events/channel hereafter),
the hadronic $W$ resonance can be observed with only a relatively small QCD background.

\begin{figure}[th]
\centering\leavevmode
\includegraphics[width=0.55\hsize,clip=]
{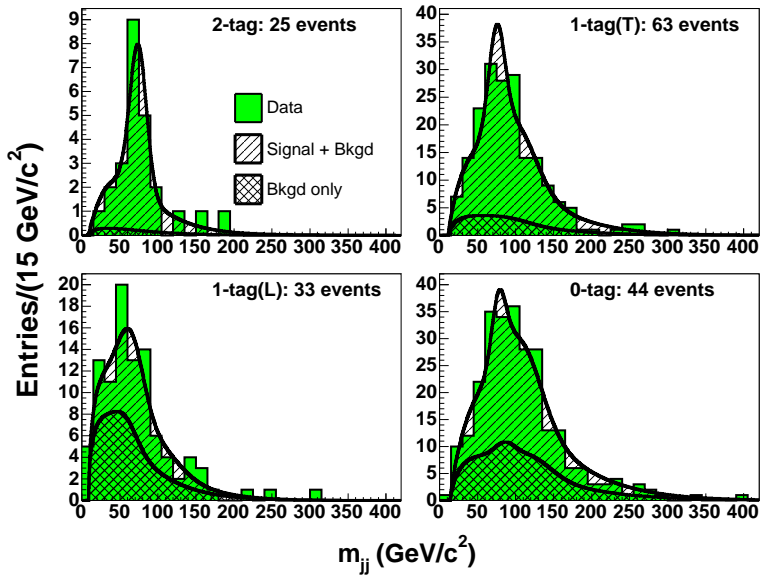} \caption{Dijet mass distributions for four subsamples
in the lepton+jets sample. The signal and background shapes corresponding to
the best fit of the jet energy scale cross-check are overlaid on the
histograms. Clear $W$ peaks can be observed, especially when one or more
$b$-tags is required. Figure from Ref.~\cite{Bhatti:2005ai}.}%
\label{fig:mwjj}%
\end{figure}

Hadronic $W$ decays in $t\bar t$ events were first observed in Run I in
lepton+jets events
in Run I~\cite{Abe:1997ev}, and in Run II these decays have been used
successfully by both the CDF and D\O \ collaborations to finely calibrate the
jet energy scale in top quark mass measurements in the lepton+jets
channel~\cite{Abulencia:2005ak,Abazov:2006bd}.
Fig.~\ref{fig:mwjj} shows the dijet mass distributions in four sub-samples of
the lepton+jet event sample; the sub-samples have been divided based on the
number of jets identified as $b$-quark jets by the standard CDF $b$-tagging
algorithm (see Section~\ref{sec:heavy_flavor_jets}).
Events with one $b$-tagged jet are further divided into two classes;
1-tag(T) refers to events with four jets with $E_{T}>15$ GeV and 1-tag(L)
refers to events with three jets with $E_{T}>15$ GeV and the fourth jet with
$8<E_{T}<15$ GeV.
In the CDF measurement~\cite{Abulencia:2005ak}, the reconstructed top mass and
dijet mass distributions are formed from $t\bar t$ MC events with various top
mass and jet energy scales ranging from -3 to +3$\sigma$ where $\sigma$ is the
total jet energy scale uncertainty described above. Fits to the data without
using the jet energy scale constraint from the standard procedure yield the
jet energy scale $[-0.25\pm1.22]\sigma$, indicating that the jet energy scale
from the aforementioned procedure is in good agreement with information
provided by the $W$ resonance peak in $t\bar t$ events.

Hadronic $W/Z$ decays in the $W/Z+\gamma$ process have been also investigated
by CDF. The observation of this signal can provide another way in which to
validate and/or constrain the jet energy scale; however, the extremely large
QCD background has made the observation of such a signal extremely difficult.
Constraining the jet energy scale with the hadronic $W$ resonance is a very
powerful technique, and the jet energy scale uncertainty from this method will
improve as more data are accumulated. However, it has to be noted that all the
detailed studies presented above are crucial for the success of this
technique, since this method relies on a good modeling of the dijet mass
distribution. Also, this technique is not able to constrain the jet energy
scale over a wide range of jet $p_{T}$.

\subsubsection{Summary}
In this section (Sec.~\ref{sec:jetcor}), 
the strategies used by the D\O\ and CDF collaborations to determine
the jet energy scale were presented.
The two collaborations used the different approaches, but achieved to
determine the jet energy scale with similar precision.
It is worth noting that, although large groups of people worked on the
jet energy scale determination in both experiments, it still took
more than three years to achieve the precision shown in
Fig.~\ref{fig:d0_cdf_sys} from the beginning of the Run II.
High quality test beam analyses, thorough detector simulation tuning
based on the test beam results, good planning for the jet energy
calibration and validation based on the in-situ data
(including the trigger implementation for the calibration/validation
datasets) will be essential for the early physics analyses at the LHC.

\subsection{Monte Carlo Tuning}
\label{sec:MCtune}

Most of the analyses at hadron colliders use parton shower Monte
Carlos in order to model the signal and background events, 
unfold the detector effects and to extract the physics quantities of
interest.
These parton shower Monte Carlo programs
have numbers of parameters that need to be tuned on the real data,
\textit{e.g.}, parameters that control the initial and final state
radiation, jet fragmentation, hadronization, and underlying event.
Well-tuned Monte Carlos are essential for a precise measurement and
proper comparison with theoretical predictions.
The good modeling of jet fragmentation properties and of
the underlying event in Monte Carlo events is crucial for corrections
of jet energies from the calorimeter level to the hadron level.
Also, in recent inclusive jet cross section measurments by
CDF~\cite{Abulencia:2005yg, Abulencia:2005jw, Abulencia:2007ez,
  craig_thesis} the effects of
underlying event and hadronization are estimated using leading-order
parton-shower Monte Carlo generators based on the variation of
the predicted jet cross sections after turning off the interaction between
beam remnants and the hadronization in Monte Carlo events.
Several measurements performed at the Tevatron, which are used for
either tuning Monte Carlos or validating the tunings, are discussed
below.

\subsubsection{Dijet Angular Decorrelations}
\begin{figure}[th]
\centering\leavevmode
\begin{tabular}
[t]{c}%
\subfigure[]{ \psfig{figure=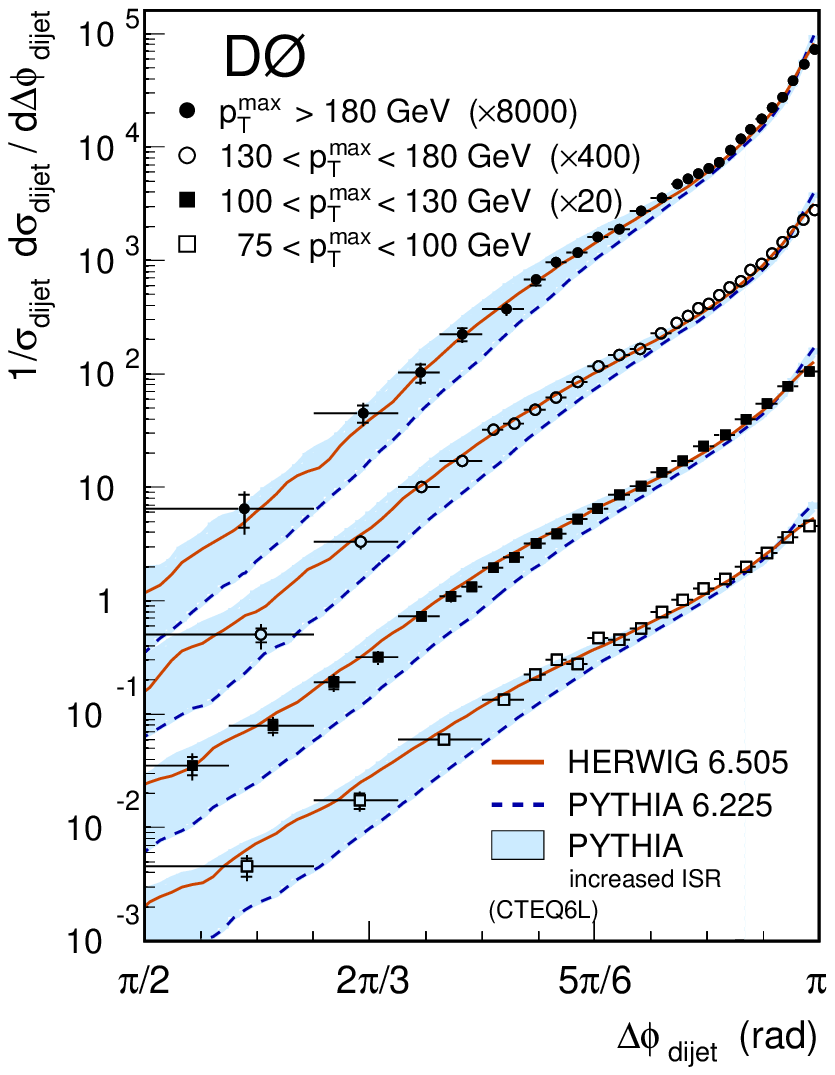,width=7.7cm}}
\subfigure[]{ \psfig{figure=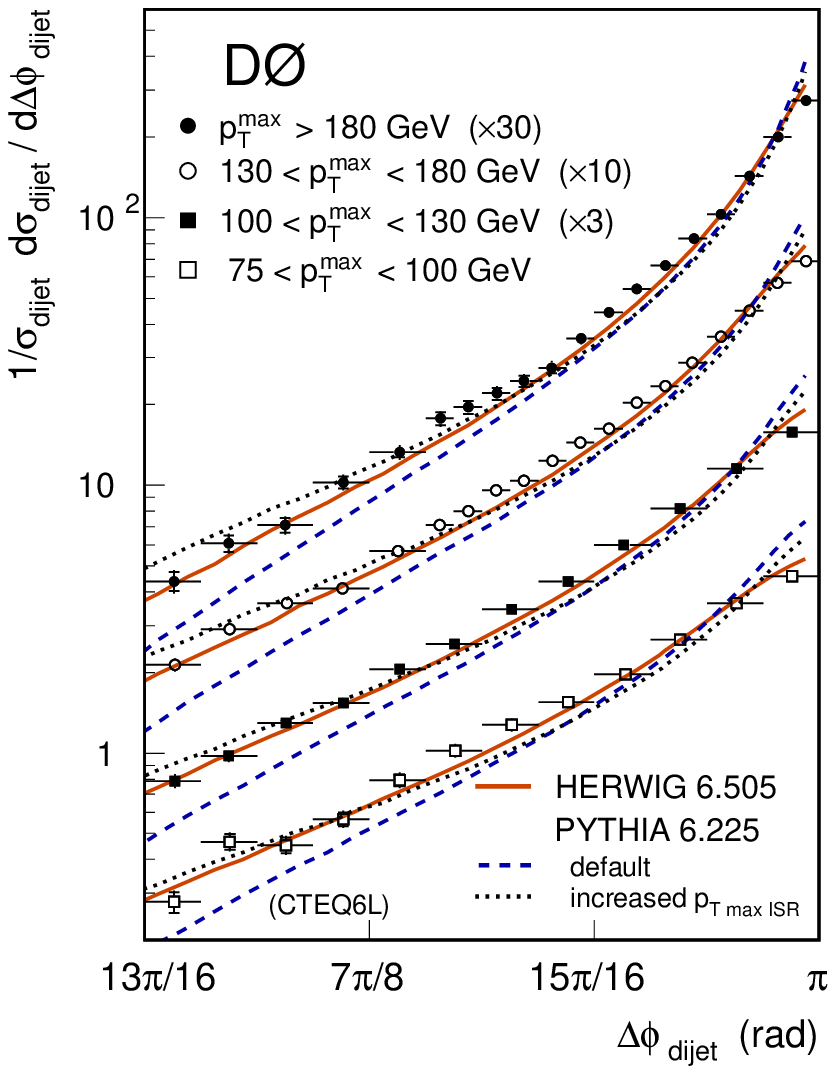,width=7.7cm}}
\end{tabular}
\caption{$\Delta\phi$ distributions in different $p_{T}^{max}$ ranges over
$\Delta\phi>\pi/2$ (a) and in the peak region $\Delta\phi>13\pi/16$ (b).
Predictions from Herwig and Pythia are overlaid for comparison. Pythia
predictions are shown for the PARP(67) parameter varied between 1.0
and 4.0. Figures from Refs.~\cite{Albrow:2006rt,Abazov:2004hm}.}%
\label{fig:dzero_dphi}%
\end{figure}

D\O \ made a measurement of the azimuthal angle between the two leading jets,
$\Delta\phi$, in Run II~\cite{Abazov:2004hm}. This provides an excellent
testing ground for the study of multiple gluon radiation effects.
Studies~\cite{Albrow:2006rt, Abazov:2004hm} have shown that such distributions
are not sensitive to underlying event and hadronization effects. Near the
$\Delta\phi\sim\pi$ peak the distribution is sensitive to soft radiation with
small $p_{T}$, and the tail at small $\Delta\phi$ is sensitive to hard
radiation with high transverse momentum. The measured $\Delta\phi$
distributions, $(1/\sigma_{jj})(d\sigma_{jj}/d\Delta\phi)$, for four different
$p_{T}^{max}$ (largest jet $p_{T}$ in an event) ranges are shown in
Fig.~\ref{fig:dzero_dphi} along with the predictions from Herwig and Pythia.
Jets are reconstructed with the Midpoint cone algorithm with a cone radius of
$R_{cone}=0.7$. The measured distributions have a sharp peak at $\Delta
\phi\sim\pi$ and the peaks are sharper at higher jet $p_{T}$ mainly due to the
running of $\alpha_{s}(p_{T})$.
The default Herwig predictions give a reasonable description of the data over
the whole $\Delta\phi$ range in all $p_{T}^{max}$ regions; however, the
default Pythia gives sharper distributions than data in all $p_{T}^{max}$
regions, and Pythia provides a
better description of the data when ISR is enhanced (see
Fig.~\ref{fig:dzero_dphi}). 
The predictions are found to be insensitive to FSR tunes, and
the measurement provides a good tool for ISR tuning.
Monte Carlo predictions from
Sherpa~\cite{Gleisberg:2003xi} and Alpgen~\cite{Mangano:2001xp,Mangano:2002ea}
have also been tested against this data, as have NLO pQCD predictions from
NLOJET++~\cite{Nagy:2001fj}. They were found to give a reasonable description
of the data.

\subsubsection{Underlying Event}
\begin{figure}[pth]
\begin{center}
\includegraphics[width=.40\textwidth]{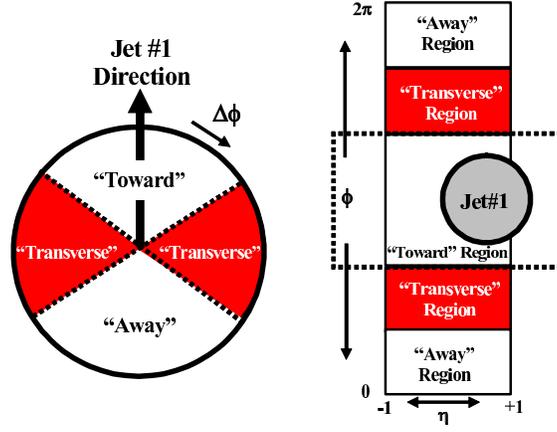}
\end{center}
\caption{Definitions of the ``toward'', ``away'' and ``transverse'' regions.
The angle $\Delta\phi=\phi-\phi_{\mathrm{jet\#1}}$ is the relative azimuthal
angle between charged particles and the direction of the leading jet. The
transverse region is defined by $60^{\circ}<|\Delta\phi|< 120^{\circ}$ and
$|\eta|<1$. Figure from Ref.~\cite{Albrow:2006rt}.}%
\label{fig:ue_ana_topology}%
\end{figure}

Many of the important observables at a hadron collider, including jets, are
sensitive to the underlying event, and so a good understanding of the
underlying event is needed for precision measurements.
A series of studies have been made on underlying events by CDF in
Run~I~\cite{Affolder:2001xt,Acosta:2004wq} and
Run~II~\cite{Albrow:2006rt,Field:2006iy}.
These studies made use of the topological structure of hadron-hadron
collisions to study the underlying event; measurements were made on event
activities in the ``transverse'' region with respect to the jet axis in jet
events; this region is most sensitive to the underlying event.
The geometry of one study is shown in Fig.~\ref{fig:ue_ana_topology}, where
the ``transverse'', ``towards'' and ``away'' regions have been defined with
respect to the direction of the leading jet.
The TransMAX (TransMIN) region refers to the transverse region containing the
highest (lowest) scalar $p_{T}$ sum of charged particles. The study made use
of two classes of events, (1) ``leading jet'' event, in which there is no
restriction on the second and third jet, and (2) ``back-to-back'' events, in
which there are at least two jets with $p_{T}>15$ GeV, the leading two jets
are nearly back-to-back ($\Delta\phi(jet1,2)>150^{\circ}$) with $p_{T}%
(jet2)/p_{T}(jet1)>0.8$, and $p_{T}(jet3)<15$ GeV/\textit{c}.

\begin{figure}[pth]
\begin{center}
\includegraphics[width=.5\textwidth]{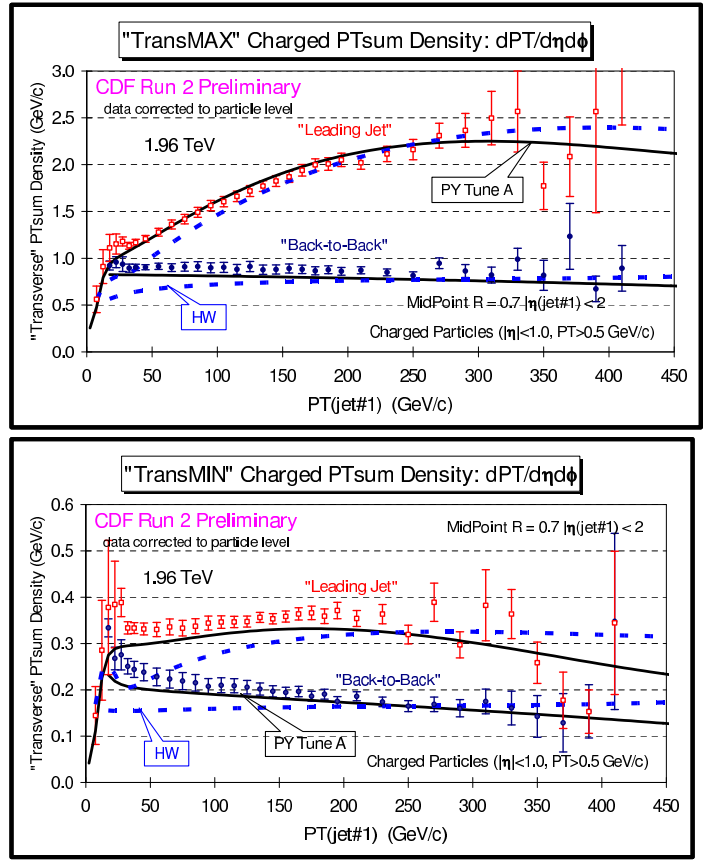}
\end{center}
\caption{ The sum $p_{T}$ of charged particles inside the TransMAX (top) and
TransMIN (bottom) regions, as a function of the leading jet $p_{T}$, in the
leading jet events and back-to-back events, in data, Pythia Tune A and Herwig
(without multiple parton interactions). Figure from Ref.~\cite{Field:2006iy}.}%
\label{fig:ue_transmax_min}
\end{figure}

The density of transverse momentum carried by charged particles, $dp_{T}/d\eta
d\phi,$ in both the TransMAX and TransMIN regions in the leading jet events
and back-to-back events is shown in Fig.~\ref{fig:ue_transmax_min}.
In the \textquotedblleft back-to-back\textquotedblright\ events, contributions
from hard components (initial and final state radiation) to the
\textquotedblleft transverse\textquotedblright\ region are suppressed, and the
sensitivity to the underlying event is increased.
In the study, the TransMAX and TransMIN regions were also used in order to
better separate the hard components from the underlying event.


We expect that the MAX region will pick up the hardest ISR or FSR, and thus
the MIN region will be more sensitive to the underlying event, which is
indicated in Fig.~\ref{fig:ue_transmax_min}; $dp_{T}/d\eta d\phi$ in the MAX
region increases with increasing leading jet $p_{T}$ in leading jet events
but, on the other hand, $dp_{T}/d\eta d\phi$ in the MIN region stays rather
flat with leading jet $p_{T}$. 
Please note that NLO can
contribute, by definition, only to the transMAX region, and not to the
transMIN.
At least in the measurments at the Tevatron, the transMIN region has a
level similar to that of an active minimum bias event, so the
results indicate that higher order
(beyond NLO) radiation effects are relatively small.
Fig.~\ref{fig:ue_transmax_min} shows that contributions from FSR and ISR
are clearly suppressed in back-to-back events compared with the leading jet
events.
The MAX and MIN $dp_{T}/d\eta d\phi$ are somewhat falling with increasing jet
$p_{T}$ at high jet $p_T$ which could be due to a saturation of the
multiple parton interactions.
All these features are fairly well described by the tuned Pythia Monte Carlo
(\textit{e.g.}, Tune A~\cite{tuneA}).

\subsubsection{Jet Shapes}

The jet shapes, \textit{i.e.}, the energy flow inside jets, in inclusive jet
data were studied by CDF in Run~I~\cite{Abe:1992wv} and
Run~II~\cite{Acosta:2005ix}.
The jet shape is dominated by gluon emissions from the primary outgoing
parton, and it depends on the type of the parton originating the jet,
\textit{i.e.}, quark or gluon.
In hadron-hadron collisions, the jet shape is also sensitive to initial state
radiation and the underlying event.
Fig.~\ref{fig:shapes} in Sec.~\ref{sec:hadron} showed the $p_{T}$ fraction in
a cone jet of radius 0.7 that lies in the annulus from 0.3 to 0.7 as a
function of jet $p_{T}$; data are compared with parton shower Monte Carlo
predictions.
Jets become narrower with increasing jet $p_{T}$ due to several different
factors: (1) power corrections that tend to broaden the jet fall as $1/p_{T}$
or $1/p_{T}^{2}$, (2) the fraction of jets originating from quarks increases
with increasing jet $p_{T}$ and (3) the probability of QCD radiation decreases
as $\alpha_{s}(p_{T})$.
The measured jet shapes are well described by the Pythia Monte Carlo with Tune
A parameters, which were obtained based on the underlying event study made in
the ``transverse'' region away from jets.
It is the good description of the jet fragmentation and underlying event
properties by the Monte Carlo events that allows the reliable evaluation of
the unfolding correction of measurements to the hadron level and also
the estimation of the hadronization and underlying event corrections.


\subsubsection{Summary}
In this section (Sec.~\ref{sec:MCtune}), a few examples of the measurements
that were useful for Monte Carlo tunings were presented.
Most of analyses at the Tevatron experiments rely on parton shower
Monte Carlo event generators, and they benefited from these studies.
At the LHC experiments, these measurements must be repeated since
the scaling of these tunings to higher center-of-mass energies are not
straightforward.

\subsection{Inclusive Jet Cross Sections}

\label{sec:incjet}

\begin{figure}[th]
\centering\leavevmode
\subfigure[]{
\includegraphics[width=0.40\textwidth]{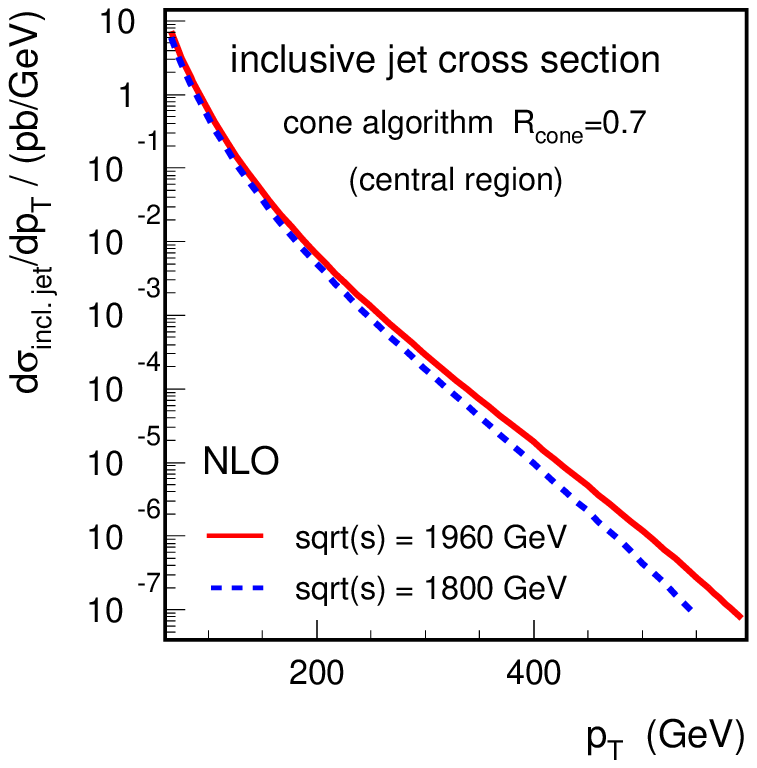}
} \subfigure[]{
\includegraphics[width=0.40\textwidth,bb=104 260 498 656]{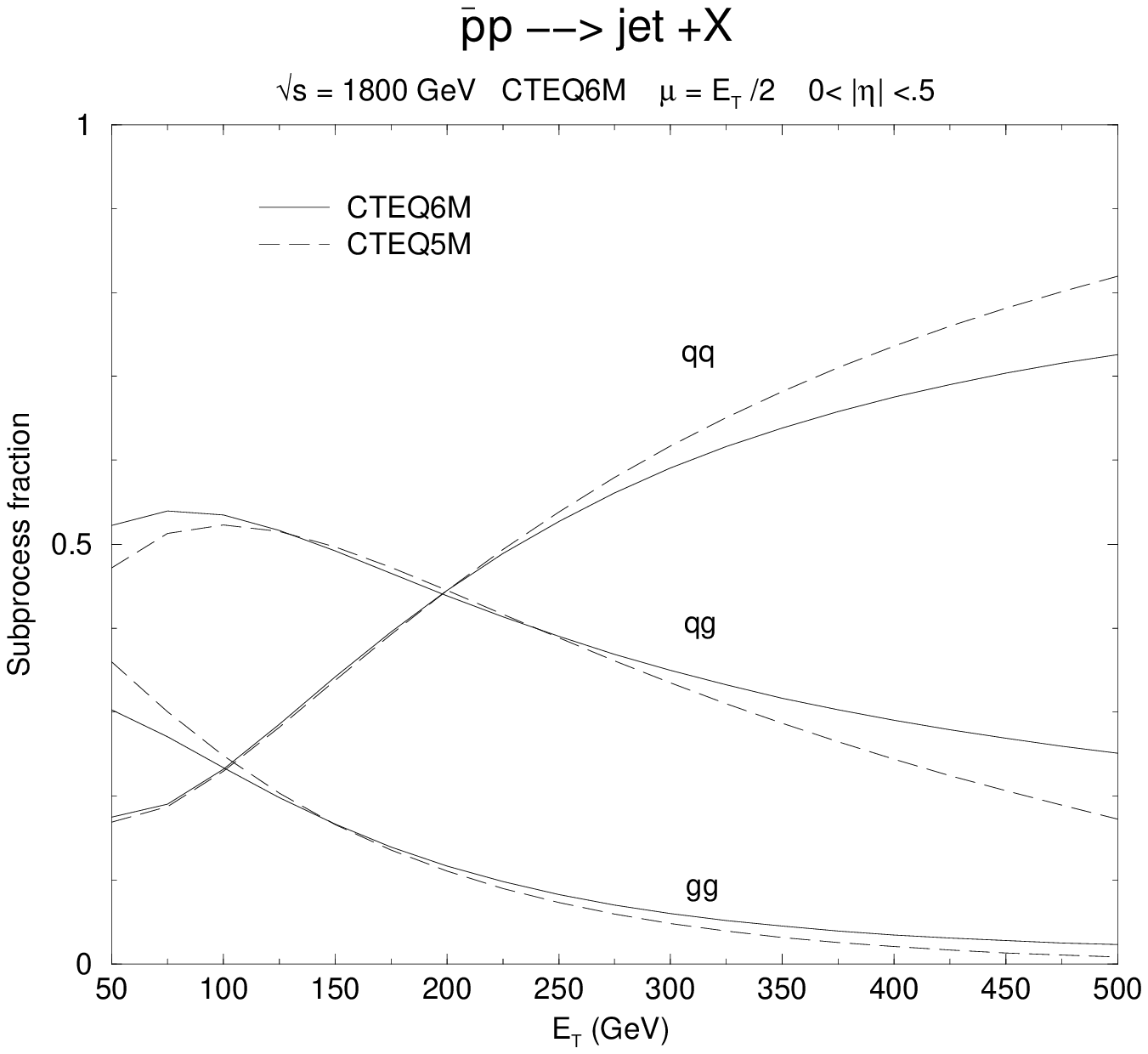}
}\caption{(a) The inclusive jet cross sections at $\sqrt{s}=1.8$ and 1.96 TeV
from NLO pQCD predictions using the CTEQ6.1M pdf (figure from
Ref.~\cite{Markus}). (b) The subprocess contributions to inclusive jet
production at the Tevatron for the CTEQ5M and CTEQ6M pdfs (figure from
Ref.~\cite{Stump:2003yu}).}%
\label{fig:incjet_run2}%
\end{figure}

The inclusive jet cross section has been extensively studied at the Tevatron
in both Run~I and Run~II. The differential inclusive jet cross section at the
Tevatron provides the highest momentum transfers currently attainable in
accelerators. It is sensitive to a wide range of new physics, such as quark
compositeness, and also tests perturbative QCD calculations over more than
eight orders of magnitude in the cross section. Due to greater statistics
compared to Run I and the higher centre-of-mass energy, the reach in
transverse momentum has increased by approximately $150$~GeV~(see
Fig.~\ref{fig:incjet_run2}(a)). Measurements of the inclusive jet cross
section have also been shown to have a large impact on global pdf
analyses~\cite{Stump:2003yu}, especially on the determination of gluon
densities at high $x$, as the inclusive jet cross section has sizable
contributions from the $qg\to qg$ subprocess, even at high jet $p_{T}$, as
shown in Fig.~\ref{fig:incjet_run2}(b). In the CTEQ6M global
fit~\cite{Pumplin:2002vw}, the full inclusive jet data sets from the Run I
measurements of CDF and D\O \ were included, resulting in the observed
enhancement of the $gq$ subprocess compared to the predictions derived from
CTEQ5M~\cite{Lai:1999wy}.

\begin{figure}[th]
\begin{center}
\includegraphics[width=12cm]{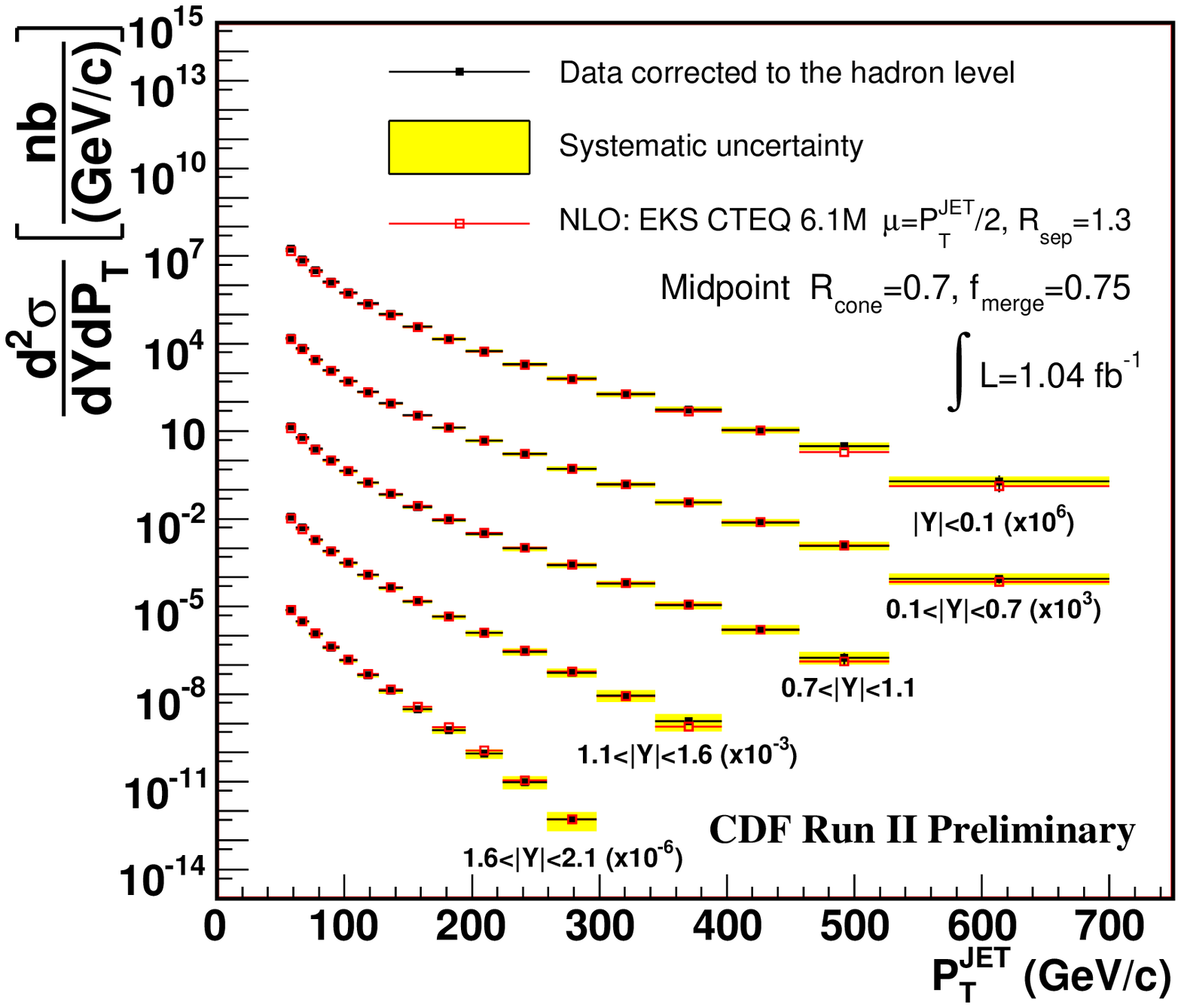}
\end{center}
\par
\vspace*{-0.5cm}\caption{ The inclusive jet cross sections measured with the
Midpoint algorithm by CDF in Run II~\cite{craig_thesis}. }%
\label{fig:cdf_jet_log}%
\end{figure}

The inclusive jet cross sections measured by the CDF collaboration in Run II
are shown in Fig.~\ref{fig:cdf_jet_log}, as a function of the jet $p_{T}$ in
five rapidity regions.
The measurement uses the Midpoint cone algorithm
with a cone radius of $0.7$~\cite{Abulencia:2005jw,
  craig_thesis}\footnote{The measurements shown here include the
search cone step covered earlier in this review, as the measurements without
the search cone are currently still underway. As discussed, the differences will be
relatively small and will be confined mostly to lower transverse momentum.}.
A similar measurement was perfomed by D\O\ as well~\cite{Voutilainen:2006yn}.
For comparisons of data to theory, the calorimeter tower energies clustered
into a jet must be first corrected for the detector response and multiple
$p\bar p$ contributions. These corrections are discussed in
Sec.~\ref{sec:jetcor}. After an additional correction that accounts for the
smearing effects due to the finite energy resolution of the
calorimeter (unfolding), the
jet cross sections are corrected to the hadron-level. For data to be compared
to a parton level calculation, either the data must be corrected from the
hadron level to the parton level or the theory must be corrected to the hadron
level. In this paper, we discuss in the latter scheme; the former just
involves the inverse corrections.

These corrections are intended to account for two effects: (1) ``underlying
event'', \textit{i.e.}, the energy not associated with the hard scattering,
and (2) ``hadronization'', \textit{i.e.}, a loss of energy outside a jet due
to the hadronization process.
In recent analyses on inclusive jet cross sections by
CDF~\cite{Abulencia:2005yg, Abulencia:2005jw, Abulencia:2007ez, craig_thesis},
the hadronization corrections are evaluated by comparing the results obtained
from Pythia at the hadron level to the results from Pythia when the underlying
event and the parton fragmentation into hadrons has been turned off. The
underlying event energy is due to the interactions of the spectator partons in
the colliding hadrons and the size of the correction depends on the size of
the jet cone. It is $\sim1.5-2$~GeV for a cone of radius $0.7$ and is similar
to the amount of energy observed in an arbitrarily placed cone of this size in
minimum bias events with a high track multiplicity.

The hadronization correction accounts for hadrons outside the jet cone
originating from partons whose trajectories lie inside the jet cone;
it does not correct for the effects of hard gluon emission outside the jet
cone, which are already accounted for in the NLO prediction\footnote{Such
corrections for hard gluon emission are often made, however, if the comparison
is to a leading order parton calculation, such as for the reconstruction of a
$t\overline{t}$ final state.}. The numerical value of the hadronization energy
is roughly constant at $1$~GeV for a cone of radius $0.7$, independent of the
jet transverse momentum. This constancy may seem surprising. However, as the
jet transverse momentum increases, the jet becomes more collimated; the result
is that the energy in the outermost annulus (the origin of the hadronization
energy) is roughly constant. The evaluation of these two effects are reliable
only if the Monte Carlo events provide a reasonable description of the
underlying event and of the jet fragmentation and hadronization properties.
Studies on underlying event and jet fragmentation are presented in
Sec.~\ref{sec:MCtune}.

\begin{figure}[thp]
\begin{center}
\includegraphics[width=7cm,angle=-90]{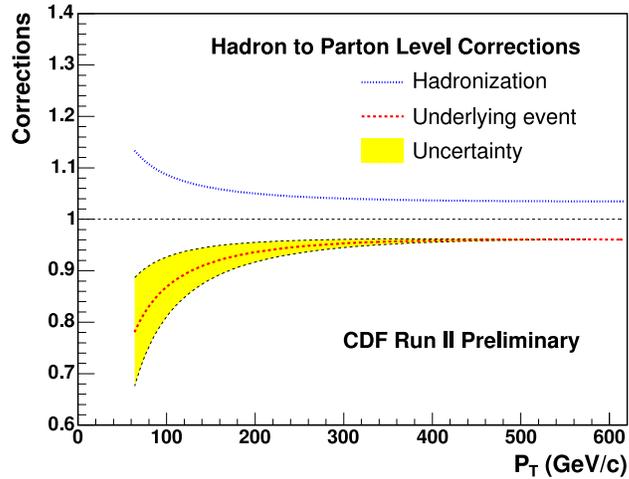}
\end{center}
\caption{ Magnitude of the fragmentation and underlying event corrections used
to correct the inclusive jet cross sections measured by CDF as a
function of jet $p_{T}$, for a cone size $R=0.7$.}%
\label{fig:cdf_jet_cor}%
\end{figure}

The two effects (underlying event and hadronization) go in opposite
directions, so there is a partial cancellation in the correction to parton
level. For a jet cone of $0.7$, the underlying event correction is larger, as
seen in Fig.~\ref{fig:cdf_jet_cor}, for the case of inclusive jet production
at CDF.
For a jet cone radius of $0.4$, the hadronization correction
remains roughly the same size but the underlying event corrections scales by
the ratio of the cone areas; as a result the two effects basically cancel each
other out over the full transverse momentum range at the Tevatron. \ Note
also, as illustrated in Fig.~\ref{fig:cdf_jet_cor}, that the magnitudes of the
fractional corrections are relatively insensitive to the value of $p_{T}$ for
jet momenta larger than 200 GeV. \ Although the fractional changes in the jet
$p_{T}$ due to the underlying event and to hadronization decrease roughly as
$1/p_{T}$, the jet cross section itself is becoming much steeper and hence
more sensitive to changes in the $p_{T}$. \ These two behaviors essentially
cancel each other and lead to the observed nearly constant correction factors.

\begin{figure}[thp]
\begin{center}
\includegraphics[width=12cm]{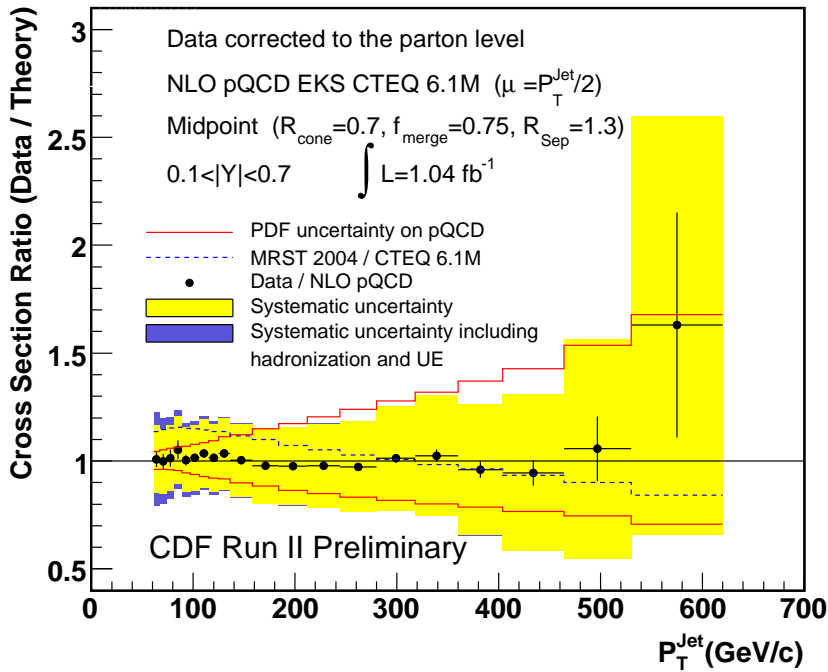}
\end{center}
\par
\vspace*{-0.5cm}\caption{The inclusive jet cross section from CDF in Run II
compared on a linear scale to NLO theoretical predictions using CTEQ6.1 and
MRST2004 pdfs.
}%
\label{fig:cdf_jet_lin}%
\end{figure}
\begin{figure}[pth]
\begin{center}
\includegraphics[width=14cm]{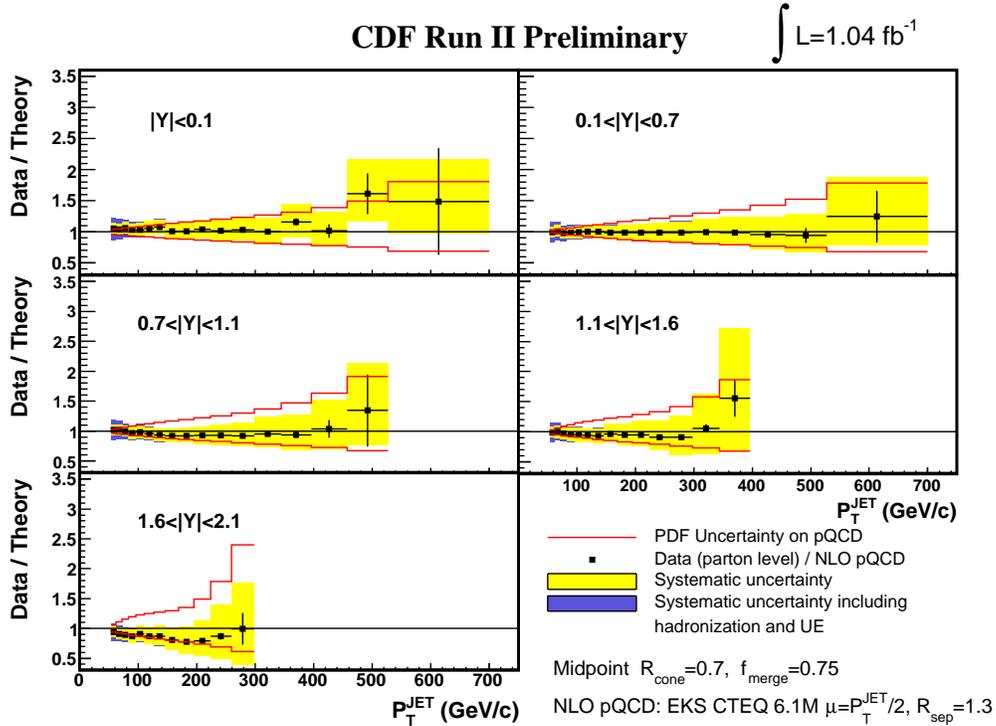}
\end{center}
\par
\vspace*{-0.5cm}\caption{ The inclusive jet cross section from CDF in Run II,
for several rapidity intervals using the Midpoint cone algorithm, compared on
a linear scale to NLO pQCD predictions using CTEQ6.1 pdfs.}%
\label{fig:cdf_jet_rap}%
\end{figure}

A comparison of the inclusive jet cross section measured by CDF in Run II with
the Midpoint cone algorithm~\cite{Abulencia:2005yg, craig_thesis} to NLO pQCD
predictions using the 
EKS~\cite{Ellis:1992en} program
with the
CTEQ6.1~\cite{Stump:2003yu} and MRST2004~\cite{Martin:2004ir} pdfs is shown in
Fig.~\ref{fig:cdf_jet_lin}.
A renormalization/factorization scale of $(p_{T}^{jet}/2)$ has
been used in the calculation. This is the scale at which the Run I jet data
were included in the global fits~\cite{Pumplin:2002vw,Stump:2003yu}, so the
same scale should be used for self-consistency. Typically, this choice of the
renormalization/factorization scale leads to the highest predictions for
inclusive jet cross sections at the Tevatron.
There is good agreement with the CTEQ6.1 predictions over the transverse
momentum range of the prediction. Note that CTEQ6.1M gluon density is already
enhanced at large $x$, as compared to previous pdfs, due to the influence of
the Run I jet data from CDF and D\O . This enhanced gluon provides a good
agreement with the high $p_{T}$ Run II measurement as well which is extended
approximately by $150$~GeV in jet $p_{T}$. The MRST2004 pdfs also contain an
enhanced higher $x$ gluon, which leads to reasonable agreement with the
measurements. The red curves indicate the pdf uncertainty for the prediction
using the CTEQ6.1 pdf error set. The yellow band indicates the experimental
systematic uncertainty, which is dominated by the uncertainty in the jet
energy scale (on the order of $3\%$ as shown in Fig.~\ref{fig:d0_cdf_sys}).
The purple band shows the effect of the uncertainty due to the hadronization
and underlying event, which is visible only for transverse momenta below
$100$~GeV. In Fig.~\ref{fig:cdf_jet_rap}, the jet cross sections measured with
the Midpoint cone algorithm are shown for the five rapidity regions of the CDF
experiment. Good agreement is observed in all rapidity regions with the
CTEQ6.1 predictions. It is also important to note that for much of the
kinematic range, the experimental systematic errors are less than pdf
uncertainties; thus, the use of this data in future global pdf fits should
serve to further constrain the gluon pdf.

\begin{figure}[pth]
\begin{center}
\includegraphics[width=14cm]{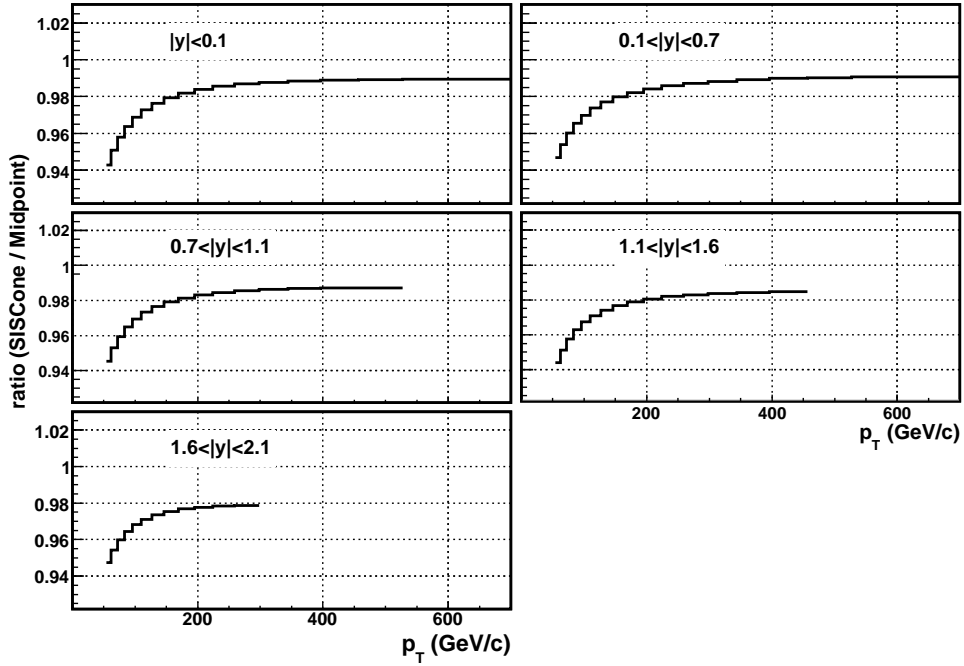}
\end{center}
\par
\vspace*{-0.5cm}\caption{The ratio of the hadron level inclusive jet cross section
with the SISCone algorithm to that with the Midpoint algorithm in five rapidity regions
in Pythia Monte Carlo samples generated with the Tune A parameters.}%
\label{fig:SISConeMidpointHadron}%
\end{figure}

\begin{figure}[pth]
\begin{center}
\includegraphics[width=14cm]{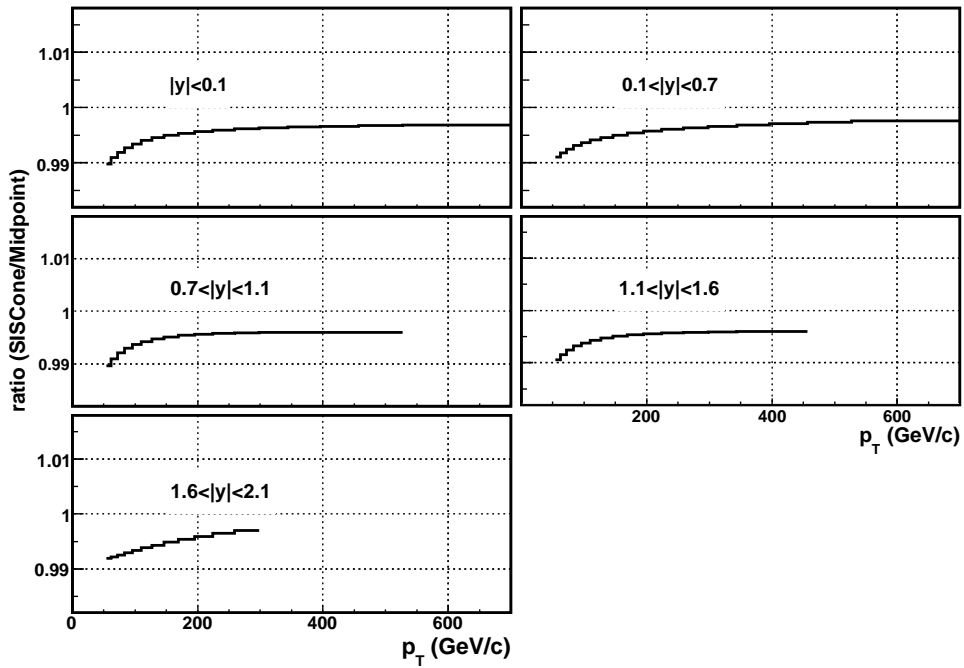}
\end{center}
\par
\vspace*{-0.5cm}\caption{The ratio of the parton level inclusive jet cross section
with the SISCone algorithm to that with the Midpoint algorithm in five rapidity regions
in Pythia Monte Carlo samples in which the underlying event has been turned off.}%
\label{fig:SISConeMidpointParton}%
\end{figure}

While this measurement has been carried out,
a new seedless cone algorithm, SISCone~\cite{Salam:2007xv} (see also
Section~\ref{sec:seedless_midpoint}), has become available.
The differences in the inclusive jet cross sections
between the Midpoint algorithm\footnote{For comparisons with the
SISCone algorithm, the Midpoint algorithm without the search cone step
is used in order to investigate the effects of seeds and slight differences
in the merging procedure of the overlapping stable cones only. Please also note
that D\O\ uses the Midpoint algorithm without the search cone step, and
a measurement using the Midpoint algorithm without the search cone step is being
finalized at CDF too.}
and SISCone algorithm are evaluated in
the Pythia Monte Carlo samples both at the hadron level and parton level
and are shown in Figs.~\ref{fig:SISConeMidpointHadron} and
\ref{fig:SISConeMidpointParton}.
The hadron level inclusive jet cross sections are obtained from
the Pythia Monte Carlo samples generated with the Tune A parameters,
and the parton level inclusive jet cross sections are obtained from
the Pythia Monte Carlo samples generated with the underlying event turned off.
The cross sections at these levels are used to determine the hadronization corrections.
Fig.~\ref{fig:SISConeMidpointHadron} indicates that the hadron level inclusive
jet cross sections are different between the SISCone and Midpoint algorithms
by 5\% at low jet $p_T$ with the differences decreasing with increasing jet $p_T$;
in the meantime, the differences are less than 1\% at the parton level
for any jet $p_T$ (see Fig.~\ref{fig:SISConeMidpointParton}).
This indicates that the hadronization correction is different between the SISCone
and Midpoint algorithms by up to 5\%; however, the impact on comparisons between the
measurement and NLO predictions is negligible (less than 1\%).

\begin{figure}[ph]
\begin{center}
\includegraphics[width=14cm]{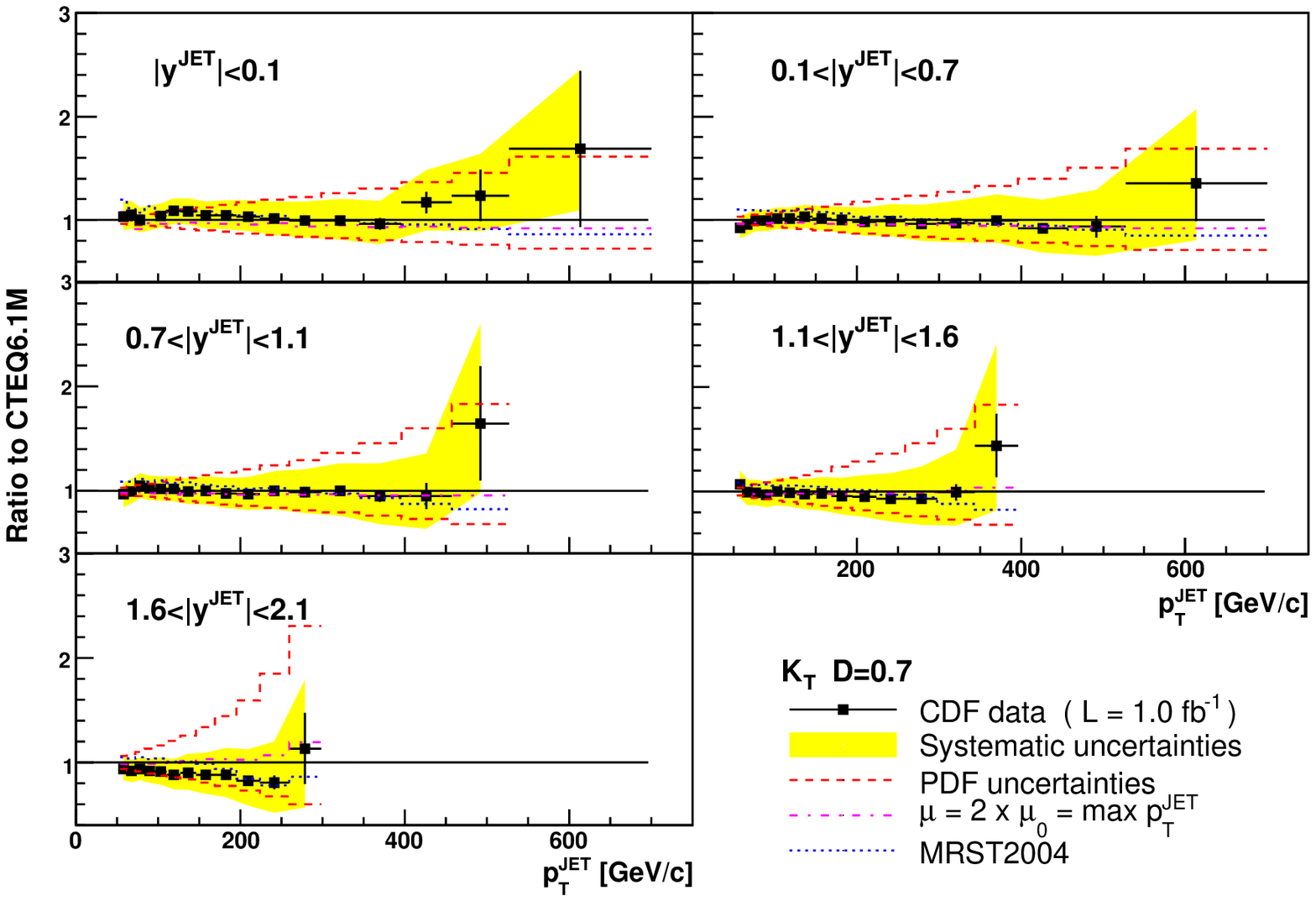}
\end{center}
\par
\vspace*{-0.5cm}\caption{ The inclusive jet cross section from CDF in Run II,
for several rapidity intervals using the $k_{T}$ jet algorithm, compared on a
linear scale to NLO theoretical predictions using CTEQ6.1
pdfs. Figure is from Ref.~\cite{Abulencia:2007ez}.}%
\label{fig:cdf_jet_kt}%
\end{figure}
\begin{figure}[thp]
\centering
\includegraphics[width=.60\hsize]
{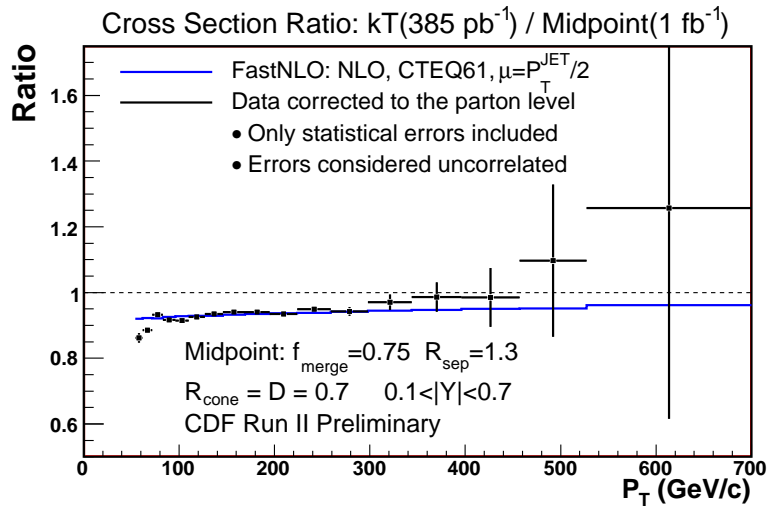}
\caption[Ratio of the inclusive jet cross
section measured with the $k_{T}$ algorithm to that measured by the Midpoint
algorithm.]{Ratio of the inclusive jet cross section measured with the $k_{T}$
algorithm to that measured by the Midpoint algorithm (black points). The
prediction of this ratio from the NLO program fastNLO is also shown in the
figure (blue line).
}%
\label{fig:kt_compare}%
\end{figure}

CDF has also made measurements of the inclusive jet cross sections with the
$k_{T}$ algorithms in Run II. In Fig.~\ref{fig:cdf_jet_kt}, the experimental
jet cross sections using the $k_{T}$ algorithm from CDF Run
II~\cite{Abulencia:2005jw,Abulencia:2007ez} are compared to NLO pQCD
predictions using the Jetrad~\cite{Giele:1994gf} program.
Similar to the measurement using the Midpoint cone algorithm, good
agreement is also observed between data and theory. A comparison of these
measurements in the central region is presented in Fig.~\ref{fig:kt_compare}%
~\cite{craig_thesis}. In order to require that the algorithms use
approximately the same size in $y-\phi$ space, the cone size for the Midpoint
algorithm and the $D$ parameter for the $k_{T}$ algorithm will both be taken
as 0.7 ($R_{cone}=D=0.7$). It is important to note that we expect different
predictions on the cross sections for jets clustered with the midpoint and
$k_{T}$ algorithms when the parameters $R_{cone}$ and $D$ are set equal.
Fig.~\ref{fig:kt_compare} shows the ratio of the measured inclusive jet cross
sections for jets clustered with the $k_{T}$ algorithm to the result for jets
clustered with the Midpoint algorithm. Only statistical errors are shown
assuming no correlation between the two measurements.
The prediction of this ratio from NLO pQCD calculations using
fastNLO~\cite{Kluge:2006xs} is also shown, and good agreement is observed. As
expected, for $D=R_{cone}$, the $k_{T}$ cross section, at the parton level, is
slightly smaller than the cross section using the cone algorithm\footnote{For
the NLO pQCD prediction for jets clustered with the Midpoint algorithm, the
$R_{sep}$ parameter of 1.3 is used. If the $R_{sep}$ parameter is set to 2,
\textit{i.e.}, $R_{sep}$ is not used, the result is a larger difference
between the two algorithms at NLO (\textit{i.e.}, setting $R_{sep}=2$ results
in a larger jet cross section as seen in Fig.~\ref{fig:allET}).}.

\begin{figure}[thp]
\centering
\includegraphics[width=.60\hsize]
{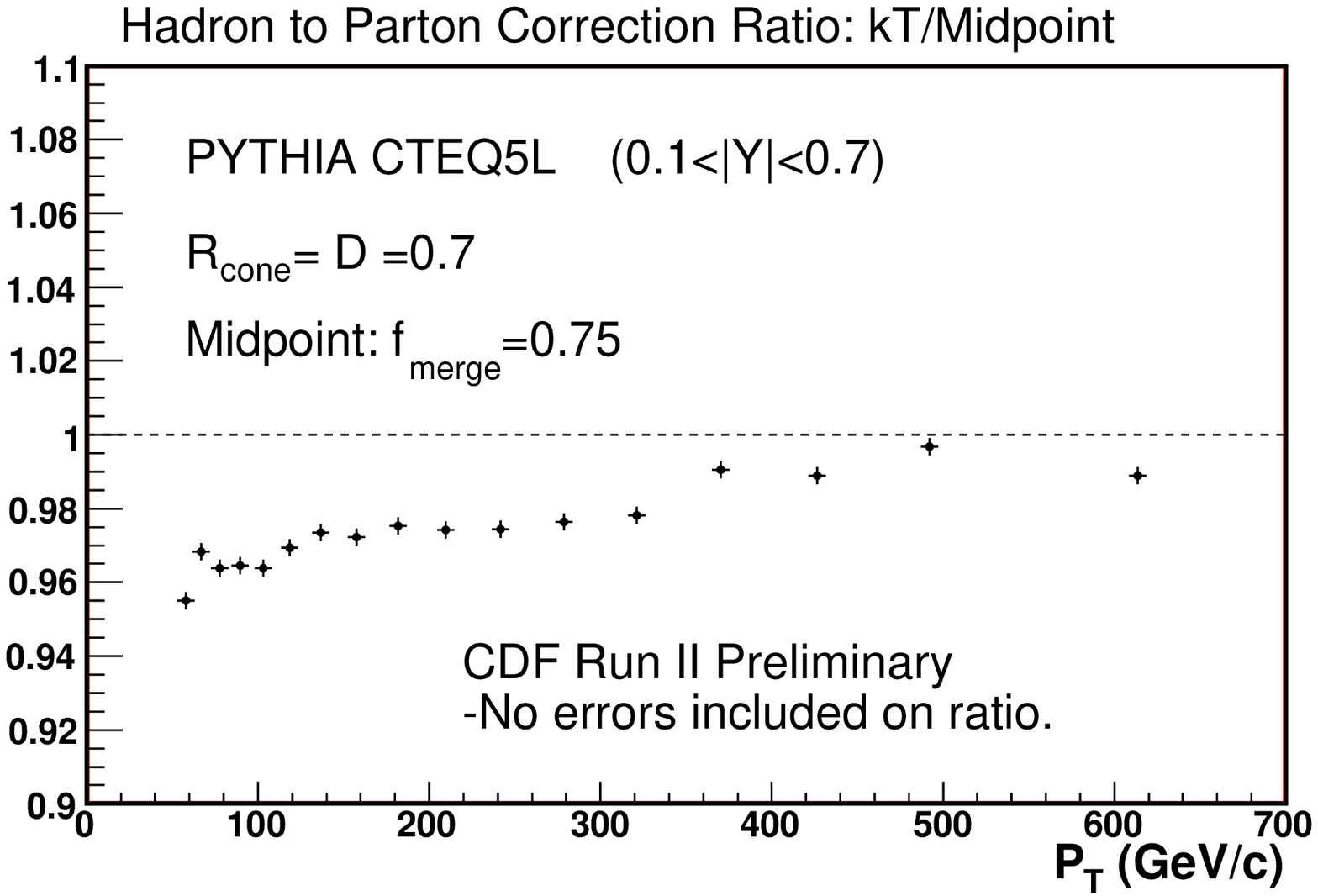}
\caption{Ratio of the hadron to parton level correction derived
with the $k_{T}$ algorithm to that derived with the Midpoint algorithm. The
multiplicative corrections are both smaller than one.
The correction derived with
the $k_{T}$ algorithm is larger (farther away from unity). The corrections
were derived from Pythia Tune A.}%
\label{fig:kt_compare_par2had}%
\end{figure}

Fig.~\ref{fig:kt_compare_par2had} shows the ratio of the hadron to parton
level correction derived with the $k_{T}$ algorithm
to the one derived with the Midpoint algorithm.
These corrections were derived from Pythia Tune A~\cite{tuneA} (see
Sec.~\ref{sec:MCtune}).
The multiplicative corrections are both smaller than one, so
the observed ratio indicates that the size of the correction obtained with the $k_{T}$
algorithm is larger (\textit{i.e.}, farther away from unity) than the
correction obtained with the midpoint algorithm.
The consistency of the data-theory comparisons between the $k_{T}$ inclusive
jet cross section measurement and the Midpoint measurement indicate the
robustness of the obtained results and adds credence to the fact that the jet
definitions are made consistently at the parton and detector levels.

\subsection{$W/Z$+Jets}

\label{sec:wzjets}

\begin{figure}[pth]
\begin{center}
\includegraphics[width=10cm]{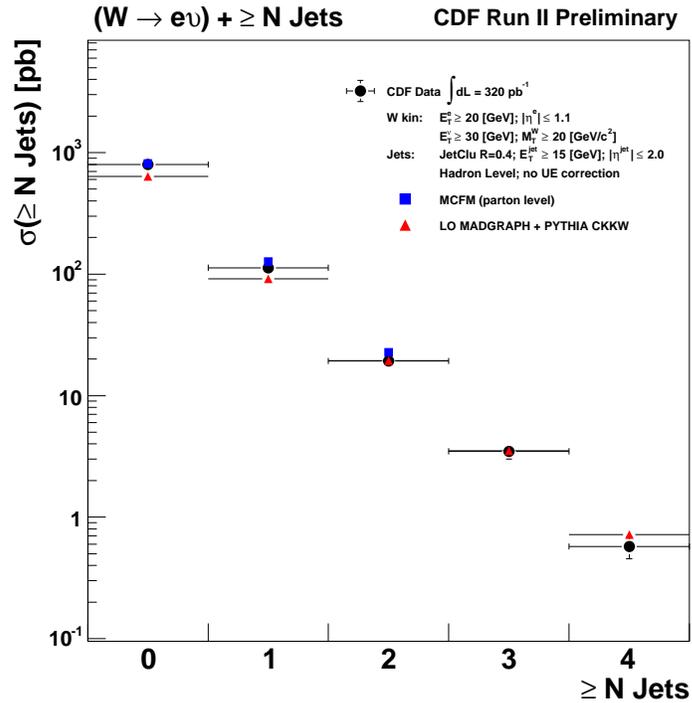}
\end{center}
\caption{ The rate of production of $W + n$~jets at CDF. The measurements and
predictions were performed with a cone jet of radius $0.4$ and with a
requirement of $15$~GeV/\textit{c} or greater. The MCFM~\cite{Campbell:2000bg}
predictions are absolutely normalized. The CKKW~\cite{Mrenna:2003if}
predictions are normalized to the first bin. A scale of
$10$~GeV/\textit{c} has been
used for the matrix element/parton shower matching.}%
\label{fig:jet_mult}%
\end{figure}

The production of a $W$ or $Z$ boson in conjunction with jets is an
interesting process in its own right as well as a background to many Standard
Model (SM) and non-SM physics signals. Jet multiplicities of up to $7$ have
been measured at the Tevatron. Production of $W$/$Z +$~jets at the Tevatron is
dominated by $gq$ initial states. The NLO cross sections have been calculated
only for $W/Z+$up to 2~jets; predictions for the higher jet multiplicity final
states are accessible through matrix element (+ parton shower) predictions and
in fact can be considered as a prime testing ground for the accuracy of such
predictions as well as for measurements of $\alpha_{s}$. The jet multiplicity
distribution for $W + n$~jets measured at the Tevatron is shown in
Fig.~\ref{fig:jet_mult}.
The measurements (and predictions) were performed with a jet cone
of radius $0.4$ and a minimum transverse energy requirement of $15$~GeV. A
smaller jet cone size is preferred for final states that may be
``complicated'' by the presence of a large number of jets. Also shown in the
figure are the NLO predictions (for jet multiplicities less than $3$), parton
shower Monte Carlo predictions and LO matrix element + parton shower Monte
Carlo predictions.
The NLO predictions are able to describe the absolute rates for jet production
(for up to 2 jets) while the LO matrix element + parton shower Monte Carlo predictions
can describe the relative jet multiplicity rates.

\begin{figure}[pth]
\begin{center}
\includegraphics[width=14cm]{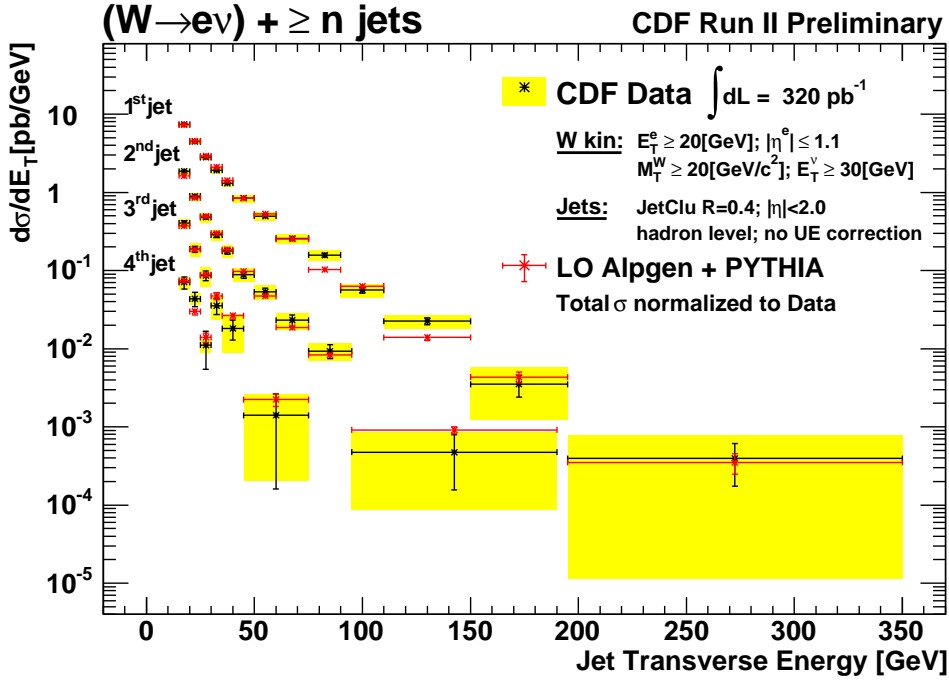}
\end{center}
\caption{ A comparison of the measured cross sections for $W+\geq n$~jets in
CDF Run II to predictions from Alpgen+Pythia. The experimental cross sections
have been corrected to the hadron level. }%
\label{fig:tev_w_mcfm}%
\end{figure}

A recent measurement of $W\rightarrow e\nu+\geq n$ jets from CDF is shown in
Fig.~\ref{fig:tev_w_mcfm}. In this analysis, the data have been
reconstructed using the JETCLU cone algorithm
with a cone radius of $0.4$. The data have been compared, at the hadron level,
to predictions using matrix element information from
Alpgen~\cite{Mangano:2001xp,Mangano:2002ea}, and parton shower and
hadronization information from Pythia~\cite{Sjostrand:2006za}. The agreement
is reasonable, although the data has a tendency to be somewhat softer than the
predictions. This data has been corrected to the hadron level; this makes it
convenient for comparison to any hadron level Monte Carlo
prediction\footnote{As mentioned before, the corrections for underlying event
and for fragmentation basically cancel each other out for a cone of radius
$0.4$, for inclusive jet production, so that the hadron level predictions are
essentially parton level predictions as well. \ It was found that the
same statement is essentially true also for $W+$ jet production.}.
Data corrections to the hadron level (and/or parton level) should be
the norm for measurements at the Tevatron and LHC, in order for the
best interplay between theory and experiment.



\subsection{Heavy Flavor Jets}
\label{sec:heavy_flavor_jets}

Many of the interesting final states at the Tevatron, such as $t\bar t$ decays
and $H\to b\bar b$, involve $b$-quark jets ($b$-jets). $W/Z+b$ and $\gamma+b$
processes are also interesting; they are major backgrounds
in Standard Model Higgs or SUSY searches, and they are also sensitive to the
$b$ content of protons. There are two main challenges in the analyses that
deal with $b$-jets; (1) the $b$-jet identification ($b$-tagging) and
(2) the energy measurement of the $b$-jet.

There are characteristics of $b$-jets that differentiate them from light
flavor and charm jets:
\begin{itemize}
\item the long lifetime of the $b$ quark
\item the large mass of $B$ hadrons
\item the energetic semileptonic decay of $B$ hadrons
\end{itemize}

The algorithms that identify $b$-jets exploit these characteristics to
separate $b$-jets from the other jets. The most-widely used algorithm to tag
heavy flavor jets at CDF is the secondary vertex algorithm, often referred to
as SecVtx~\cite{Acosta:2004hw,Neu:2006rs}. Because of their long lifetime, $b$
quarks typically decay a measurable distance from the primary interaction
point, and so the algorithm reconstructs the decay vertices (secondary
vertices) using a minimum of two or three tracks with an impact parameter
significance ($d_{0}/\sigma(d_{0})$) greater than 3.0 or 2.0, respectively.
The impact parameter ($d_{0}$) is the minimum distance between the track and
the primary vertex in the plane transverse to the beam direction and
$\sigma(d_{0})$ is its uncertainty. The two-dimensional displacement of the
secondary vertex from the primary interaction point projected along the jet
axis ($L_{2D}$) is then measured; a jet is $b$ tagged if the vertex has
$L_{2D}$ significance larger than 7.5, where the uncertainty on $L_{2D}$
includes contributions from both the primary and secondary vertex
fits~\cite{Neu:2006rs}.
\begin{figure}[th]
\centering\leavevmode
\begin{tabular}
[t]{c}%
\subfigure[]{ \includegraphics[width=0.48\hsize,clip=]
{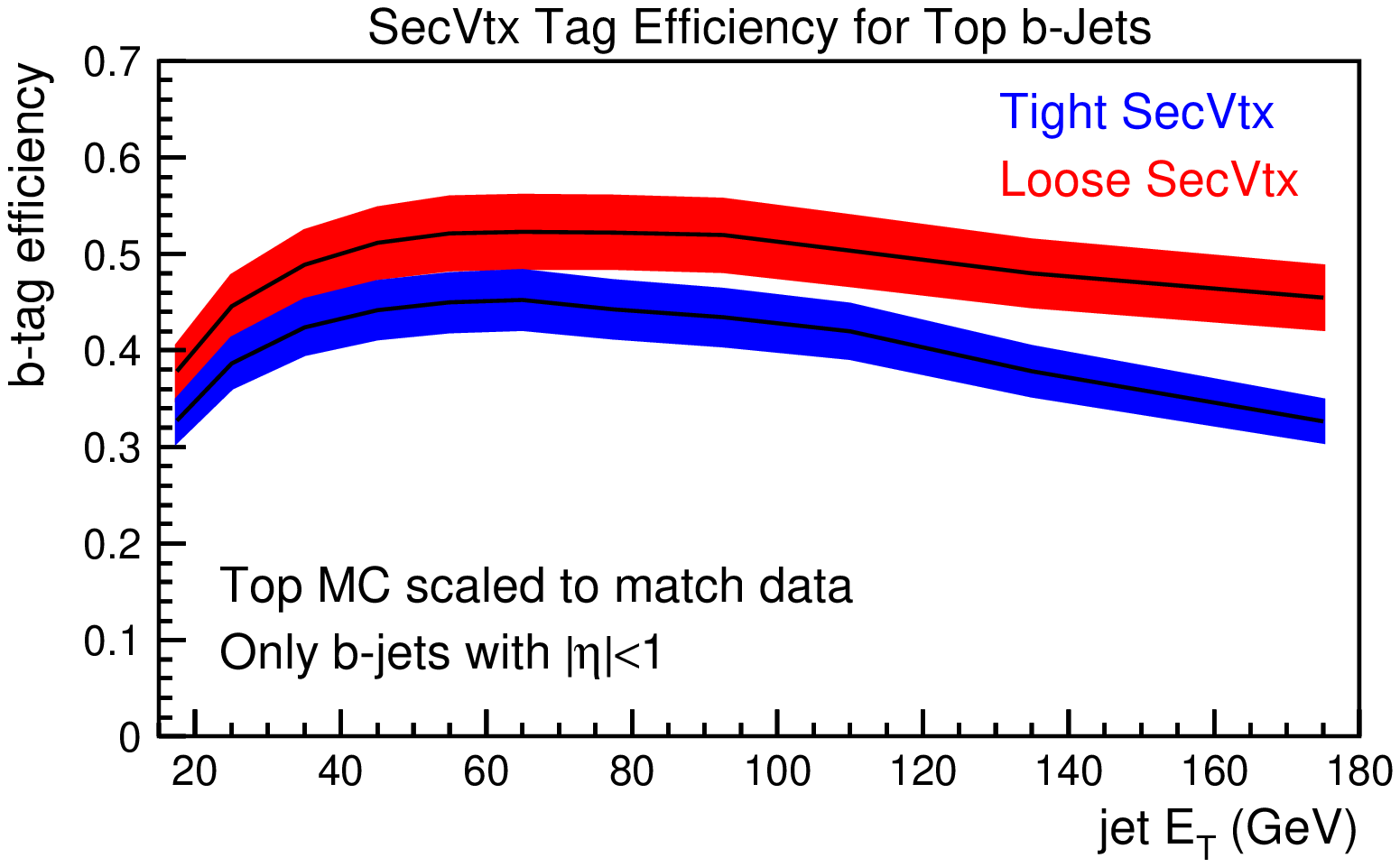} }
\subfigure[]{ \includegraphics[width=0.48\hsize,clip=]
{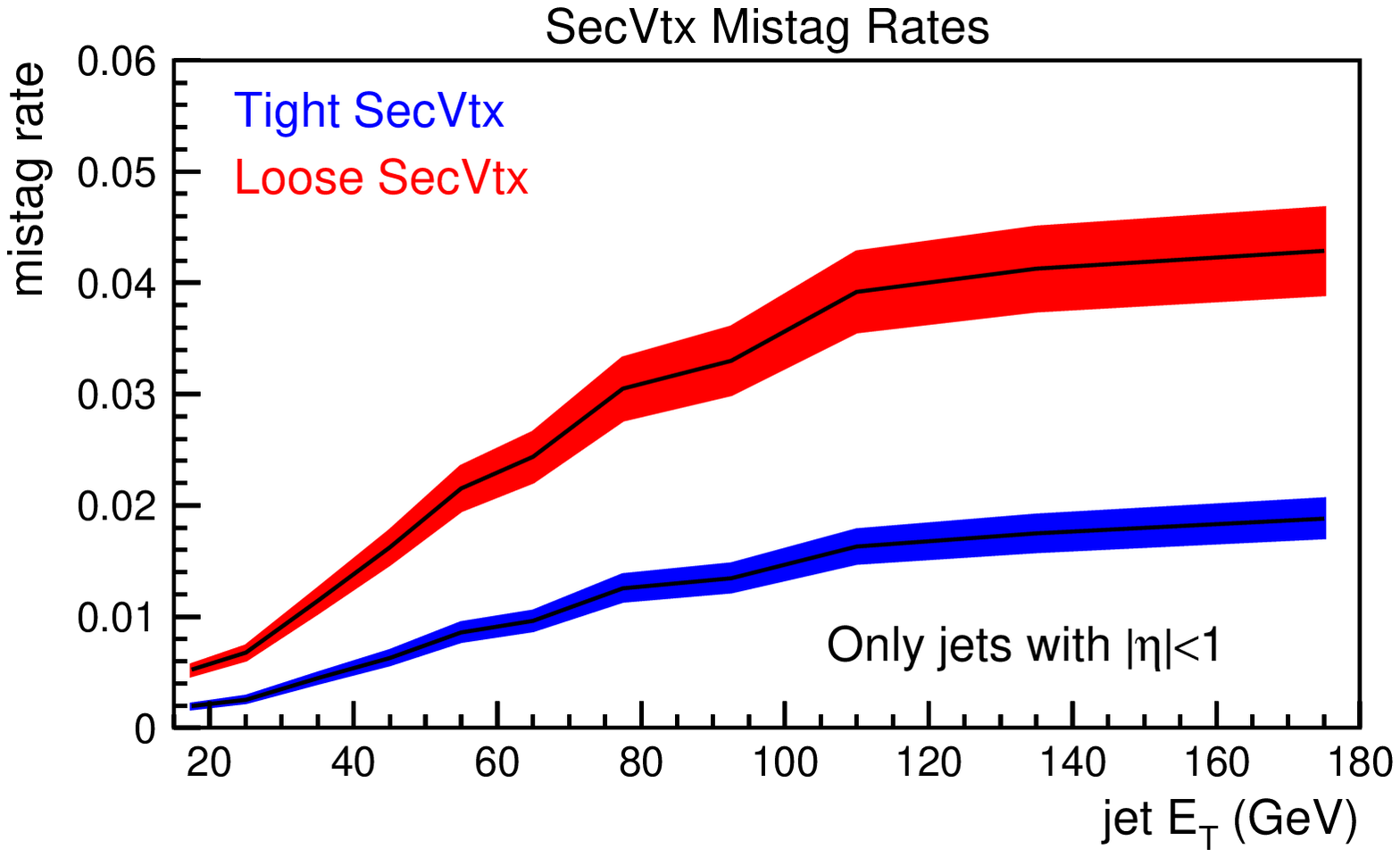} }
\end{tabular}
\caption{(a) $b$-tagging efficiency for $b$-jets in top decays and (b)
mis-tagging rate as a function of jet $E_{T}$ at CDF (figures from Ref.~\cite{Neu:2006rs}).}%
\label{fig:btag}%
\end{figure}When considering a $b$-tagging algorithm, it is important to
understand how often one tags a $b$-jet in the data (\textit{i.e.}, tagging
efficiency) and how many of the tagged jets are actually from non-$b$'s
(\textit{i.e.}, mis-tagging rate). These features for SecVtx at two operating
points are shown in Fig.~\ref{fig:btag}.

There are other $b$-tagging algorithms developed at CDF. The algorithm called
JetProbability considers $d_{0}$ of each track within a jet and constructs a
probability that a given jet is consistent with coming from a zero-lifetime
source~\cite{Abulencia:2006kv}. The soft lepton tagging algorithm identifies
$b$-jets by looking for a semileptonic $B$ hadron decay withing a
jet~\cite{Acosta:2005zd}. Efforts to combine these tagging tools using a
multivariate technique like a neural network are also underway.

Since $b$-jets have quite different characteristics from other jets
(light quark or
gluon), special care is necessary for the energy measurement. The energy
correction from calorimeter jets to hadron-level jets or to the parent parton
is different for $b$-jets than for generic jets because of different parton
shower and fragmentation properties, and also due to the presence of
semi-leptonic decays. Both CDF and D\O \ generally rely on MC simulation to
model the $b$-jet energy scale. In the CDF top quark mass
measurements~\cite{Abulencia:2005ak}, additional uncertainties are evaluated
for the $b$-jet energy scale: (1) uncertainties in energy response arising
from uncertainty in the $B$ meson semi-leptonic branching ratios, (2)
uncertainties arising from the imperfect knowledge of the fragmentation
properties of $b$-quarks, and (3) uncertainties
arising from the different color flow associated with $b$-jets produced in top
quark decay. The $b$-jet energy scale uncertainties from these sources are
evaluated by changing the relevant parameters in the MC based on the
constraints from other experiments, and they yield an additional 0.6\%
uncertainty in total.

Possible ways to test the $b$-jet energy scale in $p\bar p$ data would be to
use the photon-$b$-jet $p_{T}$ balance or the $Z\to b \bar b$ resonance.
D\O \ recently made a preliminary measurement on $b$-jet energy scale in
$\gamma$+$b$-jet events using the missing $E_T$ projection fraction
(MPF) method~\cite{Aglietti:2006ne}.
The MPF method is described in detail in Section~\ref{sec:jes_dzero}.
The study suggested that $b$-jets need additional energy corrections
of as much as 10\% at energies around 20 GeV and about 5\% at energies of 150 GeV.

\begin{figure}[th]
\centering\leavevmode
\begin{tabular}
[t]{c}%
\subfigure[]{ \includegraphics[width=0.49\hsize,clip=]
{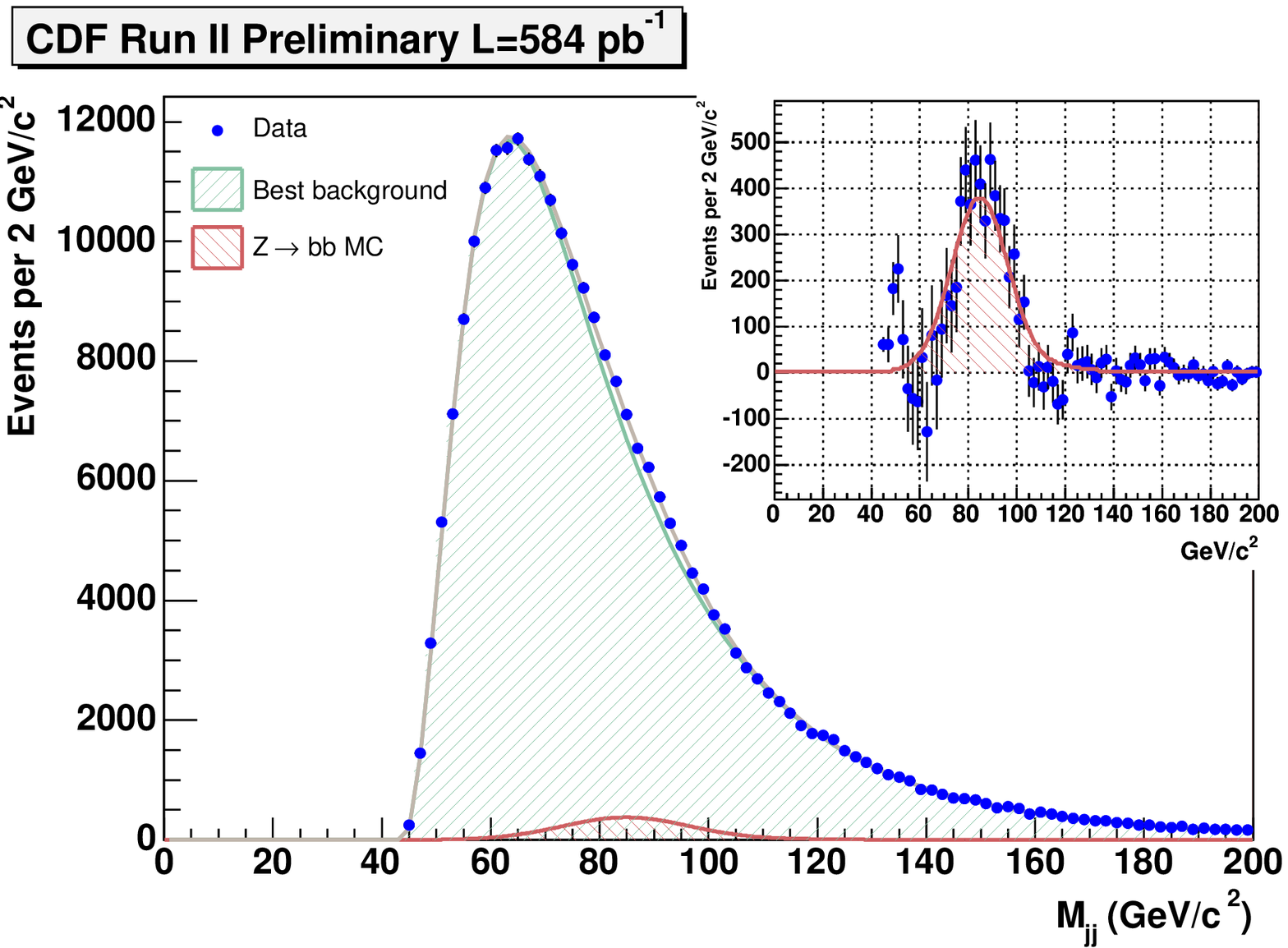} }
\subfigure[]{ \includegraphics[width=0.49\hsize,clip=]
{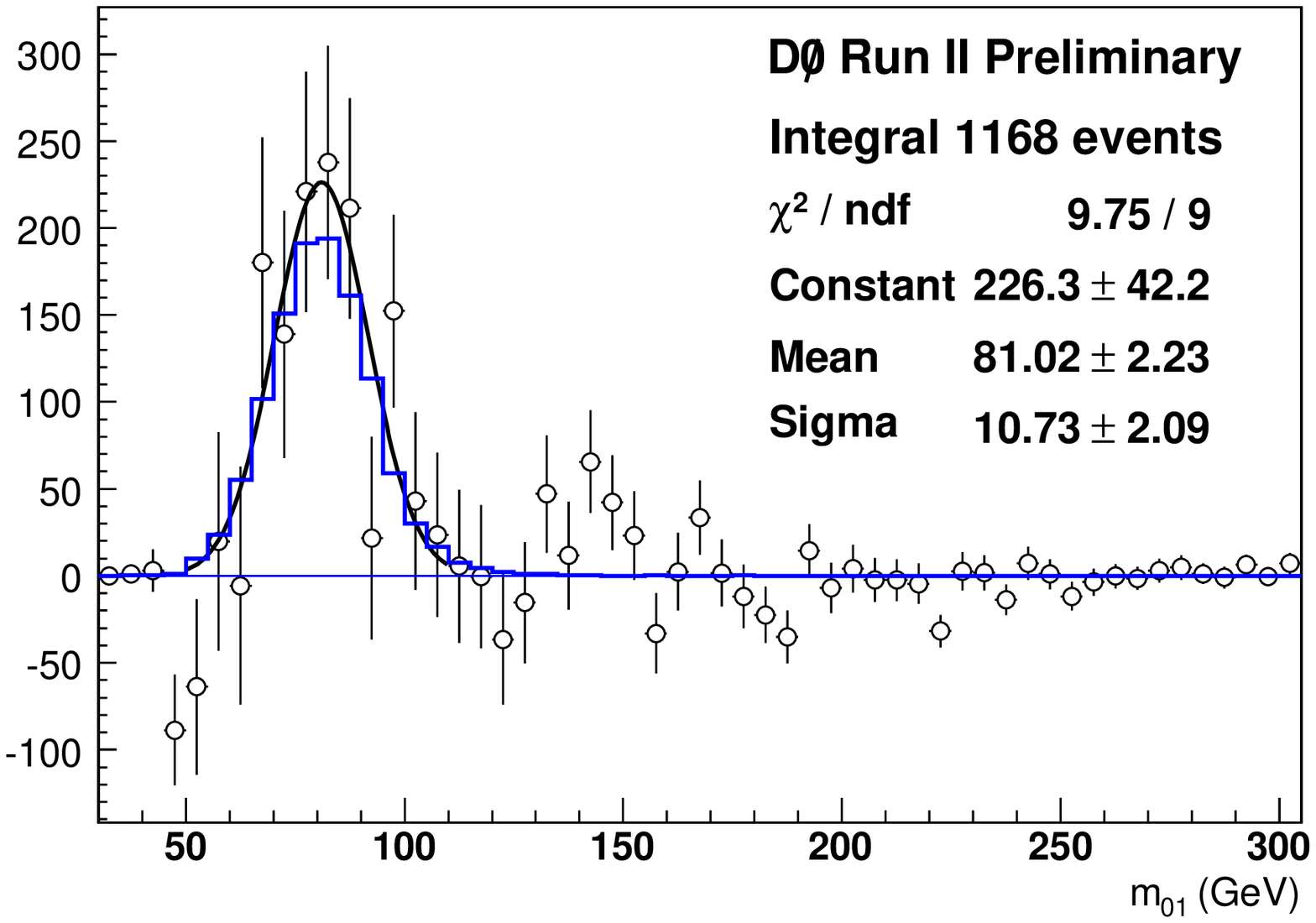} }
\end{tabular}
\caption{(a) the $b\bar b$ dijet mass distribution measured in CDF data
(points) and the estimated QCD $b\bar b$ background (green) and $Z\to b\bar b$
signal shape (red). The inset shows data minus background distributions
compared to the $Z\to b\bar b$ signal shape. (b) $Z\to b\bar b$ peak extracted
from D\O ~data (points), compared to the shape of $Z\to b\bar b$ distribution
in MC (histogram). Figures from Refs.~\cite{zbb_cdf,zbb_dzero}.}%
\label{fig:zbb_mbb}%
\end{figure}

CDF and D\O \ have also extracted the $Z\to b\bar b$ signal and are seeking to
use it to test and calibrate the $b$-jet energy scale~\cite{zbb_cdf,zbb_dzero}.
At CDF, a dedicated trigger was implemented to collect a large sample of $Z$
decays to $b$-quark pairs by requiring two tracks with displaced vertices and
two jets. Dijet events with both jets being tagged as $b$-jets were selected
offline. The signal was extracted by fitting the data with the QCD background
shape computed using untagged data passing the same kinematic selection
together with the $Z\to b\bar b$ signal and the $Z\to b\bar b$ signal shape
computed with Pythia. The extracted $Z\to b\bar b$ signals by CDF
and D\O \ are shown in Fig.~\ref{fig:zbb_mbb}. The measured data/MC scale
factor for $b$-jet energy scale for CDF is
$k=0.974+0.020-0.018(\mbox{stat}\oplus\mbox{syst})$. The achieved uncertainty
of $\sim2$\% is smaller than the convolution of the generic jet energy scale
and additional $b$-jet specific uncertainties of $\sim4$\%, and thus the
$Z\to b\bar b$ signal can provide a good constraint on the $b$-jet energy
scale. The signal is also expected to serve as a tool to test $b$-tagging
algorithms and also for improvements in the jet energy measurement algorithms
which will be discussed in Sec.~\ref{sec:pfa}.

\subsection{Particle Flow Type Approaches}
\label{sec:pfa}

In the Tevatron experiments, by default, jets are reconstructed by running a
jet clustering algorithm on energy depositions in the calorimeters,
\textit{i.e.} basically only calorimeter information is used in the standard
jet reconstruction and energy measurement.
The H1 collaboration at HERA~\cite{Adloff:1997mi} and OPAL collaboration at
LEP~\cite{Abbiendi:1999sy}, among others, have successfully used an algorithm
that incorporates tracks in jet reconstruction, and such an algorithm has been
tested at CDF and D\O\ as well.
As discussed, the energy measurement of hadronic particles by calorimeters
suffers from large fluctuations and non-linearities; tracking information can
be used to reduce such effects and to improve the energy resolution in jet
energy measurements.
There is also an attempt in CDF to further improve the
jet energy measurmenet by using the information from the shower-max
detector (a wire chamber placed near the shower maximum position in
the central electromagnetic calorimeter) to sort out the overlapping
particles like $\pi^{0}$ with $\pi^{\pm}$~\cite{Lami:2001bb}.
Although such methods are not yet mature enough to be
used in any physics analysis, they could be helpful in
achieving our goal of percent precision and should be pursued further.


\section{Jets at the LHC}

The experience gained at the Tevatron is extremely useful in the preparation
for physics analysis with jets at the LHC. However, hard scattering at the LHC
is not just \textquotedblleft re-scaled\textquotedblright\ scattering from the
Tevatron. Many of the interesting physics signatures will take place with
relatively low $x$ partons and thus there will be a dominance of gluon and sea
quark interactions, as compared to interactions involving valence quarks. In
addition, as the initial state partons are at low $x$, there is enormous phase
space for gluon emission, and so a large probability for additional jets from
initial state radiation. The underlying event is also expected to be enhanced
compared to the activity observed at the Tevatron, largely through an
increased rate of semi-hard multiple-parton interactions. These interactions
will contribute to the energy measured in jets produced in the hard scatter,
and may often lead to the production of extra low transverse momentum jets in their
own right. Thus, the LHC will be a very \textquotedblleft
jetty\textquotedblright\ environment and accurate measurements of the dynamics
of the hard scattering may be challenging~\cite{Campbell:2006wx}. \ There is
then a need for tools even more powerful than the ones used at the Tevatron to
reconstruct jets.
The most interesting tools focus on the reconstruction of jet shapes, thus
exploiting the significantly finer readout granularity of the LHC detectors.
In this section we summarize some of the related aspects presently under
study, and show some expectations for the jet final state at LHC and its
representation in the detectors.

\subsection{Expectations for Jet Final States}

As already mentioned, jets will be generated in basically all final states
expected in $pp$ collisions at $\sqrt{s} = 14$ TeV at the LHC. For most
channels they are going to be the dominant part of the detectable signal, thus
providing major input to the reconstruction of the event kinematics. Precision
requirements on the jet energy scale are high compared with the Tevatron, with
systematic uncertainties of less than $1\%$ absolute needed/expected for
jets reconstructed in $t\overline{t}$ production, or for jets generated at the
end of long decay chains in certain SUSY models.


\begin{figure}[ptb]%
\begin{tabular}
[c]{cc}%
\includegraphics[width=8cm]{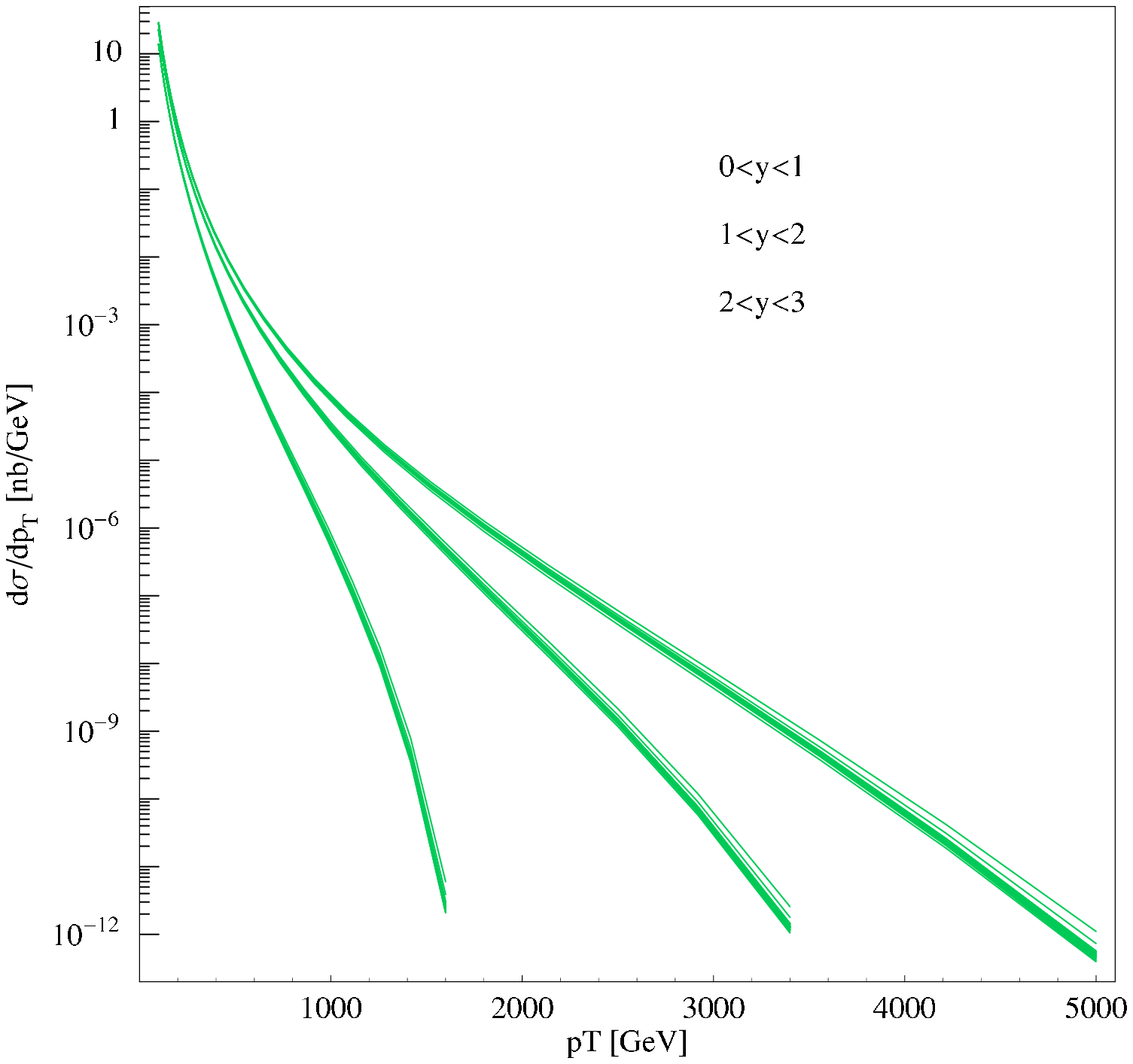} &
\includegraphics[width=10cm]{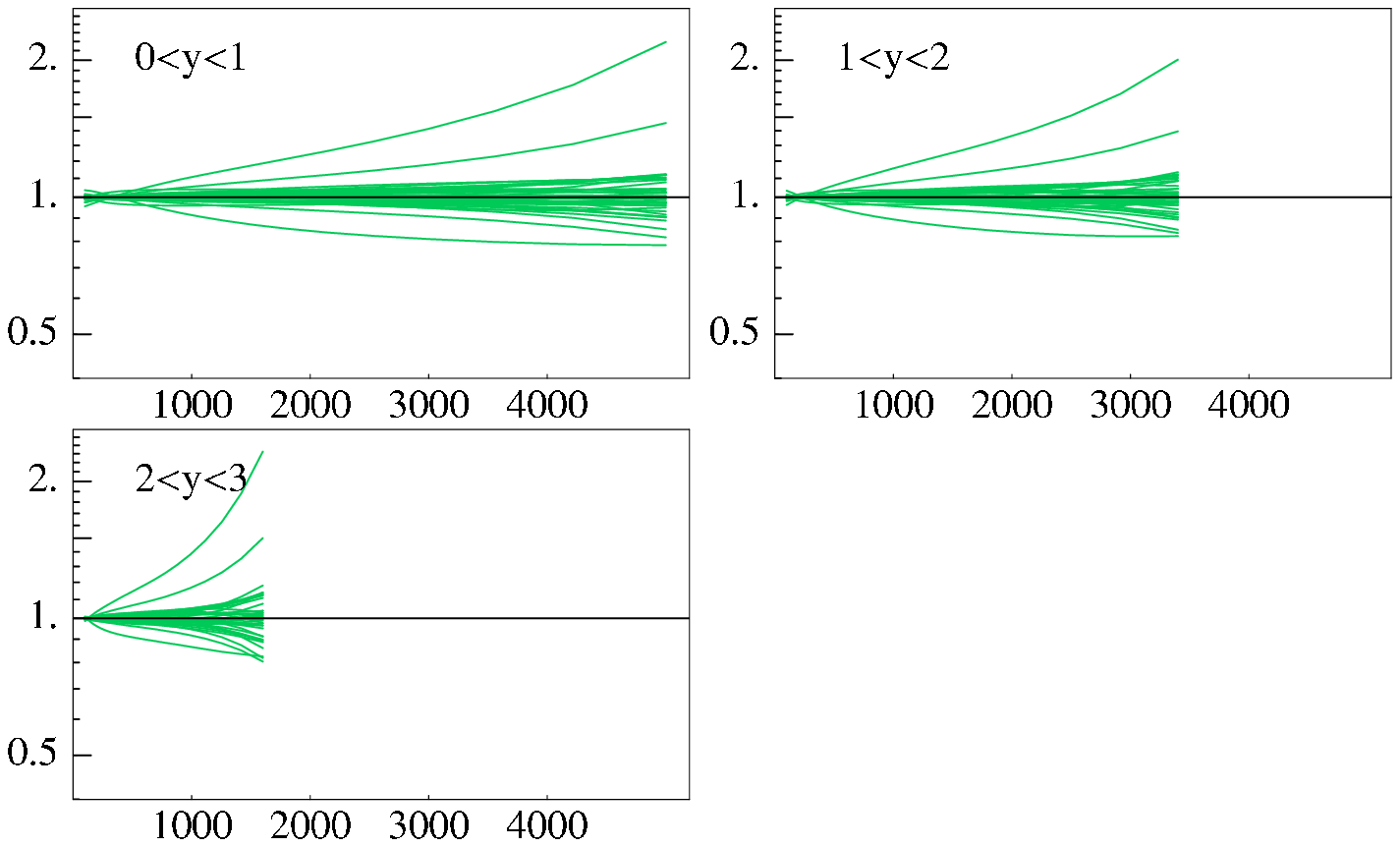}
\end{tabular}
\caption{NLO Inclusive jet cross section predictions for the LHC using the EKS
\cite{Ellis:1992en} program and the CTEQ6.1 central pdf and the $40$ error
pdf's(left); the ratios of the jet cross section predictions for the LHC using
the CTEQ6.1 error pdf's to the prediction using the central pdf (right). }%
\label{fig:jetxsect}%
\end{figure}

The kinematic reach of jets produced in QCD $2\rightarrow2$ processes at the
LHC greatly extends that possible at the Tevatron; for example, compare
Fig.~\ref{fig:incjet_run2}\ in section \ref{sec:incjet}\ with the predictions
for the LHC shown in Fig.~\ref{fig:jetxsect}. Even in the initial lower
luminosity running phase, the jet production rate is also very high, as is
summarized in Table~\ref{tab:jetcts}. It is interesting to note that the pdf
uncertainty for jets at the LHC at the highest attainable transverse momentum
is similar to that for the Tevatron at its highest attainable transverse
momentum. \begin{table}[ptb]
\caption{Expected inclusive jet production rate at LHC in one run year with an
integrated luminosity $\int\mathcal{L}dt = 10$ fb$^{-1}$.}%
\label{tab:jetcts}%
\centering
\renewcommand{\arraystretch}{1.25}
\begin{tabular}
[c]{||c|c|c||}\hline\hline
$p_{T}^{\mathit{min}}$ (TeV$/c$) & $\sigma$ (nb) & Events/year\\\hline\hline
$0.2$ & $100$ & $\approx10^{9}$\\
$1.0$ & $0.1$ & $\approx10^{6}$\\
$2.0$ & $1.0\times10^{-4}$ & $\approx10^{3}$\\
$3.0$ & $1.3\times10^{-6}$ & $\approx10$\\\hline\hline
\end{tabular}
\end{table}

\begin{figure}[ptb]
\centering
\includegraphics[width=16cm]{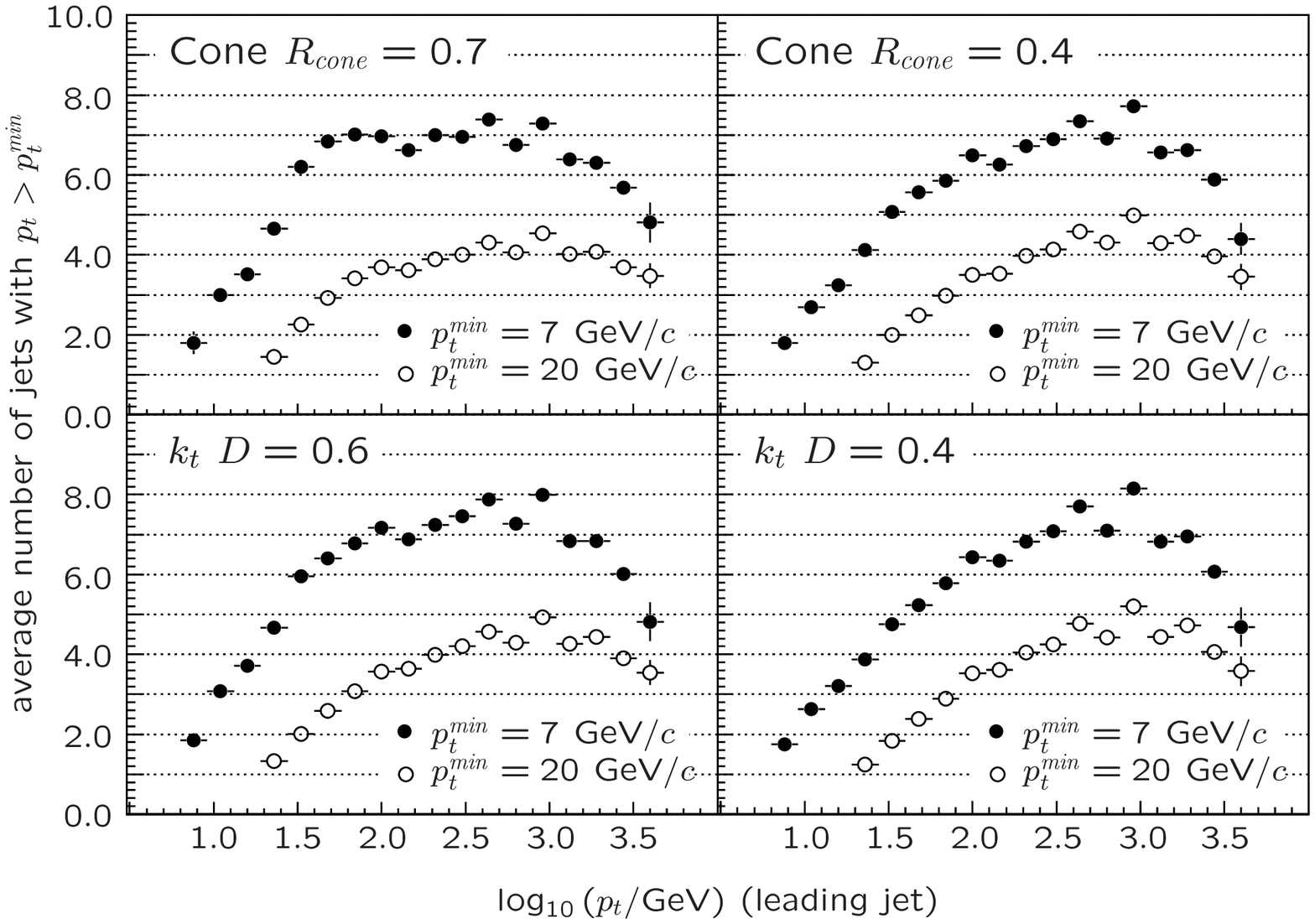} \caption{Number of hadron jets in
simulated QCD $2\rightarrow2$ LHC events, as a function of the leading jet
$p_{T}$, for various jet finders. The shapes of the curves can be understood
as a convolution of the $p_{T}$\ cut applied in jet finding, the increasing
probability for extra hard gluon emission with the increasing hardness of the
$2\rightarrow2$ scattering, the change from gluon to quark jets with
increasing $p_{T}$, and finally the restricted phase space for extra jets
accompanying very energetic primary jets. }%
\label{fig:njetsvspt}%
\end{figure}

The partonic phase space available in $pp$ collisions at the LHC allows for a
large amount of extra radiation, as can be observed in the number of jets
calculated as a function of the leading jet $p_{T}$\ in QCD $2\rightarrow2$
processes (see Fig.~\ref{fig:njetsvspt}). For low $p_{T}$\thinspace\ the
number of additional jets is suppressed by the $p_{T}$\ cut applied to jets
being non-negligible compared to the transverse momentum of the hard
scattering, while the drop towards higher $p_{T}$\ indicates that radiation is
suppressed due to the increasing dominance of $qq$ in higher partonic $x$
scatterings, and the subsequent lower color factors in the collision. There is
also an extra suppression due to the higher $x$ values of the incident
partons. 

\subsection{Jet Physics Environment at the LHC}

Jet measurements are affected by the presence of multiple semi-hard
interactions from other parton-parton pairs from the proton-proton collision
of interest, \textit{i.e.}
the underlying event (UE) of the
same collision. In addition, the physics environment at the LHC is also
affected by additional minimum bias (MB) collisions of other proton pairs in 
the
same bunch crossing. Both effects limit the efficiency for reconstructing the
hadron-level jets (and ultimately the parton-level jets) from the hard
scattering (signal) and also add complex features to the already non-trivial
detector jet signals. They are also major sources for uncertainties
in the present simulation-based performance estimates for LHC physics.

\subsubsection{The Underlying Event at the LHC}

\begin{figure}[ptb]
\centering
\includegraphics[width=8cm]{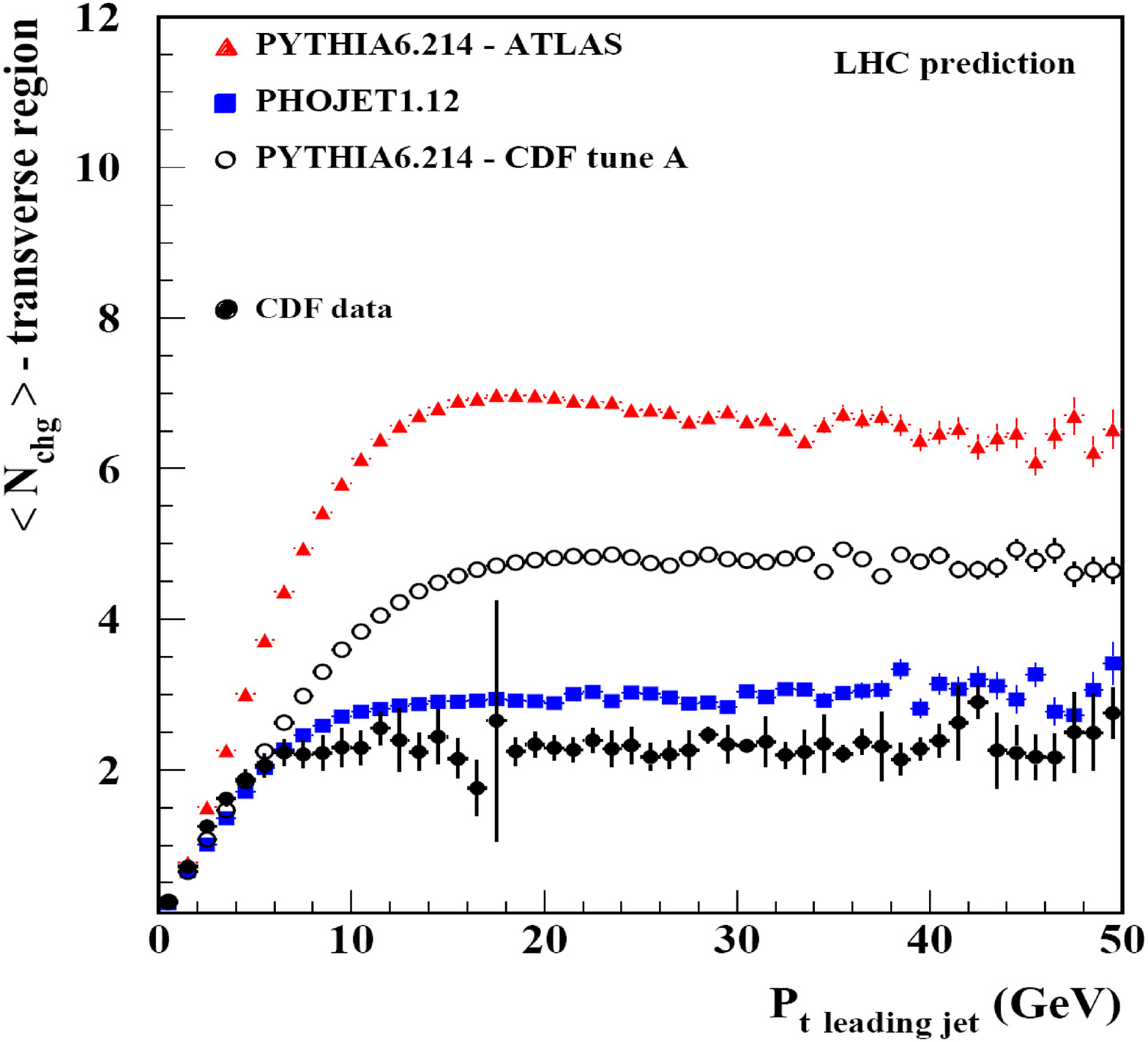} \caption{Number of charged tracks
in the transverse region of the QCD $2\rightarrow2$ interaction plane (as
defined in Fig.~\ref{fig:ue_ana_topology}) as a function of the leading jet
$p_{T}$. Data from CDF{\cite{Affolder:2001xt}} are shown together with model
predictions for the LHC (figure from {\cite{atlas:ue}}). }%
\label{fig:ue}%
\end{figure}

There is a great deal of uncertainty in the level of underlying event activity
expected for $pp$ collisions at $\sqrt{s}=14$ TeV, as can be observed in
Fig.~\ref{fig:ue}. This uncertainty is a major factor in estimating the
quality of the reconstruction of the jet signals. Estimates derived from a
recent tuning of Pythia~\cite{Sjostrand:2006za} (\textquotedblleft Pythia~6.214
CDF Tune A\textquotedblright), discussed previously in this paper,
actually predict significantly fewer tracks from the UE at the LHC than the
parameters previously used in the same Pythia version 
by ATLAS~\cite{atlas:ue} 
(\textquotedblleft Pythia~6.214 ATLAS\textquotedblright\ in 
Fig.~\ref{fig:ue}). The
determination of the level of this underlying event activity will be one of
the first measurements to take place upon startup of the LHC.

An ATLAS study similar to the one carried out by CDF (and shown in
Fig.~\ref{fig:ue_transmax_min})
indicates that the sum of the transverse momentum of the charged particles in
the transverse region (see Fig.~\ref{fig:ue_ana_topology}) will vary
from approximately 10 GeV$/c$ for
low jet transverse momentum to over 30 GeV$/c$ for jet transverse momenta larger
than 1 TeV$/c$~\cite{atlas:ue}.
%

\subsubsection{Minimum Bias Events and Pile-up}

\begin{figure}[ptb]
\centering
\begin{tabular}
[c]{cc}%
\includegraphics[width=8cm]{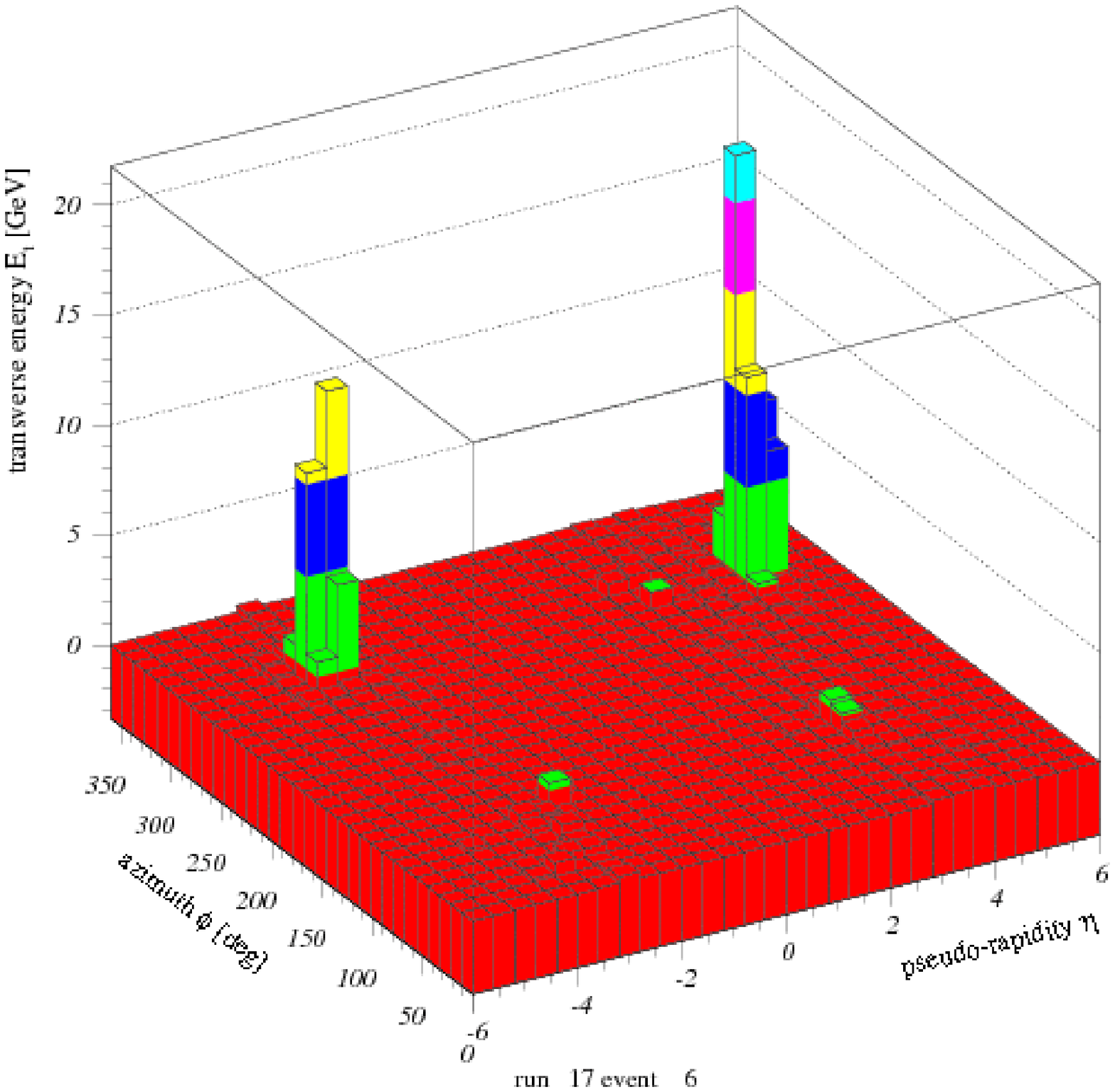} &
\includegraphics[width=8cm]{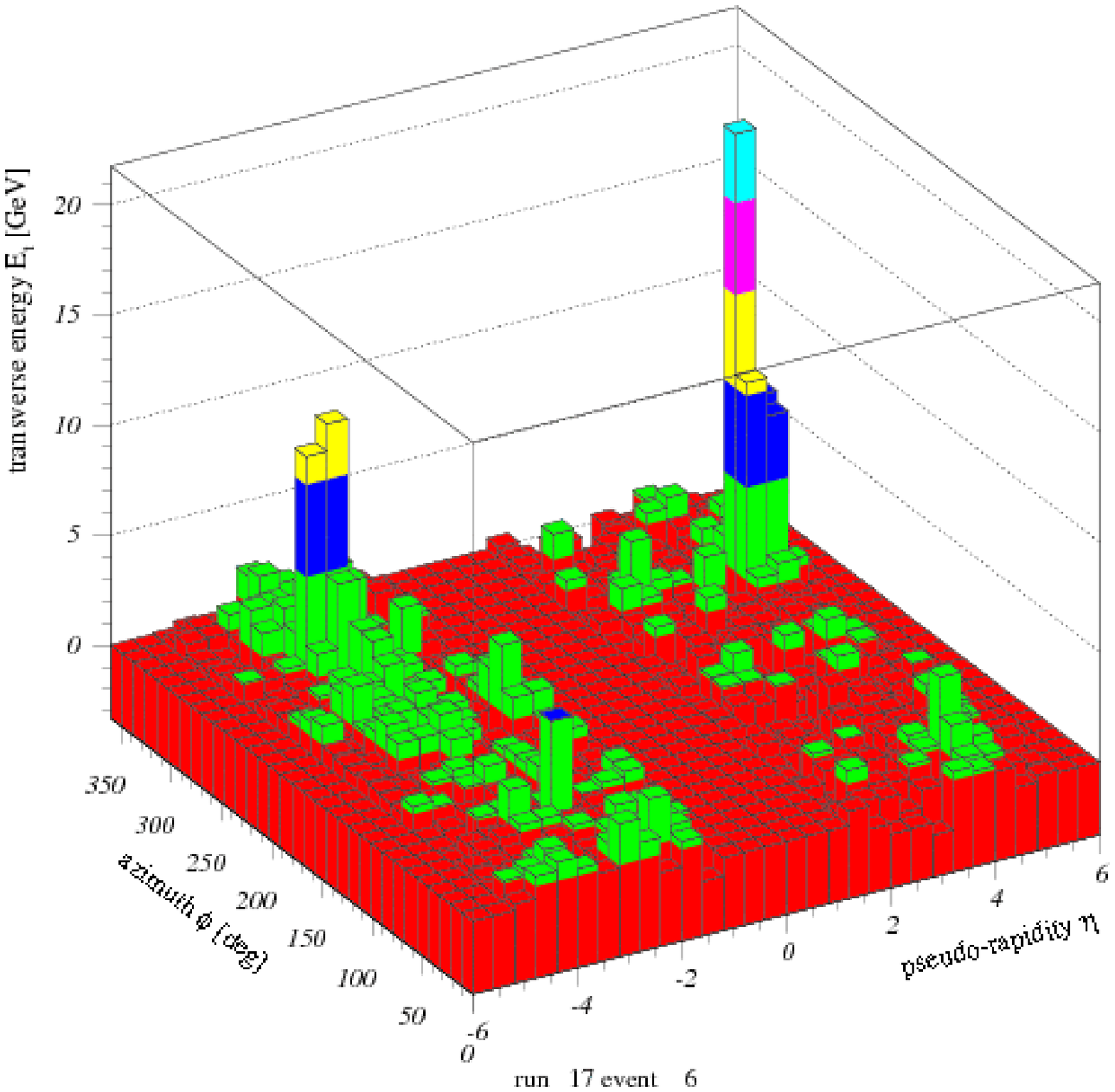}
\end{tabular}
\caption{Simulated forward going quark jets from Higgs production via $WW$
scattering, without (left) and with pile-up (right) in the ATLAS forward
calorimeter. The signals with $p_{T} < 0$ are due to the specific choice of
bi-polar signal shaping functions for this detector. The signals in the
central detector regions are omitted here for clarity. }%
\label{fig:pile-up}%
\end{figure}Contrary to the underlying event energy, the additional minimum
bias interactions present at higher luminosities provide a source for $p_{T}%
$\ flow in the LHC collisions which is not correlated at all with the
(triggered) hard scattering of the signal event. Its total contribution is
dependent on the instantaneous luminosity. Assuming a total $pp$ cross-section
of about $75$ mb and excluding single diffractive and double diffractive
collisions, one can expect an average number of MB collisions of $\approx23$
(Poisson-distributed) at the design luminosity ($L = 10^{34} \mathrm{cm}%
^{-2}\mathrm{s}^{-1} $), $\approx4.6$ at the ``initial'' luminosity $L =
2.0\times10^{33} \mathrm{cm}^{-2}\mathrm{s}^{-1} $, and $\approx0.02$ at the
LHC startup\footnote{These estimates assume that the bunch crossing
time is $25$ ns, with about $3000$ bunches in LHC. Less frequent bunch
crossings, and a smaller number of (longer) bunches at the same stored current,
as recently discussed for initial LHC running, can increase the pile-up
significantly, even at lower luminosities.}
 ($L = 10^{31} \mathrm{cm}^{-2}\mathrm{s}^{-1} $, the most recent
expectation). This activity is therefore likely negligible at the LHC start-up,
but can produce a large number of non-signal tracks/energy at the design 
luminosity.

\begin{figure}[ptb]
\centering
\includegraphics[width=6cm]{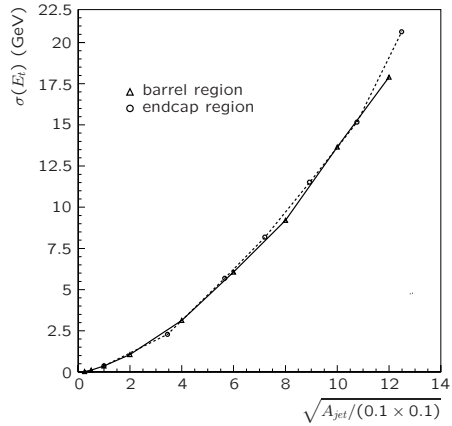} \caption{Estimated signal
fluctuations from full pile-up simulations, calculated as RMS in $p_{T}$\ for
the central and endcap region in ATLAS, as a function of the square root of
the jet cone area. For a typical cone size $R_{\mathit{cone}} = 0.7$ the
fluctuation are about $12$ GeV in both detector regions (taken from
\cite{atlas:savard}). }%
\label{fig:purms}%
\end{figure}The actual detailed effects of the minimum bias events on the jet
signal depend strongly on the calorimeter technology and readout electronics.
For example, the rather slow signal formation in the ATLAS liquid argon
calorimeters with typical charge collection times of about $500$ ns (which has
to be compared to the LHC bunch crossing time ($25$ ns)) and the rather large
$pp$ cross-section leads to a history of previous collision signals still
visible in the actual event. The effect averages to $0$ energy, due to the
specifically chosen bi-polar (canceling area) signal shaping, but the
``out-of-time'' pile-up adds to the fluctuations. Fig.~\ref{fig:pile-up} shows
the simulated response to two rather low $p_{T}$\ jets from vector
boson fusion (VBF) Higgs
production in the ATLAS forward calorimeters with and without full LHC
luminosity pile-up added. Both jets are well visible above the noise, but
their shapes, as well as their signal amplitudes, are changed, which makes
calibration in this particularly hostile region very challenging. A more
quantitative estimate for the signal fluctuations introduced by pile-up in jet
cones in shown in Fig.~\ref{fig:purms} \cite{atlas:savard}.

\subsection{Experimental Aspects of Jet Reconstruction at the LHC}

\begin{figure}[ptb]
\centering
\begin{tabular}{cc}
\includegraphics[width=0.520\hsize]{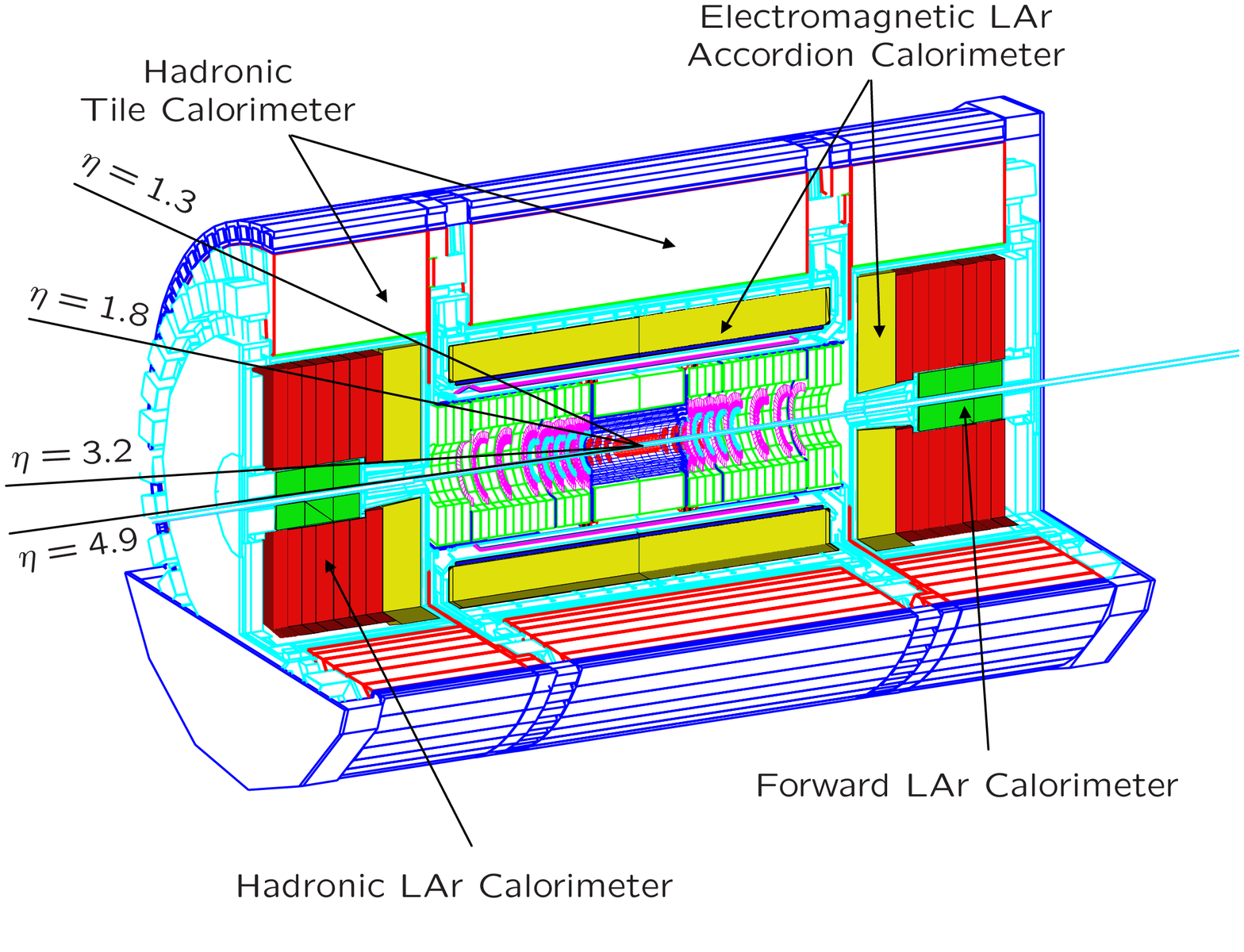}
&
\includegraphics[width=0.460\hsize, bb=0.0 -75.0 783.909180 566.268066]{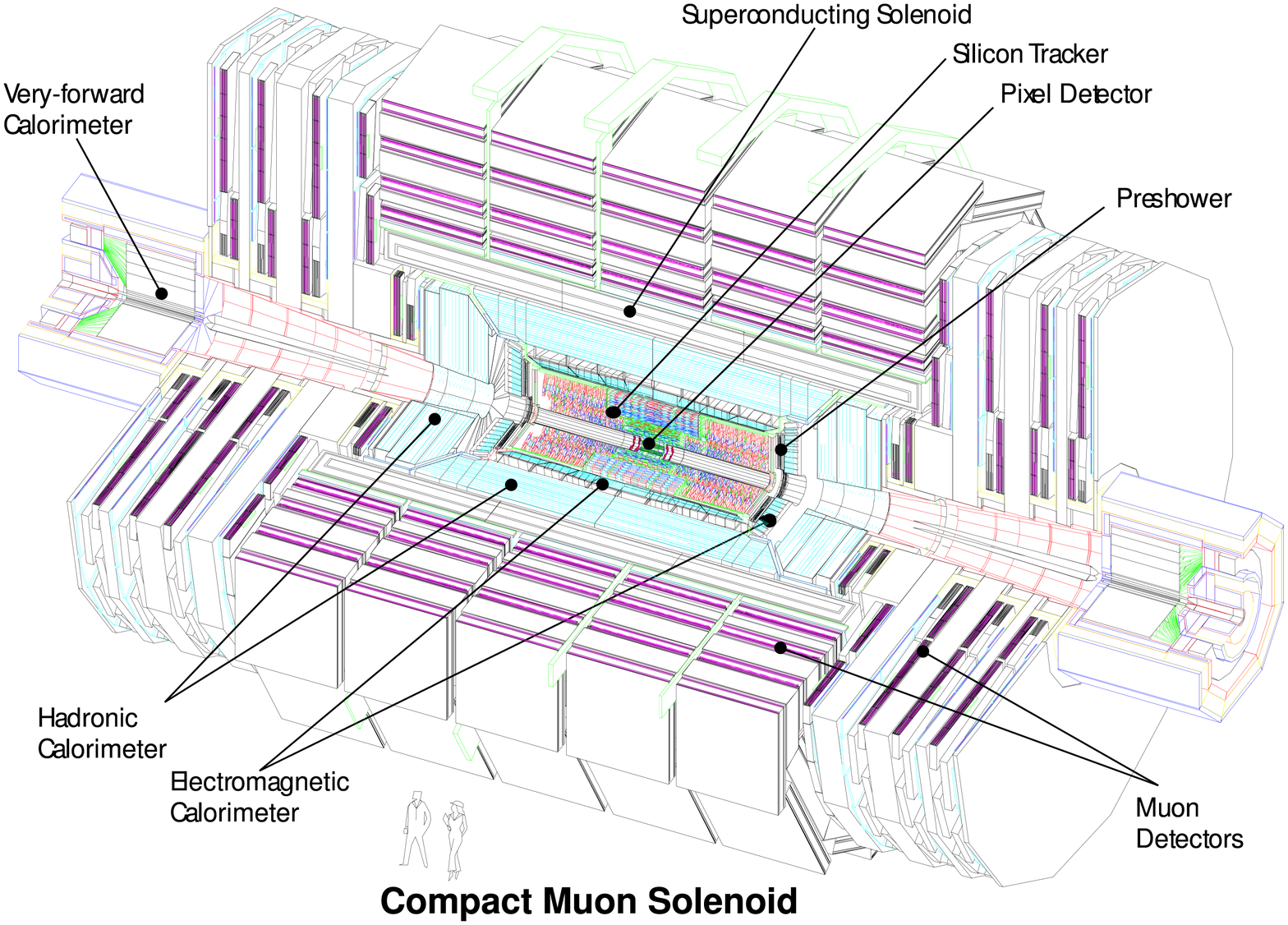}
\end{tabular}
\caption{The ATLAS calorimeter system (left) and CMS detector system (right).}%
\label{fig:atlascms}%
\end{figure}

The large calorimeter systems in ATLAS and CMS at the LHC are the basic detectors
for jet reconstruction. Both systems provide hermetic coverage up to
pseudorapidities of $\sim5$. Cell sizes and readout granularity vary widely
within each of the systems and introduce different limitations on the
calorimeter signals used for jet finding and reconstruction. In general,
though, the cell sizes are smaller than in the Tevatron calorimeters,
allowing for the development of more powerful jet clustering software. Typical
depths of the calorimeter systems exceed $8-10$ absorption lengths.
Fig.~\ref{fig:atlascms} gives an overview on these detectors, with some of the
relevant details described below. 

\subsubsection{Brief Look at the ATLAS and CMS Calorimeters}

CMS features a highly granular central electromagnetic lead-tungstate
(PbWO$_{4}$) crystal calorimeter with very small lateral cell sizes
($\Delta\eta\times\Delta\phi= 0.0174 \times0.0174$ within $|\eta|<1.479$).
The calorimeter consists of 61200 individual crystals and has a total depth of
25.6 radiation lengths ($X_{0}$). The crystals point back to the vertex,
within a very small tilt. The electromagnetic endcap calorimeters also use
PbWO$_{4}$ crystals, but with a coarser granularity and a rectangular pattern
of fixed-sized crystals. The depth of these crystals is about $24.7$ $X_{0}$.
Each endcap covers the region $1.479 <|\eta|<3.0$.

Hadronic calorimetry in CMS features tiled scintillator read-out with brass
absorbers. The central hadronic calorimeter is arranged outside the central
electromagnetic calorimeter, but still inside the solenoid magnet. It has a
granularity of $\Delta\eta\times\Delta\phi= 0.087 \times0.087 $ in one
depth segment within $|\eta|<1.4$. The hadronic outer detector, a scintillator
layer attached to the outside of the solenoid magnet, provides additional
depth coverage for hadrons. The hadronic endcap calorimeters feature the same
readout technology with decreasing granularity in $\eta$, starting with
$\Delta\eta\times\Delta\phi= 0.087\times5^{\circ}$ at $|\eta| = 1.3$ to
$\Delta\eta\times\Delta\phi= 0.35\times10^{\circ}$ at $|\eta|=3.0$. The gap
to the beam pipe is closed by a steel/quartz-fiber forward calorimeter with
typical granularity of $\Delta\eta\times\Delta\phi\approx0.175\times
10^{\circ}$. All CMS calorimeters are non-compensating and require specific
calibrations for hadrons and jets. More details on the CMS detector can be
found in \cite{cms:tdr}.

The ATLAS calorimeter system features a central highly granular
electromagnetic liquid argon/lead accordion calorimeters with a pointing
readout geometry. Cell sizes vary from $\Delta\eta\times\Delta\phi= 0.003
\times0.1 $ in the first depth segment to $\Delta\eta\times\Delta\phi=
0.025 \times0.025 $ in the second, and $\Delta\eta\times\Delta\phi= 0.5
\times0.025 $ in the third depth sampling. Depending on $\eta$, the depth
varies from $26$ to $36$ $X_{0}$. The central electromagnetic calorimeter is,
contrary to the setup in CMS, located outside the solenoid magnet. The
calorimeter covers a region of $|\eta|<1.475$.

The electromagnetic endcap calorimeter in ATLAS features a liquid argon/lead
accordion-type calorimeter with the absorber folded like a Spanish fan. It
covers $1.375<|\eta|<3.2$, with three depth samplings up to $|\eta|=2.5$, and
two in $2.5<|\eta|<3.2$. The lateral size of the pointing cells is $\Delta
\eta\times\Delta\phi= (0.003-0.006) \times0.1$\ in the first sampling,
$\Delta\eta\times\Delta\phi= 0.025 \times0.025$\ in the second, and
$\Delta\eta\times\Delta\phi= 0.05 \times0.025$\ in the third sampling, all
up to $|\eta|=2.5$. Beyond that, the two remaining samplings of the
electromagnetic endcap calorimeter have $\Delta\eta\times\Delta\phi= 0.1
\times0.1$.

Hadronic calorimetry in ATLAS is provided by the steel/scintillator tile
calorimeter in $|\eta|<1.7$, which has three samplings with quasi-projective
cells of $\Delta\eta\times\Delta\phi= 0.1 \times0.1$\ in the first two, and
$\Delta\eta\times\Delta\phi= 0.2 \times0.1$\ in the last depth sampling.
The endcap hadronic calorimeter is a liquid argon/copper parallel plate
calorimeter with four depth samplings and quasi-projective cells with
$\Delta\eta\times\Delta\phi= 0.1 \times0.1$\ in $1.5<|\eta|<2.5$, and
$\Delta\eta\times\Delta\phi= 0.2 \times0.2$\ in $2,5<|\eta|<3.2$. The net
result is that the ATLAS calorimeters have six or seven depth samplings for
hadrons, depending on the particle direction.

The ATLAS forward calorimeter covers $3.1<|\eta|<4.9$ and consists of three
modules. The first (electromagnetic) module is a liquid argon/copper
calorimeter featuring tubular thin gap electrodes. The two hadronic modules
have a tungsten absorber with a similar electrode geometry. The readout of the
forward calorimeter is organized in non-projective rectangular cells, with an
approximate cell size of $\Delta\eta\times\Delta\phi= 0.2 \times0.2$. The
total number of calorimeter channels in ATLAS is close to $200,000$. And like
in CMS, all ATLAS calorimeters are non-compensating as well. For more details,
see \cite{atlas:tdr}.

\subsubsection{Calorimeter Jet Basics} \label{sec:calojetbasics}

CMS uses projective calorimeter towers on a grid of $\Delta\eta\times
\Delta\phi= 0.1 \times0.1$\ as input to jet finding. Only towers with
$p_{T} > 1$ GeV$/c$ are considered. 
The towers correspond to massless 4-vectors by definition,
\textit{i.e.}, their kinematic contribution is fully specified by their
transverse momentum $p_{T}$, their rapidity $y$, and azimuth $\phi$~\cite{cms:tdr:2}.
\begin{figure}[ptb]
\centering
\includegraphics[height=18cm]{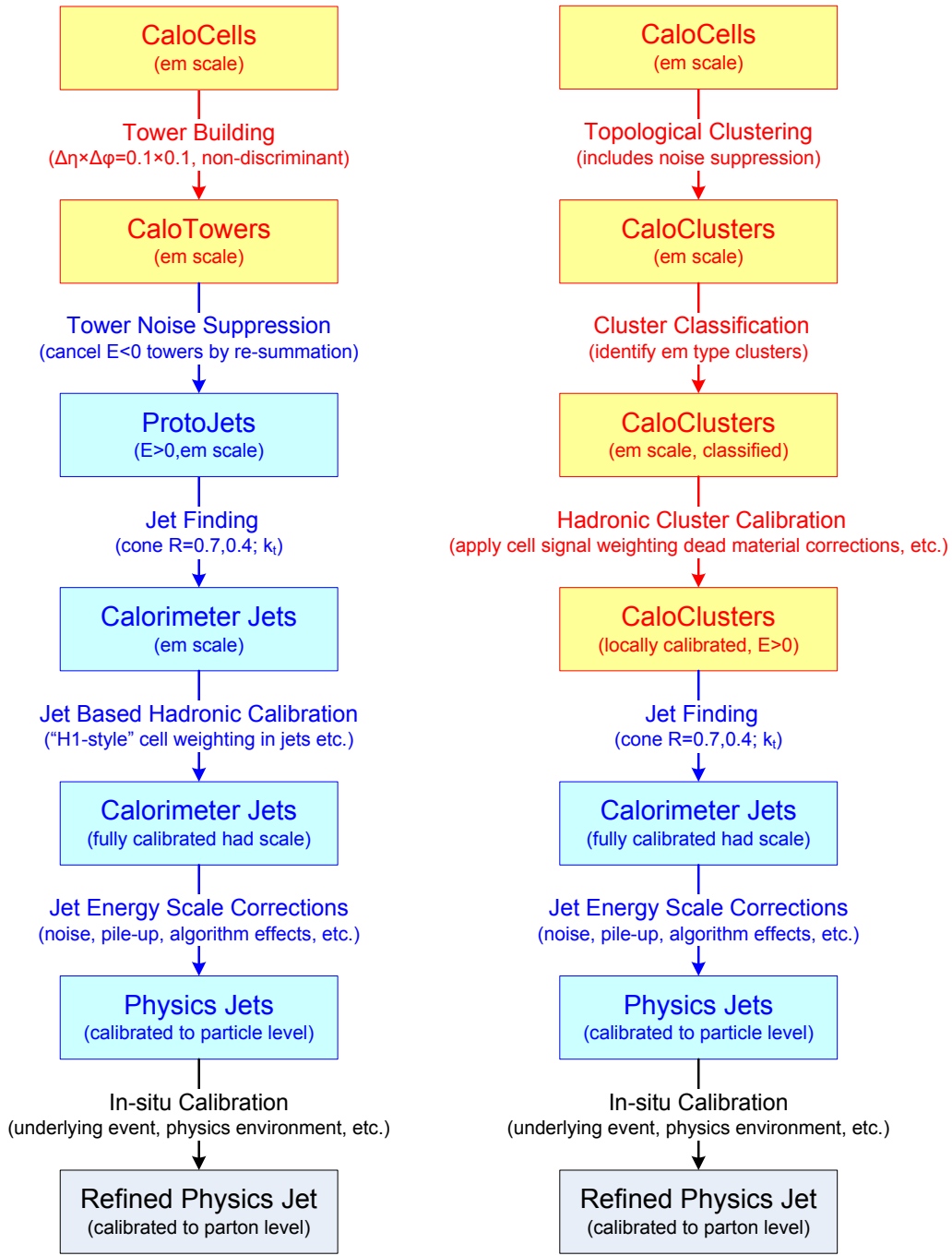} \caption{Jet
reconstruction flow in ATLAS. The left diagram shows the sequence for the
tower based reconstruction, while the right diagram shows the cluster based
reconstruction sequence. }%
\label{fig:jetflow}%
\end{figure}
In ATLAS, two different calorimeter signal definitions are used as
input for jet reconstruction. As in CMS, projective cell towers with
$\Delta\eta\times\Delta\phi= 0.1 \times0.1$ are used, but without any
restriction on the actual value of $p_{T}$. This means it is possible to have
towers with a non-physical four-momentum like $E < 0 \Rightarrow p_{T} < 0$, 
with $\eta$, $\phi$, and $m = 0$ fixed by location and/or definition. The 
negative tower
signal is generated by electronic noise as well as by signal fluctuations from
the MB (pile-up) events.

On the other hand, the negative energy towers cannot be accepted by the jet
finders, as the four-momentum recombination of protojets (the towers) requires
legitimate, physically allowed kinematic variables. This introduces the need
for noise compensation or suppression. In ATLAS, a noise compensation based on
the pre-summation of towers into protojets has been introduced, which is based
on the idea that negative signal towers are merged with neighboring positive
signal towers until the total energy in the corresponding protojet is above
$0$. Negative towers without any close-by positive signal towers are dropped.

ATLAS also uses three-dimensional topological calorimeter cell clusters as
input to jet finding. The basic idea of this signal definition is the attempt
to reconstruct calibrated \emph{energy blobs} in the calorimeter using the
energy flow and shower development correlation between neighboring cells. The
result makes optimal use of the fine granularity in ATLAS, especially of the
longitudinal segmentation. The clustering algorithm is based on cell signal
significance, as measured by the signal-to-noise ratio $S/\sigma$, where the
noise $\sigma$ can include pile-up fluctuations. 
Different thresholds are requested for cells seeding a cluster
(primary seeds, $\left|S/\sigma\right|>4$), cells defining the growth
of a cluster (secondary seeds, $\left|S/\sigma\right|>2$), and cells
to be included because they are direct neighbors of any one of these
two seeds ($S/\sigma>0$). 
Note that large negative fluctuations can seed a cluster as well. This has been
introduced to have an average cancellation of positive noise contributions.
Naturally clusters with negative total signal are not used in jet
reconstruction, but negative signal cells within a cluster with positive total
signal contribute to the jet signal. The algorithm initially puts all
topologically connected cells into one cluster. In a second pass, the clusters
are analyzed with respect to local signal maxima. Clusters with more than one
local signal maximum are split, and the energy in the cells between the two
signal peaks is typically shared by the resulting two clusters~\cite{sven}.

Most clusters have measurable shapes with respect to location, longitudinal
and lateral extension, energy sharing in calorimeter compartments, etc.
Variables describing these shapes are used to fully calibrate the clusters to
a local hadronic energy scale. This procedure includes the attempt to first
classify each cluster with respect to the character of its generating
particle(s), followed by the application of hadronic weighting functions for
``hadronic-looking'' clusters. These functions are typically parametrized
using shape and location variables (\textit{e.g.}, the depth of cluster center
in the calorimeter) sensitive to hadronic shower development. Next,
corrections for dead material losses are applied to all kinds of clusters
found near cracks and dead material in the calorimeter. The final 
cluster correction then attempts to recover signal losses outside of clusters,
as introduced by the noise cuts used in the cluster formation algorithm 
discussed above. Note that none of these procedures use a jet context at all.
All calibrations and
corrections here are derived from single pion and electron signals
alone, and can thus be bench-marked with experimental test-beam data.

Fig.~\ref{fig:jetflow} shows a schematic overview on calorimeter jet 
reconstruction from towers and clusters in ATLAS, respectively. Both 
calorimeter signal definitions have their specific advantages and 
disadvantages when used in jet reconstruction. 
Any particular choice of the calorimeter signal has serious effects on
jet reconstruction and calibration, though, as discussed more in 
the following sections.  

\subsubsection{Calorimeter Jet Calibration}

Several jet calibration models are under investigation in ATLAS. The most
commonly used model is based on a modified cell signal weighting technique,
following the original suggestions from H1~\cite{Adloff:1997mi}. It can
be applied to jets from towers as well jets from uncalibrated topological
clusters.

In a first step, the cell content of the tower or cluster jet is retrieved.
Then, a cell signal weighting function $w_{c}$
is applied for each cell, depending on
its location\footnote{The cell location is typically indicated by
calorimeter module and sampling identifiers, together with a cell index, 
rather than absolute coordinates, for example.}
$\vec{x}_{\mathit{cell}}$ and the cell signal density
$\rho_{\mathit{cell}} = E_{\mathit{cell}}/V_{\mathit{cell}}$, where
$E_{\mathit{cell}}$ is the cell signal on an initial (electromagnetic)
energy scale, and $V_{\mathit{cell}}$ is the physical volume of the
cell.
Finally, the jet
kinematics is re-calculated using the now calibrated cell signals:
\[
P_{\mathit{jet}} = \sum_{\mathit{cells} \in\mathit{jet}} w_{c}(\vec
{x}_{\mathit{cell}},\rho_{\mathit{cell}}) P_{\mathit{cell}},
\]
with $P_{\mathit{cell}} = (E_{\mathit{cell}},\vec{p}_{\mathit{cell}})$ and
$|E_{\mathit{cell}}| = |\vec{p}_{\mathit{cell}}|$. Note that in this scenario 
the direction of the original jet can change.

The weights can be determined using the particle ``truth'' jet in simulations
by adjusting them such that the reconstructed jet energy on average is
identical to the matched particle jet energy. In this scenario the weights
reflect all corrections needed to reconstruct the particle jet. In particular,
energy lost in particles which do not reach the calorimeter due to the
magnetic field is compensated. In the same sense, upstream energy losses in
dead material are corrected by this normalization choice.

Using the calibrated topological clusters as input for jet finding makes the
cell signal-based calibration in the jet context, as discussed above, 
obsolete. 
In this approach the cluster signals are already fully calibrated in a less 
biased context, as for example the jet algorithm choice does not enter into
the hadronic calibration at this level (see discussion at the end of 
section~\ref{sec:calojetbasics}).
To first order, the reconstructed jet kinematic is then given by the sum of
the calibrated cluster four-momenta. Note that the cluster calibration does not
account for energy losses due to the magnetic field and some
dead material, as obviously only energy losses with some correlation to
the shape or magnitude of 
a nearby cluster signal can be corrected. Especially energy lost away from 
any cluster, and energy losses due to the loss of small signals by 
the implicit noise suppression applied in the cell clustering algorithm,
still need to be corrected for in the (larger) jet context. An attempt is
under study in ATLAS to correct for these effects jet by jet, possibly using 
measurable
jet shapes and detailed cluster information, but not all sensitivities and 
useful variables are yet fully understood to achieve a particle level
calibration this way.

Important validation signals for jet calibration at the LHC, as at the
Tevatron, are prompt photons and hadronic $W$ decays. For systems which have a
photon balancing one or more jet(s), corrections can be extracted from the
$p_{T}$\ balance. Uncertainties in this procedure mostly arise from initial
and final state radiation, and the underlying event activity, all of which may
limit the applicability of these corrections in different collision
topologies. The $W$ mass in $W\rightarrow q\bar{q}$ can be used for the same
purpose. Again, care is required when applying the corrections derived from
the mass constraint to other physics topologies. At the LHC, most of the
hadronically decaying $W$s in the recorded data will be reconstructed in the
context of $t\bar{t}$ events, which are very busy final states with
potentially large amounts of energy \textquotedblleft
accidentally\textquotedblright\ scattered into the jet(s). Also, the $W$ is
color-disconnected from the rest of the event, which changes the general
$p_{T}$\ flow around its decay jets.

\subsubsection{Use of Tracks in Jets}

Reconstructed tracks from the inner tracking detectors in both ATLAS and CMS,
in principle, can be used to better calibrate and characterize a given
reconstructed jet. \ Classical energy flow based reconstruction techniques
combine a track with the calorimeter response (typically one cluster) and make
use of the feature that the track provides much better energy resolution than
the calorimeter for lower energy particles. \ The application of these
techniques promises considerable improvement in the kinematic reconstruction
of a single isolated particle like an electron or $\tau$; the application to
jets is still under study. In particular, the \textquotedblleft
subtraction\textquotedblright\ of the charged response from the calorimeter
jet signal is much more challenging due to the already mentioned overlap of
showers and generally high tracker occupancy in the jet case. There are
indications, however, that the fraction $f_{\mathit{trk}}$ of the jet energy 
carried by (reconstructed) charged
tracks into a jet is already a useful variable to refine the jet calibration,
even without individual track/cluster matching. If $f_{\mathit{trk}}$ is large
for a given jet, the corresponding calorimeter response has a larger 
contribution from hadronic showers and may deserve additional calibration
corrections to improve the jet energy scale, especially with a precision 
requirement at the level of $1\%$.

\subsubsection{Jet Algorithms}

Both ATLAS and CMS use the iterative seeded cone jet finder with
$R_{\mathit{cone}}=0.4$ (ATLAS) and $R_{\mathit{cone}}=0.7$ (ATLAS and CMS),
and the $k_{T}$\ jet finder with $D=0.4$ and $D=0.6$ (ATLAS), and $D=1$ in
CMS. Other jet algorithms are generally available and under study in
particular with respect to certain physics needs (see the discussion in
Section \ref{sec:Sparty}). The software is implemented such that exactly the
same code runs on all possible input objects (partons and hadrons from Monte
Carlo generators, detector signals like clusters, towers, reconstructed
tracks, etc.), as long as those represent a full four-momentum measure.

\subsection{Jet Signal Characteristics at the LHC}

\begin{figure}[ptb]
\centering
\includegraphics[width=14cm]{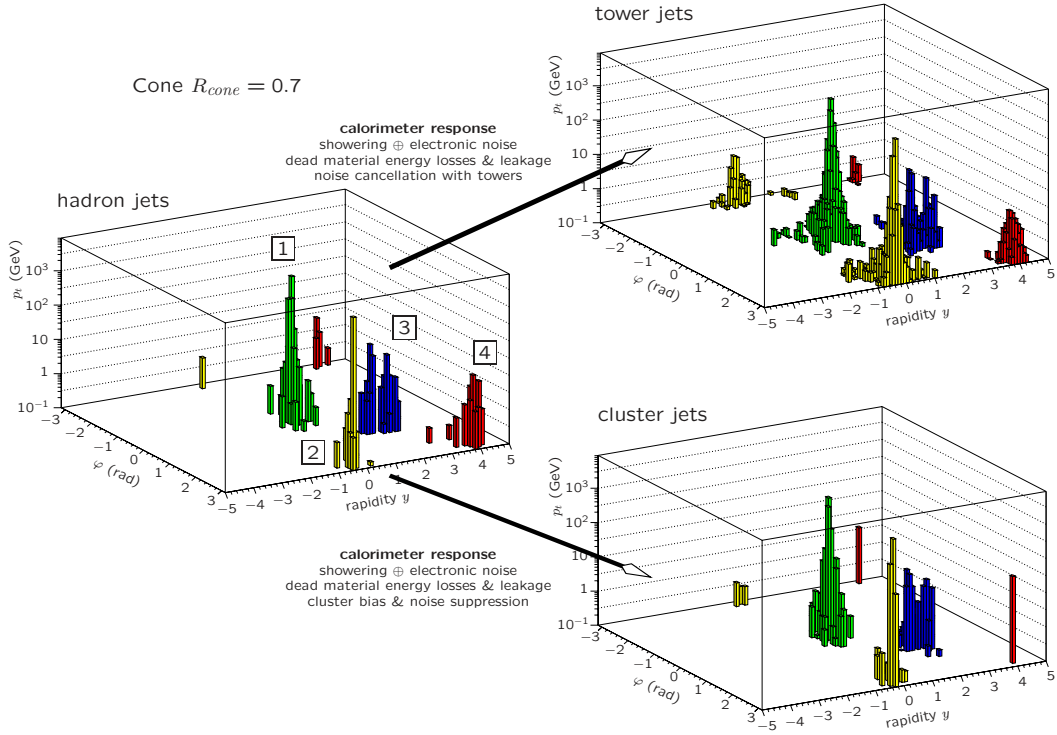}
\caption{A simulated QCD $2\rightarrow2$ event with two hard cone jets ($1$
and $2$) with energies around $2$ TeV each and two softer jets with energies
around $20$ GeV ($3$ and $4$) in ATLAS. Same colored bins belong to the same
jet. The change of the jet shape clearly depends on the calorimeter signal
definition. }%
\label{fig:jetshape}%
\end{figure}

The jet topology as it unfolds in the calorimeters is basically driven by the
combination of calorimeter
absorption characteristics, the chosen
signal definition and jet algorithm choice. This can be seen in the Pythia
event in
Fig.~\ref{fig:jetshape}. The tower picture of this particular final state
shows rather large jets in general, especially when compared to the 
hadron jet. This is partly a consequence of the re-summation discussed above,
but also due to the fact that towers especially in the endcap and forward
regions are filled with signals more generated by the lateral shower
development (see jet $4$ in Fig.~\ref{fig:jetshape}) than the 
hadron energy flow. Topological clusters, on the other hand, collect spatially
distributed cell signals, as can again be seen very well in jet $4$. In the
cluster picture this jet is very collimated in rapidity, and consists of only
two clusters. This reflects the rather coarse cell readout granularity of the
ATLAS forward calorimeters, which suppresses cluster splitting due to lack of
resolvable (lateral) signal structures. In the highly granular central and
endcap regions the cluster jet shapes match very well the hadron jet shape, as
can be seen from jets $1$ to $3$ in this figure. \begin{figure}[ptb]
\centering
\begin{tabular}
[c]{cc}%
\includegraphics[height=12cm]{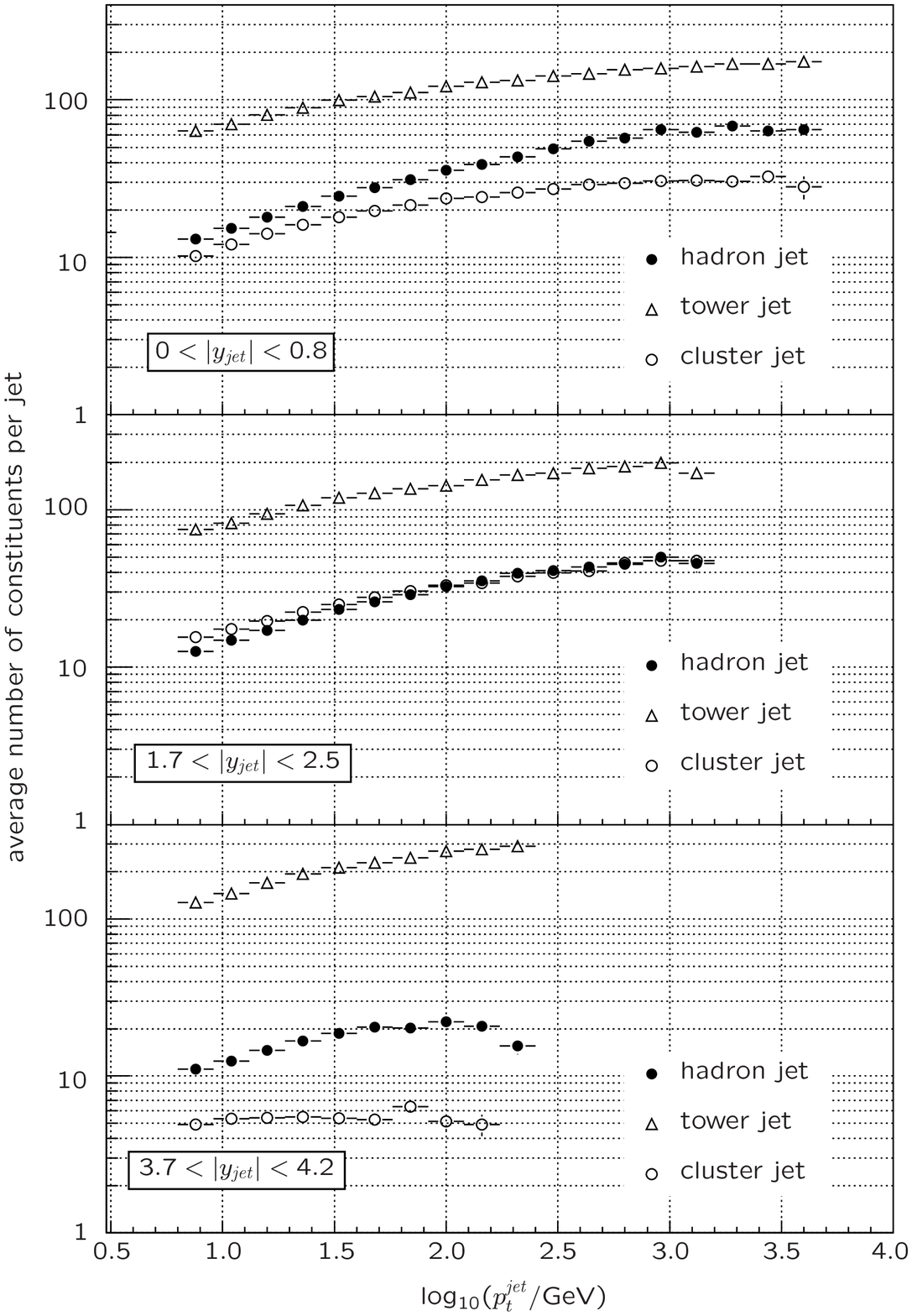} &
\includegraphics[height=12cm]{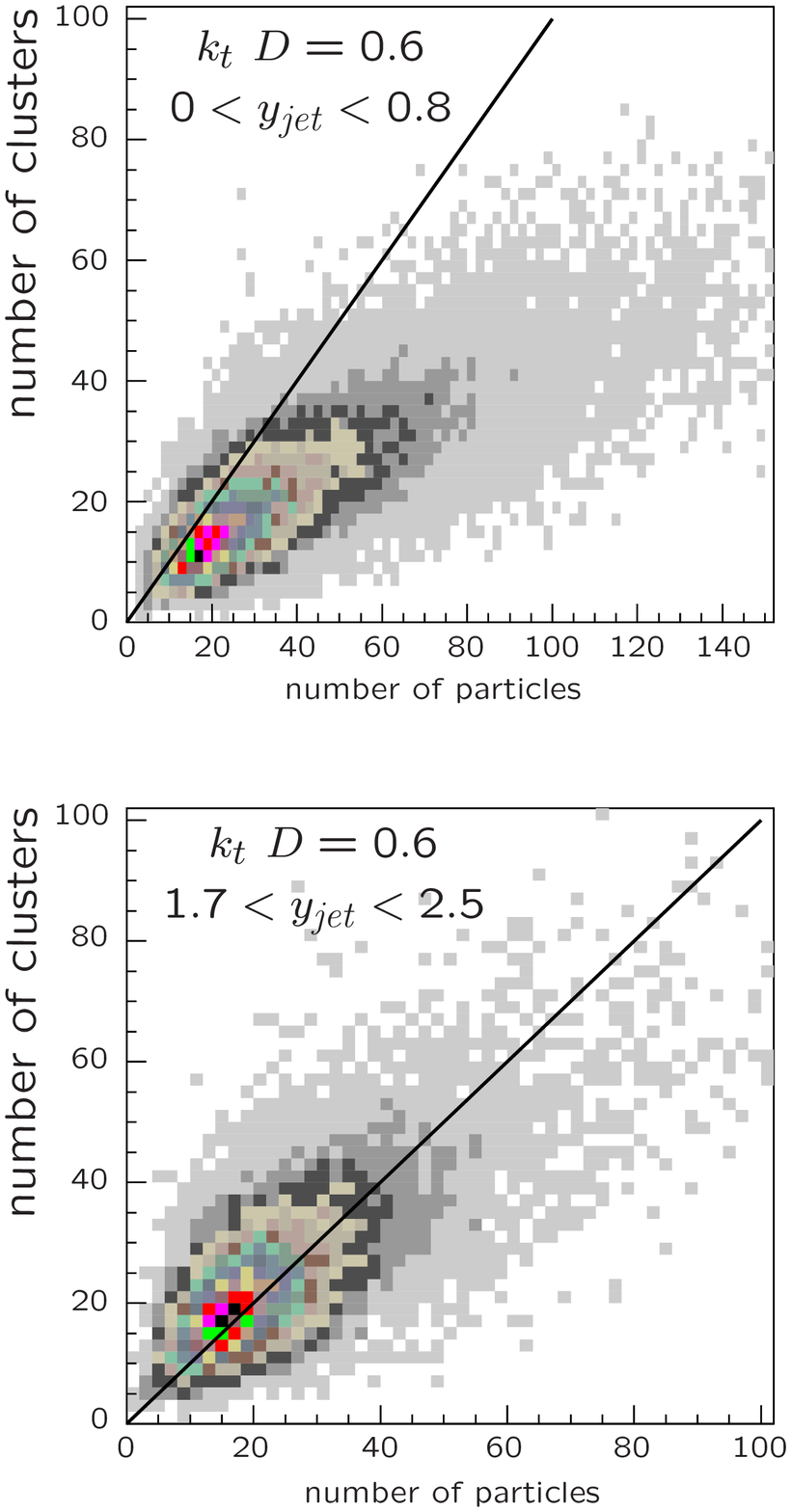}
\end{tabular}
\caption{The number of constituents for hadron, cluster, and tower $k_{T}%
$\ jets ($D=0.6$) as a function of the jet $p_{T}$\ for simulated central
events in ATLAS, in various regions of rapidity $y$ (left). The right figure
shows the number of clusters versus the number of particles in matched cluster
and hadron jets in the central and endcap region, from the same simulated
data. }%
\label{fig:nc}%
\end{figure}

The particular choice of the calorimeter signal definition to be used in jet
reconstruction affects the ability to reconstruct some of the jet kinematics,
such as the jet mass.
For example, for cone jets with $R = 0.7$ made from calorimeter towers with
$\Delta\eta\times\Delta\phi= 0.1 \times0.1$, the number of constituents
$N_{c}$ is given by
\[
N_{c} \approx\frac{\pi R_{\mathit{cone}}^{2}}{\Delta\eta\times\Delta\phi=
0.1 \times0.1 } \approx154,
\]
independent of the jet direction and energy. Calorimeter signals such as the
topological clusters, which are more sensitive to the hadron level composition
of the jet, typically generate jets with $N_{c}$ at least indirectly related
to the number of incoming particles (see Fig.~\ref{fig:nc}). Naturally, the
incoming particle energy flow is convoluted with possibly overlapping shower
developments and distributed onto a finite readout granularity. The relation
between cell sizes and electromagnetic and hadronic shower sizes puts
limitations on the reconstruction of the original incoming particles, and
defines the image of the jet in the calorimeter. Fig.~\ref{fig:nc} indicates
that in the central region of ATLAS the cell sizes, even though small in
$y/\eta$ and $\phi$, are comparably big with respect to shower sizes, while
in the endcap region the cell sizes get sufficiently small, thus improving the
structural resolution power of the cluster jets.

\begin{figure}[ptb]
\centering
\includegraphics[width=12cm]{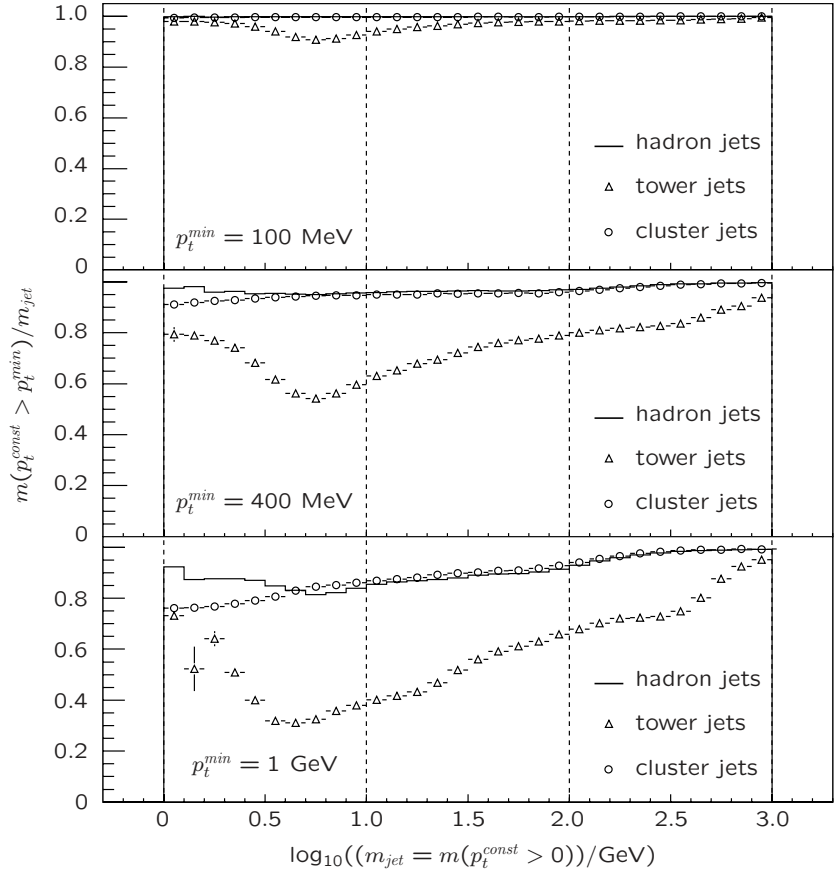}
\caption{The change of the jet mass
for a constituent selection based on their transverse momentum $p_{T}%
^{\mathit{const}}$ ($p_{T}^{\mathit{const}}>p_{T}^{\mathit{min}}$), for $k_{T}%
$\ jets from particles, towers, and clusters ($D=0.6$), as a function of the
jet mass calculated from all constituents.}%
\label{fig:massevol}%
\end{figure}


One of the more interesting jet variables is its mass. The ability to
reconstruct this mass within reasonable precision has considerable impact on
the general reconstruction of heavily boosted systems like the top quark,
where all final state decay products may end up within a typical jet cone. In
this case the jet mass is the only reconstructable observable giving any
indication of the nature of the decaying system. The mass reconstruction is
still nearly perfect at the hadronization level, at least if all final state
particles belonging to a given parton are efficiently collected by the chosen
jet algorithm. Reconstructing the mass from calorimeter signals is much more
challenging in that not only the showering and the resulting signal overlap in
the calorimeters smear the mass measurement considerably, but also the fact
that the solenoidal fields in front of the calorimeters ($\sim2$ T in ATLAS,
$\sim4$ T in CMS) bend charged particles with $p_{T} < (400-800)$ MeV$/c$ away
from the detectors, \textit{i.e.} outside the jet cone. In addition, the
unavoidable amount of dead material typically introduced by the inner
detectors and their services as well as calorimeter support structures and
cryostat walls (ATLAS only), can significantly reduce the low energy photon
signal. This effect has been addressed in a brief study for ATLAS, where the
jet mass variation $\delta m/m$ introduced by excluding jet constituents below
certain thresholds, starting from $100$ MeV up to $2$ GeV, is calculated from
QCD di-jet simulations for particles, and the corresponding simulated
calorimeter tower and cluster signals with the same cuts applied.
Fig.~\ref{fig:massevol} summarizes the results of this study, which indicate
that the cluster signals follow the effect at particle level quite well, if
compared to tower jets. \begin{figure}[ptb]
\centering
\includegraphics[width=0.8\textwidth]{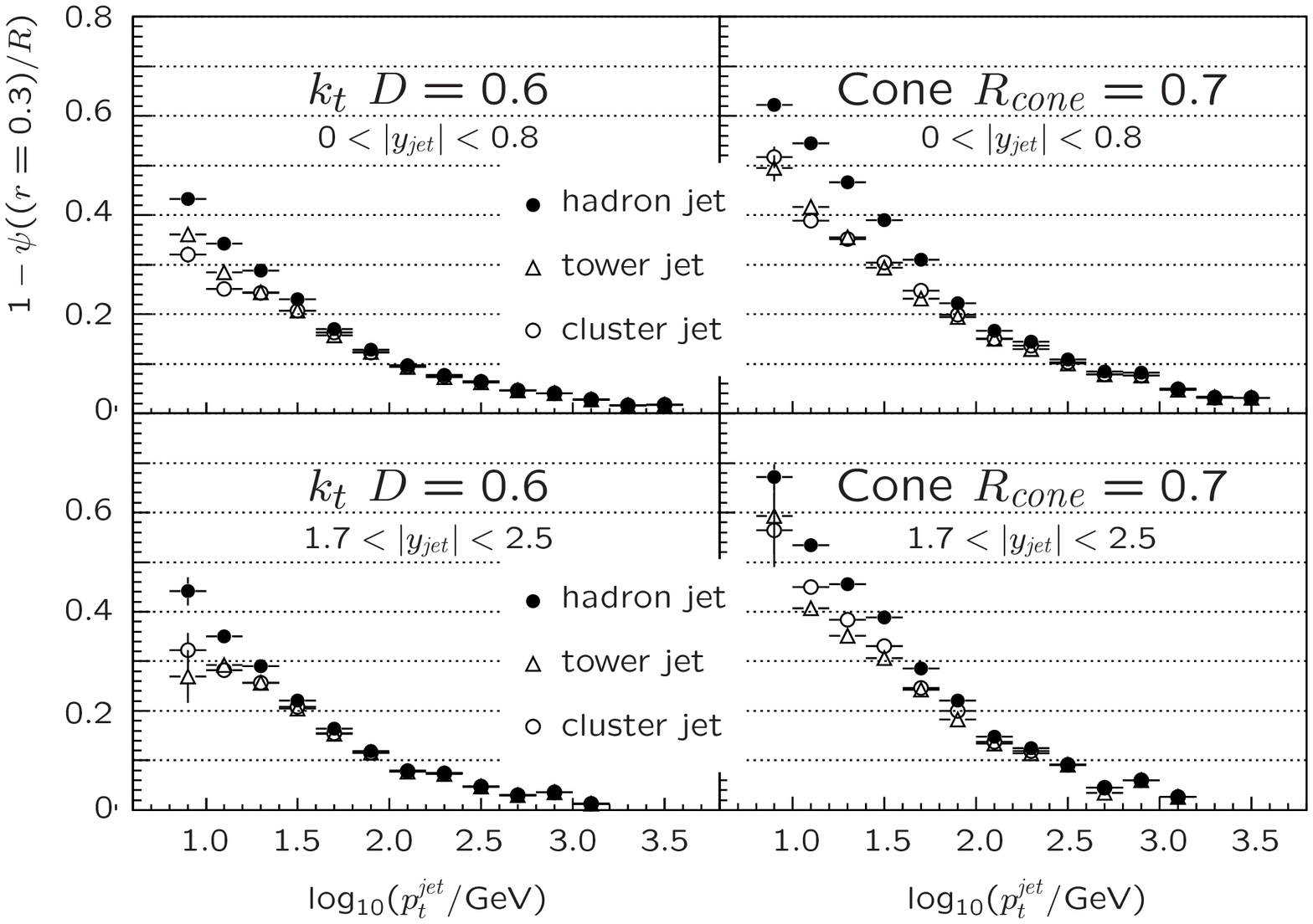}
\caption{Estimates from simulations for the ATLAS detector for the energy
fraction outside of a cone with $r = 0.3$ around the jet axis, as a function
of the jet $p_{T}$\ for $k_{T}$ (left) and cone (right) jets, in two different
regions of rapidity $y$, for jets in QCD $2 \rightarrow2$ processes.}%
\label{fig:rho}%
\end{figure}

Jet reconstruction is therefore in general affected by magnetic field effects,
the upstream energy losses from dead materials, showering, leakage,
calorimeter regions of low efficiency (cracks), and the underlying and 
pile-up event activity. Any particular sensitivities of a given jet algorithm 
to any of these effects can be enhanced or suppressed by a specific
calorimeter signal choice, \textit{e.g.} towers or clusters.
The jet energy scale calibrations and corrections applied to recover the
corresponding energy losses 
are as much as possible factorized for better control of systematic 
uncertainties.
The magnitude of the contribution of any of these effects to the
jet energy scale and jet shape reconstruction varies, depending on the 
combination of the calorimeter signal choice and the chosen jet algorithm. 

One example for
an experimentally accessible observable reflecting some of these sensitivities
is the jet
energy density measure $\psi(r)$, which is the fraction of energy contained in
a cone of radius $r$ within a jet. Fig.~\ref{fig:rho} shows that towers and
clusters in the ATLAS calorimeter are expected to
produce very similar densities for high
$p_{T}$\ jets, but show some differences for jets with $p_{T}\lesssim100$
GeV$/c$. \ Hadron jets at lower $p_{T}$\ are significantly broader,
\textit{i.e.}, have a larger fraction of the jet energy away from the center
of the cone, than either tower or cluster jets. \ In general the 
shape of $\psi(p_{T})$ in different rapidity regions changes, depending on 
the jet algorithm choice, in a similar way for all three kinds of jets.


\section{SpartyJet}

\label{sec:Sparty}

As we have emphasized throughout this review, jets, unlike photons or
electrons, are complex objects and the resultant reconstructed 4-vectors may
depend on the details of the jet clustering algorithm. Each algorithm has its
own strengths and weaknesses and a more robust understanding of the physics of
an event can be obtained by examining the result of reconstruction with more
than one jet algorithm~\cite{Seymour:2006vv}.

\begin{figure}[t]
\begin{center}
\includegraphics[width=14cm]{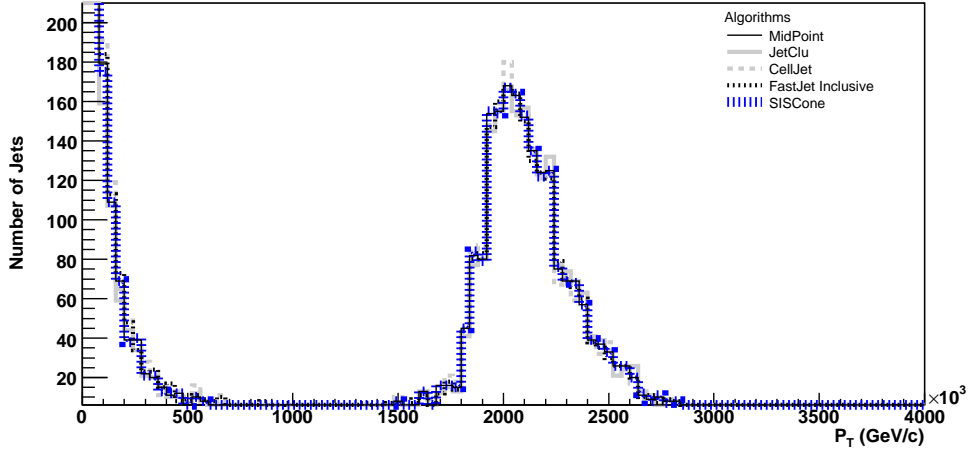}
\end{center}
\par
\vspace*{-0.5cm}\caption{ The inclusive jet cross section for the LHC with a
$p_{T,min}$ value for the hard scattering of approximately 2 TeV$/c$, using
several different jet algorithms with a distance scale ($D=R_{cone}$) of 0.7.
The first bin has been suppressed. }%
\label{fig:J8_pt}%
\end{figure}

Past experience at the Tevatron has been that only one algorithm is typically
used for any physics process, at least partially because of the limitations of
the analysis machinery. In order to foster a more flexible experimental
philosophy, a collection of jet routines was created
(SpartyJet)~\cite{SpartyJet} that can allow the reconstruction of the jets
from either data or Monte Carlo using multiple algorithms/varied parameters.
The routine makes use of the FastKt package for the $k_{T}$ jet algorithm and
the seedless cone algorithm SISCone in addition to other algorithms used by
experiments at the Tevatron and LHC. The program can run either in the ROOT~\cite{ROOT}
format, for example inside an ntuple, or it can run in stand-alone fashion on
a collection of 4-vectors. As an example, in Fig.~\ref{fig:J8_pt}, is shown
the results of running SpartyJet on a sample of Monte Carlo events generated
for the ATLAS experiment. The sample consists of dijet events with a
$p_{T,min}$ of approximately 2 TeV$/c$ and the inputs to SpartyJet are the topological
calorimeter clusters discussed in the previous section. The cone algorithms
are run with a cone radius of 0.7 and a split/merge criterion of 0.75. The
$k_{T}$ algorithms are run with a $D$ parameter of 0.7 as well. The dijet
character of the events can be seen by the clear peak at approximately 2
TeV$/c$, and the impact of hard gluon radiation off the initial and final states
can be observed in the sizeable lower tail at lower transverse momenta. The
jets found by the 2nd pass algorithm for the Midpoint algorithm can also be
observed at the lower transverse momentum values. Note that on this plot the
cone and $k_{T}$ algorithms with similar scale parameters give similar results
for the cross section. Any differences need more detailed comparisons to
become apparent.

\begin{figure}[t]
\begin{center}
\includegraphics[width=14cm]{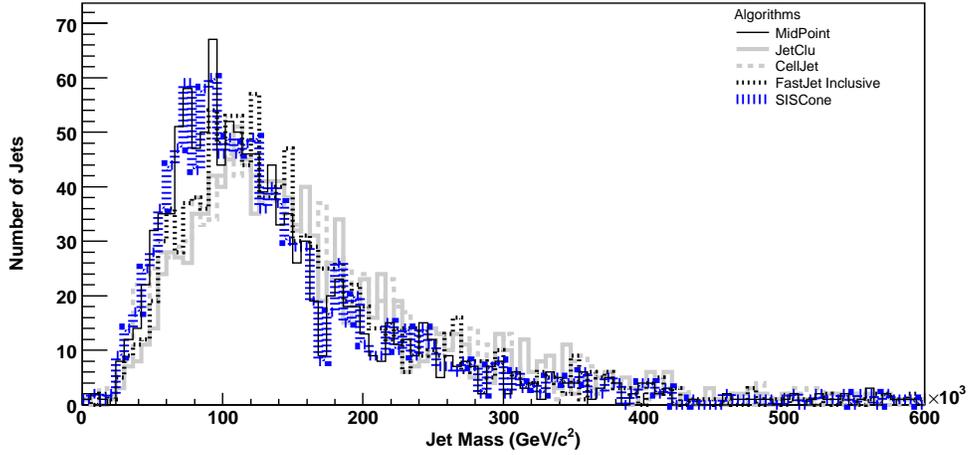}
\end{center}
\par
\vspace*{-0.5cm}\caption{ The jet mass distributions for an inclusive jet
sample generated for the LHC with a $p_{T,min}$ value for the hard scattering
of approximately 2 TeV$/c$, using several different jet algorithms with a
distance scale ($D=R_{cone}$) of 0.7. The first bin has been suppressed. }%
\label{fig:J8_mass}%
\end{figure}

Another interesting variable to plot using SpartyJet is the distribution of
jet masses. As discussed in Section~\ref{sec:jetmass}, the mass of 
typical QCD (gluon or light quark) jets is generated primarily by
perturbative gluon emission.  This is to be contrasted with a jet arising from a heavy quark (such as a top quark) that also has an
intrinsic mass from the heavy quark.
The distribution of jet masses from a typical QCD event sample
(restricting the transverse momentum range to $1.8-2.2$ TeV$/c$) is
shown in Fig.~\ref{fig:J8_mass}. There is a Sudakov suppression of low
jet masses, which can arise only if there is little or no gluon
radiation from the short-distance final state partons.  At jet masses
above
the peak (here at approximately 125 GeV$/c^{2}$) the jet mass distribution
falls slowly, roughly between $1/m$ and $1/m^{2}$, with the average
jet mass at a value above the peak (approximately 150 GeV$/c^{2}$ in
this sample).  There is also a suppression of jets with large masses
due to the tendency of the jet algorithms to split jets in which the
energy is widely dispersed. Note that for these very high $p_{T}$ jet
events
that a jet mass of the order of 175 $\mbox{GeV}/c^{2}$ (purely from
gluon emission) is not uncommon, so this is a caveat for the naive use
of the mass of a jet to search for highly boosted top quarks.

\begin{figure}[t]
\begin{center}
\includegraphics[width=14cm]{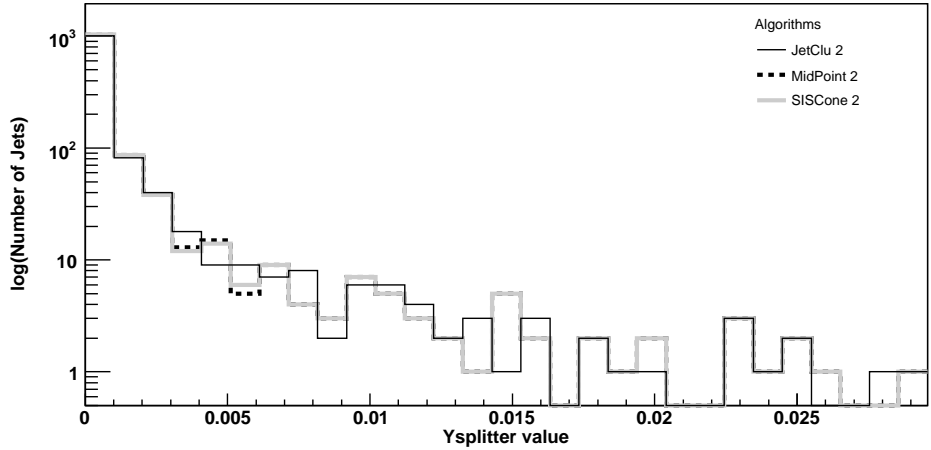}
\end{center}
\par
\vspace*{-0.5cm}\caption{ The $y$-scale distributions
for the jets from the $1.8-2.2$ TeV$/c$ jet sample to be split into two subjets.
Several different jet algorithms with a distance scale ($D=R_{cone}$) of 0.7
are used. }%
\label{fig:ysplit}%
\end{figure}

One can gain further information about the origin of a jet by
examining the $y$-scale~\footnote{
$y_2=\min(p^2_{T,1},p^2_{T,2})\cdot d_2^2/p^2_{T,jet}$, where $p^2_{T,1} and
p^2_{T,2}$ are the transverse momentum values of the two subjets and
$d_2$ is the distance scale at which the jet is divided into two
subjets.} (using the $k_T$ algorithm) at which the jet can
be split into two subjets. This ability has been implemented into
SpartyJet by the use of the $y$-splitter
routine~\cite{Butterworth:2002tt}. This scale will tend to be larger
for highly boosted massive objects (like a high $p_T$ $W$ or top quark) than
for QCD jets at the same transverse momentum.  For example, for a
boosted $W$, the $y$ scale for resolving the $W$-jet into two subjets
should be on the order of $m_W/p_{T,jet}$, while for jet structures produced
by QCD radiation, the scales should be much smaller.
The $y$-scale distribution
for the jets from the $1.8-2.2$ TeV$/c$ jet sample to be split into
two subjets is shown in Fig.~\ref{fig:ysplit}. Low scales dominate as
expected for a QCD jet sample. 

\begin{figure}[thp]
\begin{center}
\includegraphics[width=14cm]{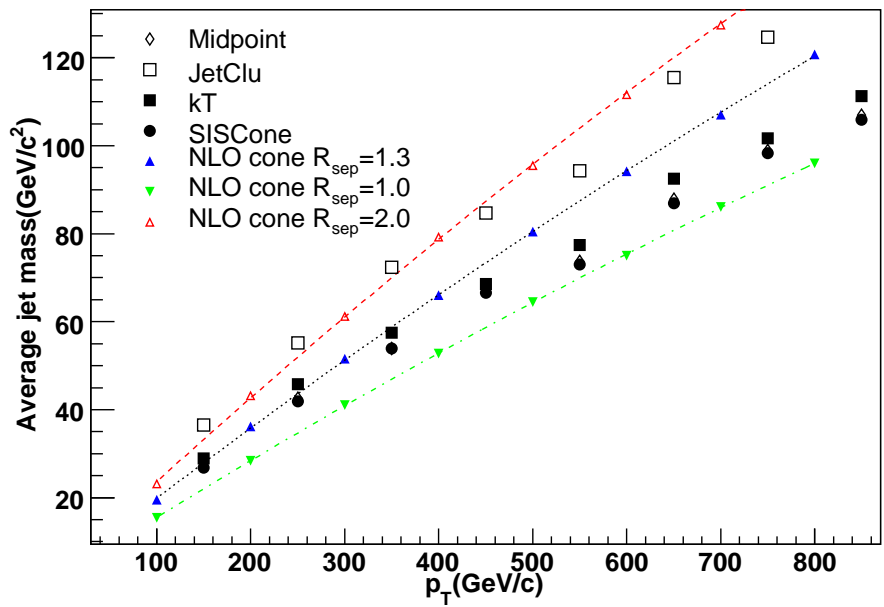}
\end{center}
\par
\vspace*{-0.5cm}\caption{ The average jet mass is plotted versus the transverse momentum of the jet using
several different jet algorithms with a distance scale ($D=R_{cone}$) of 0.7.
 }%
\label{fig:jet_mass}%
\end{figure}

The average jet mass is plotted in Fig.~\ref{fig:jet_mass} versus the
transverse momentum of the jet for several jet algorithms for an
inclusive jet Monte Carlo sample with transverse momenta from 100
GeV$/c$ to 1 TeV$/c$. The average mass increases roughly linearly with the
jet transverse momentum.  The reconstructed Monte Carlo jet mass
values are in reasonable agreement with the NLO perturbative
predictions discussed in Section~\ref{sec:jetmass}. Thus, the jet
mass, just as the jet shape, can be reasonably described by a NLO
partonic calculation, The average Monte Carlo jet mass at high $p_T$
tends to approach the NLO prediction carried out with the use of an
$R_{sep}$ value of 1 (compared to the canonical value of 1.3), perhaps
due to the impact of the very narrow jet profiles at these high
transverse momenta.  All of the jet algorithms result in a similar
average jet mass, although JetClu tends to give larger results, due to
the effects of ratcheting.  As discussed in Section~\ref{sec:jetmass},
the average jet mass is expected to scale as
$R\cdot p_T$, where $R$ is the size parameter for the jet.
It is interesting to note that variations in theoretical ($R_{sep}$)
and in experimental jet algorithms (ratcheting) have little impact on
the magnitude of the inclusive jet cross section, but do have a
noticeable effect on the jet mass.

\begin{figure}[thp]
\centering\leavevmode
\begin{tabular}
[t]{c}%
\subfigure[]{
  \includegraphics[width=0.49\hsize,bb=0 0 530
  365]{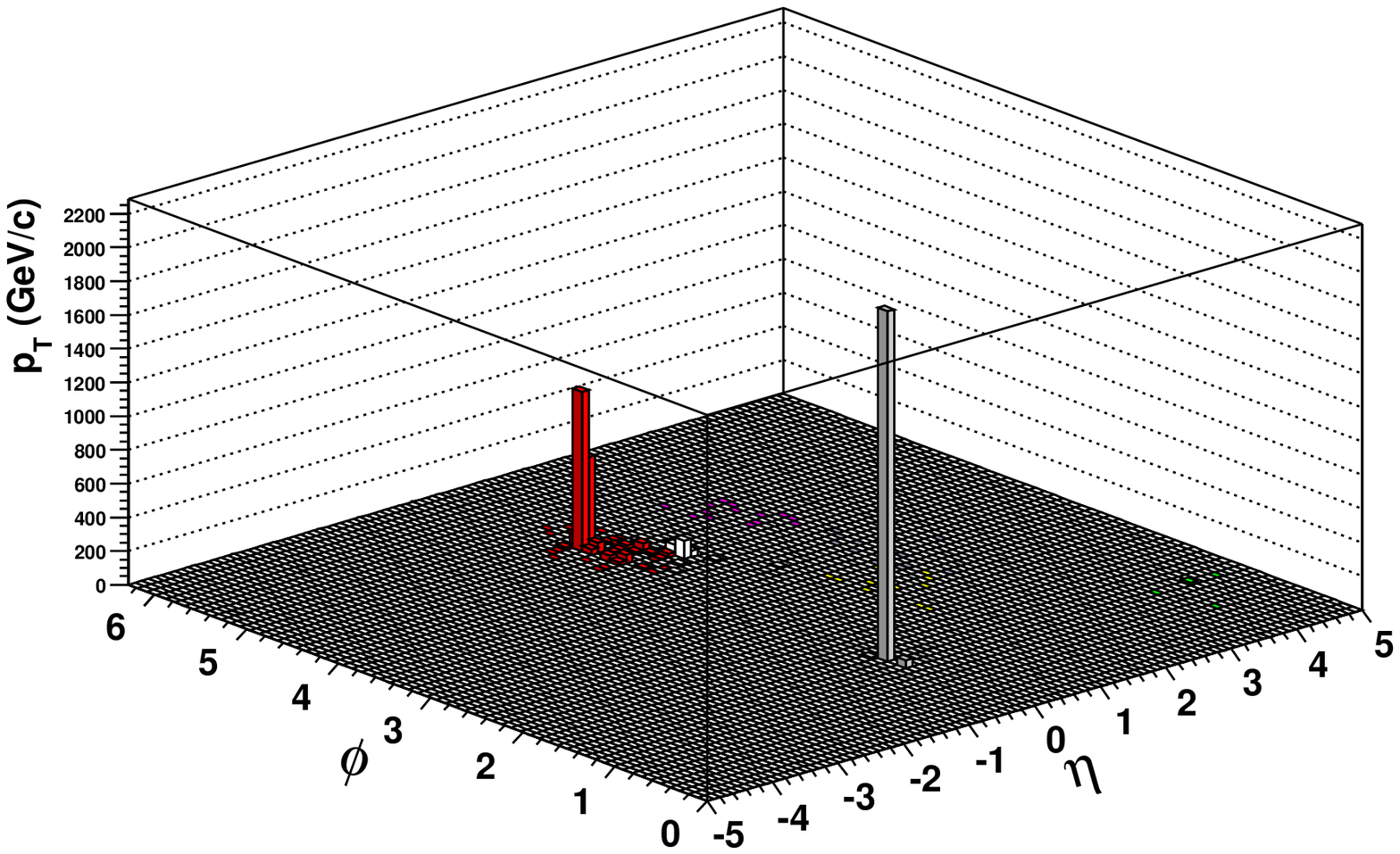}}
\subfigure[]{
  \includegraphics[width=0.49\hsize,bb=0 0 530
  365]{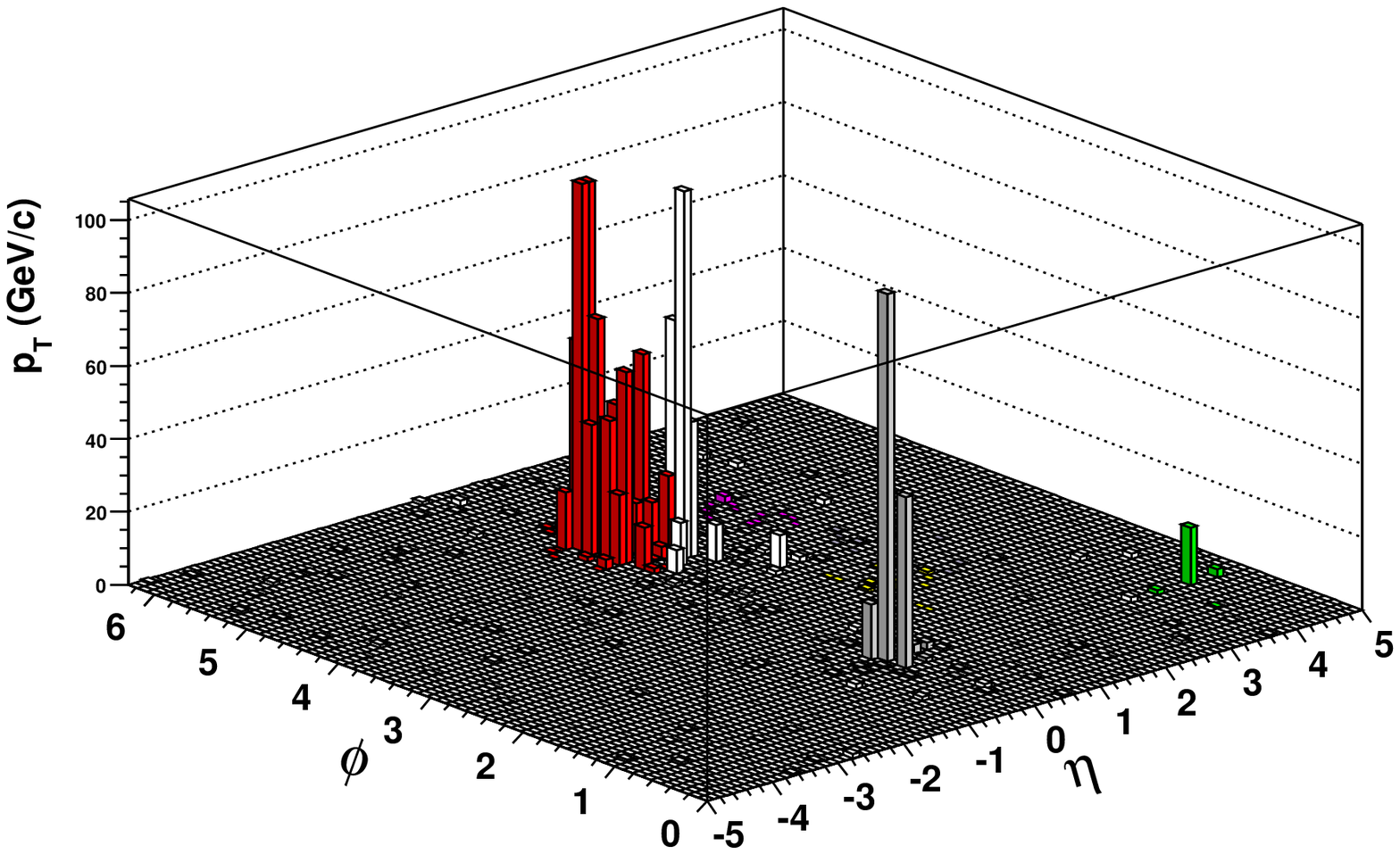}}
\end{tabular}
\vspace*{-0.5cm}\caption{(a) A lego plot for an event from the inclusive jet
sample generated for the LHC with a $p_{T,min}$ value for the hard scattering
of approximately 2 TeV$/c$. (b) A lego plot for the same event;
however, the $p_T$ scale has been cut off at 100 GeV$/c$ to give a
clear view of low $p_T$ jets. }%
\label{fig:lego}%
\end{figure}

A lego plot of a single event from the high $p_T$ jet sample is shown
in Fig.~\ref{fig:lego}(a).
Again, there is a clear dijet structure in which the jets have very collimated
cores but the presence of such a large scale in the event results in there
being a several extra jets of quite sizeable transverse momentum in their own
right. This becomes more apparent when we change the transverse momentum scale
in the lego plot as shown Fig.~\ref{fig:lego}(b). Note that the jet colored in
white is a 2nd pass jet, having been missed by the Midpoint algorithm in the
first pass due to the effects discussed earlier.




SpartyJet is currently in use by ATLAS, CDF and CMS.


\section{Conclusions}

Jets are present in nearly every final state measured at the Tevatron.
This will be true as well at the LHC, with the environment for jet
measurements being even more challenging.
Thus, it is crucial both to improve our understanding of the
subtleties of jet reconstruction as well as to continue the
development of new tools.  In this article, we have tried to remove some of
the mystique regarding
the measurement of jets at hadron-hadron colliders by pointing out the
connections between the theoretical predictions and the experimental
measurements, and the similarities and differences between the cone and
$k_{T}$ algorithms. In the process we have reviewed the history of jets
at the Tevatron, including both the experimental and theoretical  successes,
and the surprises and mis-steps. Looking to the future we have outlined the
issues expected to be important at the LHC.  
Specifically, we have discussed jet reconstruction and jet energy
calibration experience at the Tevatron experiments and the on-going work at
LHC experiments.  LHC experiments benefit from the experience at the
Tevatron (and also other experiments
such as the HERA experiments), and, due to excellent capabilities of the
ATLAS and CMS calorimeters, more advanced schemes are being explored such as
the use of topological calorimeter cell clusters as inputs to jet
clustering.  An essential payoff of measurements performed at the
Tevatron is the ability to accurately tune Monte Carlo event simulations to
improve the modeling of jets.  As we have discussed it is important that
these experimental measurements be presented at the hadron level to
facilitate the model tuning process.
These measurements must also be repeated in the new environment of the LHC
experiments.

In this context of past experiences and future expectations we have made
several recommendations that we feel will play an essential role in the
successful analysis of the data from the LHC.  These include:
\begin{itemize}
\item the use of a variety of jet algorithms for physics analyses with
continuous cross-checking of results
\item the use of 4-vector kinematics, including evaluation of the jet mass,
to characterize a jet
\item the use of seedless algorithms (or correction back to seedless) in
cone-based jet clustering
\item the correction (where possible) of jets back to the hadron level in
experimental analyses
\end{itemize}
In addition, we have presented a framework (SpartyJet) that facilitates the
use of multiple jet algorithms in both experimental and theoretical studies.

We close by applauding the 20 years of highly successful jet physics at the
Tevatron and looking forward to an equally exciting application of jets to
physics beyond the Standard Model at the LHC.


\section{Acknowledgements} We would like to thank Anwar Bhatti, Albert de Roeck, Lance Dixon, Rick Field, Kurtis
Geerlings, Craig Group, Aurelio Juste, 
Christophe Royon, Gavin Salam and Markus Wobisch for useful conversations and supply of
figures.
We also would like to thank the members of the ATLAS jet performance
working group and the members of the ATLAS simulation production team for 
providing the simulated data for this experiment. In particular, we are 
grateful for help from Walter Lampl, Chiara Paleari and Rolf Seuster,
especially concerning ATLAS jet reconstruction software issues.


\end{document}